\documentclass[fleqn,usenatbib]{mnras}

\usepackage[T1]{fontenc}

\DeclareRobustCommand{\VAN}[3]{#2}
\let\VANthebibliography\thebibliography
\def\thebibliography{\DeclareRobustCommand{\VAN}[3]{##3}\VANthebibliography}

\usepackage{threeparttable}
\usepackage{float}
\usepackage{subcaption}

\usepackage{graphicx}	
\usepackage{amsmath}	
\usepackage{amssymb}	

\newcommand{\usmg}{USMg\textsc{ii}}
  
\newcommand{\lya}{Ly$\alpha$ }

\newcommand{\zabs}{$z_{abs}$}
\newcommand{\kms}{$km s^{-1}$}

\newcommand{\HI}{\mbox{H\,{\sc i}}}

\newcommand{\OII}{\mbox{[O\,{\sc ii}]}}
\newcommand{\OIII}{\mbox{[O\,{\sc iii}]}}

\newcommand{\CIV}{\mbox{C\,{\sc iv}}}

\newcommand{\SiIV}{\mbox{Si\,{\sc iv}}}

\newcommand{\MgII}{\mbox{Mg\,{\sc ii}}}
\newcommand{\MgI}{\mbox{Mg\,{\sc i}}}

\newcommand{\FeII}{\mbox{Fe\,{\sc ii}}}

\newcommand{\CaII}{\mbox{Ca\,{\sc ii}}}

\newcommand{\Magiicat}{\textsc{MAGIICAT}}

\title[Host galaxies of \usmg\ at $z \sim 0.7$]{Host Galaxies of Ultra-Strong \MgII\ Absorbers at $z \sim 0.7$}

\author[Guha et al.]{
Labanya K. Guha,$^{1}$\thanks{E-mail: labanya@iucaa.in (LKG)}
Raghunathan Srianand,$^{1}$\thanks{E-mail: anand@iucaa.in (RS)}
\& Patrick Petitjean$^{2}$\thanks{E-mail: petitjean@iap.fr (PPJ)}
\\
$^{1}$IUCAA, Postbag 4, Ganeshkhind, Pune 411007, India\\
$^{2}$ Institut d'Astrophysique de Paris, Sorbonne Universit\'e and CNRS, 98bis boulevard Arago, 75014 Paris, France\\
}

\date{Accepted XXX. Received YYY; in original form ZZZ}

\pubyear{2015}

\begin{document}
\label{firstpage}
\pagerange{\pageref{firstpage}--\pageref{lastpage}}
\maketitle

\begin{abstract}

We report spectroscopic identification of the host galaxies of 18  ultra-strong \MgII\ systems (\usmg) at $0.6 \leqslant z \leqslant 0.8$. We created the largest sample by merging these with 20 host galaxies from our previous survey within $0.4 \leqslant z \leqslant 0.6$. Using this sample, we confirm that the measured impact parameters ($\rm 6.3\leqslant D[kpc] \leqslant 120$  with a median of 19 kpc) are much larger than expected, and the \usmg\ host galaxies do not follow the canonical $\rm W_{2796}-D$ anti-correlation. We show that the presence and significance of this anti-correlation may depend on the sample selection. 
The $\rm W_{2796}-D$ anti-correlation seen for the general \MgII\ absorbers show a mild evolution at low $\rm W_{2796}$ end over the redshift range $0.4 \leqslant z \leqslant 1.5$ with an increase of the impact parameters. Compared to the host galaxies of normal \MgII\ absorbers, \usmg\ host galaxies are brighter and more massive for a given impact parameter. While the \usmg\ systems preferentially pick star-forming galaxies, they exhibit slightly lower ongoing star-forming rates compared to main sequence galaxies with the same stellar mass, suggesting a transition from star-forming to quiescent states. For a limiting magnitude of $m_r < 23.6$, at least 29\% of the \usmg\ host galaxies are isolated, and the width of the Mg~{\sc ii} absorption in these cases may originate from gas flows (infall/outflow) in isolated halos of massive star-forming but not starbursting galaxies. We associate more than one galaxy with the absorber in $\ge 21\%$ cases where interactions may cause wide velocity spread.

\end{abstract}




\section{Introduction}
\label{sec:intro}
The formation and evolution of galaxies is one of the most fundamental problems in extragalactic astronomy. Galaxies evolution is thought to be governed by the joint action of accretion of pristine metal-poor gas from the intergalactic medium (IGM), in-situ star formation in the galactic disk, and feedback from star-formation-driven metal-rich galactic winds -- a complex process known as the `cosmic baryon cycle' \citep{Angles-Alcazar, Peroux2020b}. The interplay between galactic winds and gas accretion from the IGM creates a metal-enriched gaseous halo surrounding galaxies out to a few virial radii known as the circumgalactic medium (CGM). The kinematically complex, multi-phase CGM, which serves as an interface between the  galactic disk (interstellar medium, ISM) and the IGM, controls the competition between IGM inflow and galactic feedback processes, which in turn controls how the galaxy evolves \citep{Faucher-Giguere2023, Tumlinson2017}.

Due to the low density of most of the gas in the CGM \citep[see for example,][]{Zehedy2019}, it is difficult to capture the physical processes taking place in the CGM using the emission from this gas. Diffuse \lya\ emission is frequently detected from high-$z$  (i.e., $z\ge2$) galaxies which provide important information on the gas density and velocity field around galaxies \citep{Wisotzki2018}. However, such studies are rare for low-$z$, although there has been a steady attempt to map the CGM in \MgII\ emission either directly \citep{Chisholm2020, Burchett2021} or in the stacked images \citep[see for example,][]{Dutta2023}. As a result, absorption line spectroscopy remains the best means to understand the gas kinematics and physical processes that take place in the CGM. While the existence of large-scale gas flows (inflows/outflows) in the starforming galaxies is well-established by the ``down-the-barrel'' studies \citep{Rubin2010, Rubin2014, Tremonti2007}, the distance of the flowing gas from the galaxy (which is crucial for deriving wind parameters such as mass outflow rate) cannot be precisely determined with respect to the stellar discs. In contrast, the projected distance of the absorbing gas from the host galaxy is well measured in CGM studies using bright background sources like galaxies and quasars. However, drawing a clear connection between star formation and gas absorption is challenging, particularly at large impact parameters.

Historically \MgII\ doublet absorption seen in the spectra of quasars is used to trace the metal-enriched CGM around low-$z$ galaxies \citep{bergeron1991}.  Probing the physical condition of the gas around galaxies using background quasar sight lines at very low impact parameters (e.g., within the regions influenced by winds over a characteristic time-scale of star formation) can provide vital clues on the role played by large-scale winds in shaping the physical conditions of the CGM and how it regulates galaxy evolution. For the past couple of years, we have been carrying out a systematic study of the CGM of $z\sim 0.7$ galaxies at low impact parameters by (i) searching for the host galaxies of Ultra-Strong \MgII\ absorbers (\usmg) and (ii) studying the nature of \MgII\  galaxies that happen to lie within very small impact parameters (i.e. Galaxies On Top of Quasars; GOTOQs) to background quasars \citep[see][]{Guha2022a, Guha2022b}.

Here we concentrate on \usmg\ systems again.
Following \citet{Nestor_2007}, we define the \usmg\ systems as the ones having the rest equivalent width of the \MgII\ $\lambda\, 2796$ line ($W_{2796}$) more than or equal to 3\AA. Given the canonical anti-correlation between $\rm W_{2796}$ and the impact parameter (D) for the general population of \MgII\ absorbers \citep{Chen_2010, Nielsen_2013}, the impact parameters for the \usmg\ host galaxies are expected to be extremely low ($\lesssim$ 10 kpc).
The measured $W_{2796}$ using low dispersion spectra is well known to be linked to the number of absorbing clouds and the velocity dispersion between them, rather than the column density \citep{Petitjean1990}.
For a fully saturated \MgII\ line a 3\AA\ limit on $W_{2796}$  corresponds to a minimum velocity width of 320 \kms. Such large velocity spreads could be caused by (i) galactic-scale outflows \citep{Nestor_2011}, (ii) filamentary accretion onto galaxies \citep{rubin2012}, and (iii) galaxy mergers \citep{Rubin2010} and intra-group gas \citep{Gauthier2013, Nielsen2022, Zou2018}. In such instances, one could use the measured metallicities  \citep{Lehner2013} and galaxy orientations relative to quasar sightlines to differentiate between the various possibilities \citep{kacprzak2012}. To measure metallicity, one requires high-resolution spectra covering H~{\sc i} and metal absorption lines. For accurately measuring the orientation of host galaxies, one needs high spatial-resolution images.

In a previous paper \citep{Guha2022a}, we studied the nature and environment of a well-defined sample of \usmg\ host galaxies at $z \sim 0.5$. We find that the impact parameters are larger than that predicted by the $\rm{W_{2796}}$ versus the D relationship of the general population of \MgII\  absorbers. The \usmg\ host galaxies seem to form a distinct population in the $\rm W_{2796}-D$ plane. \usmg\ host galaxies are found to be massive and bright compared to those of the relatively weak \MgII\ absorption systems for the same impact parameters. At least $~$33\% of the \usmg\ host galaxies (with a limiting magnitude of $m_r < 23.6$) are isolated, and the large $\rm W_{2796}$ in these cases may originate from gas flows (infall/outflow) in a single halo of a massive but not a starbursting galaxy. We also find galaxy interactions could be responsible for large velocity widths in at least $~17$\% of the cases. 

In the present study, we proceed further with a slightly higher redshift ($z \sim 0.7$) sample to identify any redshift evolution associated with these systems. This paper is organized as follows. In section \ref{sec:sample}, we discuss our \usmg\ sample at $z \sim 0.7$. Section \ref{sec:observations} discusses the observations and data reduction procedures. In section \ref{sec:results}, we describe the data analysis and the results from this survey. The discussions and  summary \& conclusions are provided in Section \ref{sec:discuss} and \ref{sec:summary}, respectively. Throughout this paper, we assume a flat $\Lambda$CDM cosmology with
$H_0 = 70\, km\, s^{-1}\, Mpc^{-1}$ and $\Omega_{m,0} = 0.3$.

\section{The \usmg\ sample at $\lowercase{z}\, \sim\, 0.7$}
\label{sec:sample}

\begin{table*}
     \begin{threeparttable}
     \caption{Details of our \usmg\ sample. Columns (2), (3), and (4) respectively provide the quasar name, emission redshift ($z_{qso}$), and the absorption redshift ($z_{abs}$). Column (5) to (8) provides the REW of \MgII\ $\lambda$ 2796, \MgII\ $\lambda$ 2803, \FeII\ $\lambda$ 2600, \MgI\ $\lambda$ 2803 lines respectively.}
     \centering
     \begin{tabular}{lccccccc}
     \hline
     No. & Quasar & $z_{qso}$  & $z_{abs}$  & $W_{2796}\,(\text{\AA})$  & $W_{2803}\,(\text{\AA})$ & $W_{2600}\,(\text{\AA})$ & $W_{2853}\,(\text{\AA})$\\
       (1)&(2)&(3)&(4)&(5)&(6)&(7)&(8)\\
     \hline
     \hline
1 & J001030.81+012203.43 & 2.197 & 0.6433 & 3.19 $\pm$ 0.45 & 3.22 $\pm$ 0.46 & 2.01 $\pm$ 0.56 & $<0.60$\\
2 & J002022.66+000231.98 & 2.749 & 0.7700 & 3.18 $\pm$ 0.28 & 3.03 $\pm$ 0.27 & 2.11 $\pm$ 0.37 & 0.87 $\pm$ 0.21 \\
3 & J002839.24+004103.05 & 2.493 & 0.6565 & 3.33 $\pm$ 0.19 & 3.29 $\pm$ 0.19 & 2.93 $\pm$ 0.19 & 1.21 $\pm$ 0.37 \\
4 & J003336.04+013851.06 & 2.658 & 0.7188 & 3.80 $\pm$ 0.30 & 3.11 $\pm$ 0.24 & 1.35 $\pm$ 0.10 & 0.42 $\pm$ 0.13 \\
5 & J005554.25$-$010058.62 & 2.363 & 0.7150 & 4.02 $\pm$ 0.67 & 3.80 $\pm$ 0.63 & 2.90 $\pm$ 0.89 & 1.74 $\pm$ 0.59 \\
6 & J010543.52+004003.86 & 1.078 & 0.6489 & 3.41 $\pm$ 0.21 & 2.87 $\pm$ 0.18 & 1.69 $\pm$ 0.17 & $<0.48$ \\
7 & J012711.11$-$055020.95 & 2.137 & 0.6838 & 3.19 $\pm$ 0.51 & 2.54 $\pm$ 0.40 & 3.52 $\pm$ 0.75 & $<0.54$ \\
8 & J014115.32$-$000500.98 & 2.130 & 0.6106 & 2.77 $\pm$ 0.38 & 2.02 $\pm$ 0.27 & 1.05 $\pm$ 0.28 & $<0.42$ \\
9 & J014258.83+094942.43 & 0.981 & 0.7858 & 3.90 $\pm$ 0.26 & 3.27 $\pm$ 0.22 & 2.43 $\pm$ 0.15 & 0.82 $\pm$ 0.17 \\
10 & J015007.91$-$003937.09 & 2.730 & 0.7749 & 4.34 $\pm$ 0.53 & 4.08 $\pm$ 0.50 & 3.04 $\pm$ 0.26 & $<0.57$ \\
11 & J015049.39+060432.42 & 2.676 & 0.6707 & 3.34 $\pm$ 0.22 & 2.69 $\pm$ 0.18 & 2.38 $\pm$ 0.28 & 0.77 $\pm$ 0.12 \\
12 & J025607.25+011038.56 & 1.3490 & 0.7255 & 3.29 $\pm$ 0.20 & 2.91 $\pm$ 0.18 & 2.29 $\pm$ 0.33 & 0.82 $\pm$ 0.12 \\
13 & J090805.76+072739.90 & 2.414 & 0.6123 & 5.30 $\pm$ 0.58 & 4.38 $\pm$ 0.48 & 4.24 $\pm$ 0.94 & 1.45 $\pm$ 0.30 \\
14 & J095619.49+001800.34 & 2.172 & 0.7820 & 6.34 $\pm$ 0.46 & 6.35 $\pm$ 0.46 & 5.43 $\pm$ 0.55 & 2.23 $\pm$ 0.51 \\
15 & J103325.92+012836.35 & 2.180 & 0.6709 & 3.11 $\pm$ 0.17 & 2.82 $\pm$ 0.15 & 1.53 $\pm$ 0.15 & 0.44 $\pm$ 0.08 \\
16 & J104642.70+045731.96 & 2.542 & 0.7849 & 4.47 $\pm$ 0.66 & 4.40 $\pm$ 0.65 & 2.19 $\pm$ 0.22 & $<0.42$ \\
17 & J111627.65+050049.96 & 2.571 & 0.7208 & 3.78 $\pm$ 0.48 & 3.47 $\pm$ 0.44 & 1.93 $\pm$ 0.27 & $<0.41$ \\
18 & J115026.11+090048.40 & 2.492 & 0.7568 & 3.39 $\pm$ 0.24 & 2.76 $\pm$ 0.19 & 1.66 $\pm$ 0.34 & 1.13 $\pm$ 0.11 \\
19 & J120139.57+071338.24 & 1.205 & 0.6842 & 4.55 $\pm$ 0.18 & 4.08 $\pm$ 0.16 & 2.30 $\pm$ 0.24 & 0.83 $\pm$ 0.11 \\
20 & J121727.80$-$011548.57 & 2.624 & 0.6642 & 4.02 $\pm$ 0.45 & 3.64 $\pm$ 0.41 & 2.88 $\pm$ 0.28 & 1.57 $\pm$ 0.44 \\
21 & J132200.79$-$010755.70 & 2.160 & 0.7226 & 3.24 $\pm$ 0.24 & 3.20 $\pm$ 0.24 & 3.05 $\pm$ 0.30 & 1.09 $\pm$ 0.14 \\
22 & J133653.73+092221.23 & 2.531 & 0.7059 & 3.06 $\pm$ 0.09 & 2.86 $\pm$ 0.08 & 2.03 $\pm$ 0.10 & 0.69 $\pm$ 0.08 \\
23 & J140017.69$-$014902.40 & 2.555 & 0.7928 & 4.23 $\pm$ 0.25 & 4.02 $\pm$ 0.24 & 3.45 $\pm$ 0.25 & 1.04 $\pm$ 0.34 \\
24 & J141930.09+034643.73 & 2.316 & 0.7250 & 3.30 $\pm$ 0.40 & 2.80 $\pm$ 0.34 & 1.52 $\pm$ 0.22 & 1.50 $\pm$ 0.40 \\
25 & J144936.18$-$011650.46 & 0.772 & 0.6620 & 3.66 $\pm$ 0.26 & 2.99 $\pm$ 0.21 & 1.37 $\pm$ 0.13 & $<0.54$ \\
26 & J145108.53$-$013833.06 & 2.390 & 0.7407 & 3.29 $\pm$ 0.25 & 2.96 $\pm$ 0.23 & 1.94 $\pm$ 0.31 & $<0.62$ \\
27 & J235639.31$-$040614.47 & 2.880 & 0.7707 & 3.77 $\pm$ 0.48 & 3.68 $\pm$ 0.47 & 1.81 $\pm$ 0.21 & $<0.68$\\

\hline
\end{tabular}
\label{tab:sample}
\end{threeparttable}
\end{table*}

In this work we extend our study of \usmg\ absorbers \citep[][which focused on \usmg\ absorbers at $0.4\leqslant z\leqslant 0.6$]{Guha2022a} to slightly higher redshifts. For this, we compiled a sample of \usmg\ absorption system candidates from the twelfth data release (DR12) of the Sloan Digital Sky Survey \citep[SDSS,][]{Alam2015, York2000} \FeII / \MgII\ absorbers catalog \citep[][]{Zhu2013} that are accessible to the South African Large Telescope \citep[SALT,][]{buckley_2005} (i.e., with declination, $\delta \leqslant 10^\circ$) in the redshift range $0.6 \leqslant z_{abs} \leqslant 0.8$. Our preliminary search resulted in a total of 151 \usmg\ absorption system candidates along the line of sight towards 148 different background quasars. However, a careful visual inspection of the SDSS spectrum of each of these 151 \usmg\ absorption systems revealed that a total of 88 are basically \CIV / \SiIV\ broad absorption lines (BAL) misidentified as \usmg\ systems. Additional 36 systems are wrongly identified as \usmg\ because of line blendings, poor SNR, or false identification of \MgII\ absorption doublets, leaving behind only 27 secure \usmg\ systems in the redshift range $0.6 \leqslant z_{abs} < 0.8$ with $\delta \leqslant 10^\circ$. Details of all these 151 absorption systems are provided in the appendix (Table A1 of Appendix A) where we explicitly mention whether the absorption system is a secure \usmg\ absorption system or falsely identified as a \usmg\ absorption system. 

To confirm that the selected 27 \usmg\ systems are indeed bona fide \usmg\ absorption systems, we measure the rest equivalent widths  of the \MgII\ $\lambda\lambda\, 2796,\, 2803$  absorption lines. For $W_{2796}$, we first approximate the continuum around the \MgII\ absorption lines with a smooth polynomial and then fit the absorption doublet using a pair of Gaussian profiles on top of this smooth polynomial. While fitting, we impose the redshift and velocity width of both the Gaussian profiles to be the same. This exercise confirms that within $1\sigma$ uncertainty all the selected 27 systems are indeed \usmg\ absorption systems. Similarly, we fit the associated \FeII\ $\lambda$ 2600 and \MgI\ $\lambda$ 2853 absorption lines each with a single Gaussian profile in addition to a smooth polynomial continuum, and compute their rest equivalent widths. Upon visual inspection of the fitted profiles for five of the \usmg\ systems (J0127$-$0550, J0150+0604, J0256+0110, J0908+0727, J0956+0018), even in the low-resolution SDSS spectra, we identify sub-structures in the absorption profiles of \MgII\ and \FeII. Subsequently, these absorption profiles are not very well characterized by a single Gaussian profile with velocity offsets between absorption components ranging from 100 to 300 \kms. The details of these 27 \usmg\ systems along the emission redshifts, absorption redshifts, and the obtained rest equivalent widths are given in Table \ref{tab:sample}.  In the case of non-detections, we provide the 3$\sigma$ upper limit on the REW.

Combining these findings for \usmg\ systems at $z=0.6-0.8$  with that of Guha et al. (2022) for \usmg\ systems at $z=0.4-0.6$ we can summarise that there are 260 \usmg\ candidates at $\delta<10$ deg in the catalog of \citet{Zhu2013} with $0.4\le z \le 0.8$ and out of which we confirm only 54 of them to be secured \usmg\ absorbers. 

\section{SALT Observations and data reduction}
\label{sec:observations}
\begin{table*}
    \caption{The observational log. Column (2) corresponds to the \usmg\ systems observed. The observations date, total exposure, the slit position angle from the north, the grating angle, and the wavelength range are provided in columns (3) - (7), respectively. }
    \centering
    \begin{tabular}{lcccccc}
    \hline
    No. & Quasar   &  Date & Exposure (s)  &  PA (deg.) & Grating     &  Wavelength  \\
        &          &       &               &            & Angle (deg) &  Range (\AA)\\
    (1) & \multicolumn{1}{c}{(2)} & (3) & (4) & (5) & (6) & (7) \\
    \hline
    \hline
    1 & J002022.66+000231.98 & 2021-07-11 & 2400 & 17 & 20 & 6000 - 8985 \\
      &                      & 2021-08-03 & 2400 & 17 & 20 & 6000 - 8985 \\
    
    2 & J002839.24+004103.05 & 2020-07-29 & 2560 & 124 & 18.125 & 5310 - 8320 \\
      &                      & 2020-09-15 & 2560 & 124 & 18.5   & 5450 - 8450 \\
      &                      & 2022-08-04 & 2560 & 67 & 18.125  & 5310 - 8320 \\
    
    3 & J003336.04+013851.06 & 2020-06-24 & 2560 & 29  & 19.625 & 5870 - 8850 \\
      &                      & 2020-07-24 & 2560 & 29  & 19.25  & 5865 - 8865 \\
      &                      & 2021-07-16 & 2560 & 155 & 20     & 6000 - 8985 \\
      &                      & 2021-08-06 & 2280 & 155 & 20     & 6000 - 8985 \\
    
    4 & J005554.25$-$010058.62 & 2021-08-10 & 2400 & 171 & 19.625 & 5870 - 8850 \\
    
    5 & J010543.52+004003.86 & 2020-07-29 & 2560 & 141 & 18.5   & 5450 - 8450 \\
      &                      & 2020-10-13 & 2560 & 141 & 18.125 & 5310 - 8320 \\
      &                      & 2022-08-27 & 2560 & 106 & 18.5   & 5450 - 8450 \\
    
    6 & J012711.11$-$055020.95 & 2020-07-01 & 2560 & 41 & 19.25 & 5865 - 8865 \\
      &                        & 2020-10-14 & 2560 & 41 & 18.875 & 5585 - 8585 \\
      &                        & 2022-08-20 & 2560 & 68 & 18.875 & 5585 - 8585 \\
      
    7 & J014258.83+094942.43 & 2021-09-12 & 2200 & 86 & 20.375 & 6140 - 9120 \\
      &                      & 2021-08-04 & 2000 & 86 & 20     & 6000 - 8985 \\

    8 & J015007.91$-$003937.09 & 2022-09-21 & 2560 & 0  & 20 & 6000 - 8985 \\
    
    9 & J015049.39+060432.42 & 2020-10-16 & 2360 & 90 & 18.875 & 5585 - 8585 \\
      &                      & 2022-07-17 & 2360 & 90 & 18.5   & 5450 - 8450 \\
      &                      & 2022-09-23 & 2360 & 146 & 18.5  & 5450 - 8450 \\
    
    10 & J025607.25+011038.56 & 2020-11-12 & 2580 & 133 & 20    & 6000 - 8985 \\
      &                      & 2021-08-09 & 2560 & 133 & 19.25 & 5865 - 8865 \\
      &                      & 2022-09-03 & 2560 & 95  & 19.25 & 5865 - 8865 \\
      &                      & 2021-08-10 & 2560 & 133 & 19.625 & 5870 - 8850 \\
    
    11 & J090805.76+072739.90 & 2021-02-05 & 2300 & 73  & 18.875 & 5585 - 8585 \\
      &                      & 2021-02-06 & 2300 & 73  & 18.5   & 5450 - 8450 \\
    
    12 & J095619.49+001800.34 & 2021-02-08 & 2580 & 135 & 20    & 6000 - 8985 \\
       &                      & 2021-04-07 & 2580 & 135 & 20.375 & 6410 - 9120 \\
    
    13 & J103325.92+012836.35 & 2022-02-07 & 2440 & 21  & 18.875 & 5585 - 8585 \\
    
    14 & J104642.70+045731.96 & 2021-04-14 & 2200 & 161 & 20.375 & 6140 - 9120 \\
        
    15 & J111627.65+050049.96 & 2022-02-07 & 2300 & 87 & 19.625 & 5870 - 8850 \\
    
    16 & J115026.11+090048.40 & 2022-02-06 & 2200 & 41 & 20 & 6000 - 8985 \\
    
    17 & J120139.57+071338.24 & 2021-01-21 & 2300 & 149 & 19.25 & 5865 - 8865 \\
       &                      & 2021-04-12 & 2300 & 149 & 19.25 & 5865 - 8865 \\
    
    18 & J121727.80$-$011548.57 & 2021-03-12 & 2580 & 139.25 & 19.25 & 5865 - 8865\\
       &                        & 2023-01-23 & 2560 & 139    & 19.25 & 5865 - 8865\\

    19 & J132200.79$-$010755.70 & 2023-02-23 & 2560 & 132 & 19.625 & 5870 - 8850 \\
       &                        & 2023-02-24 & 2560 & 132 & 19.625 & 5870 - 8850 \\
    
    20 & J133653.73+092221.23 & 2022-04-30 & 2200 & 162 & 19.625 & 5870 - 8850\\
       &                      & 2023-02-22 & 2560 & 163 & 19.625 & 5870 - 8850\\
    
    21 & J140017.69$-$014902.40 & 2021-05-13 & 2560 & 0 & 20.375 & 6140 - 9120 \\
       &                      & 2021-06-13 & 2560 & 90 & 20.375 & 6140 - 9120 \\
    
    22 & J141930.09+034643.73 & 2021-04-18 & 2300 & 142 & 20 & 6000 - 8985 \\
       &                      & 2021-05-17 & 2400 & 132 & 20 & 6000 - 8985 \\
       &                      & 2021-06-13 & 2400 & 67  & 20 & 6000 - 8985 \\
    
    23 & J144936.18$-$011650.46 & 2020-06-16 & 2560 & 180 & 18.5 & 5450 - 8450 \\
       &                      & 2020-06-24 & 2560 & 180 & 18.875 & 5585 - 8585 \\
    
    24 & J145108.53$-$013833.06 & 2020-07-22 & 2560 & 31 & 20 & 6000 - 8985 \\
       &                      & 2021-05-11 & 2560 & 61 & 19.625 & 5870 - 8850 \\
    
    25 & J235639.31$-$040614.47 & 2017-06-05 & 2500 & 45 & 20.375 & 6140 - 9120 \\
       &                      & 2017-08-19 & 2350 & 315 & 20.375 & 6140 - 9120 \\
       &                      & 2020-09-08 & 2560 & 90 & 20 & 6000 - 8985 \\
    
    \hline
    \end{tabular}
    \label{tab:observation_log}
\end{table*}

\begin{figure*}
    \begin{subfigure}{0.195\textwidth}
    \centering\includegraphics[width=\textwidth]{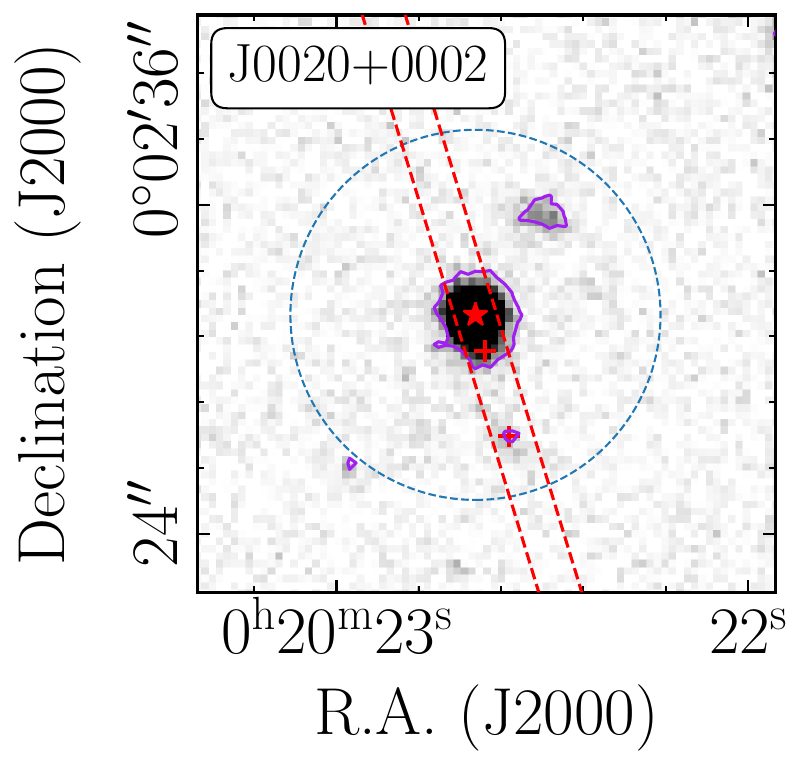}
  \end{subfigure}
  \begin{subfigure}{0.195\textwidth}
    \centering\includegraphics[width=\textwidth]{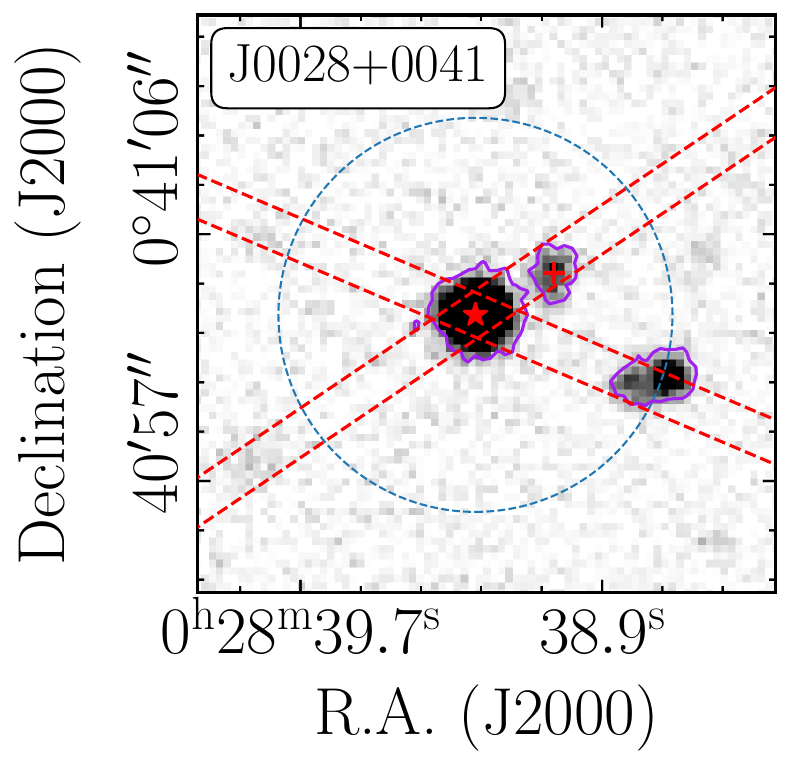}
  \end{subfigure}
  \begin{subfigure}{0.195\textwidth}
    \centering\includegraphics[width=\textwidth]{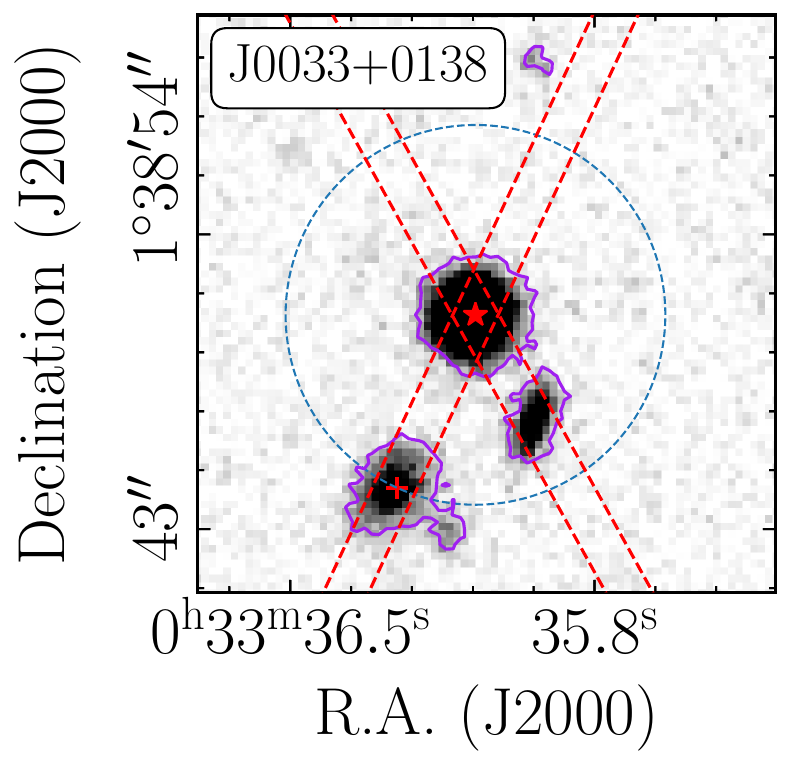}
  \end{subfigure}
  \begin{subfigure}{0.195\textwidth}
    \centering\includegraphics[width=\textwidth]{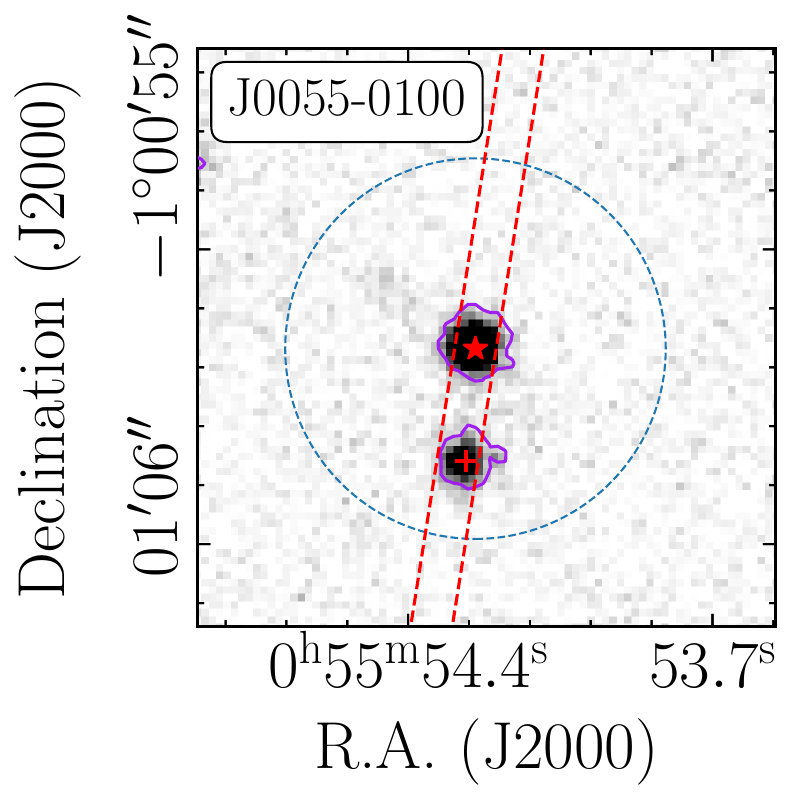}
  \end{subfigure}
  \begin{subfigure}{0.195\textwidth}
    \centering\includegraphics[width=\textwidth]{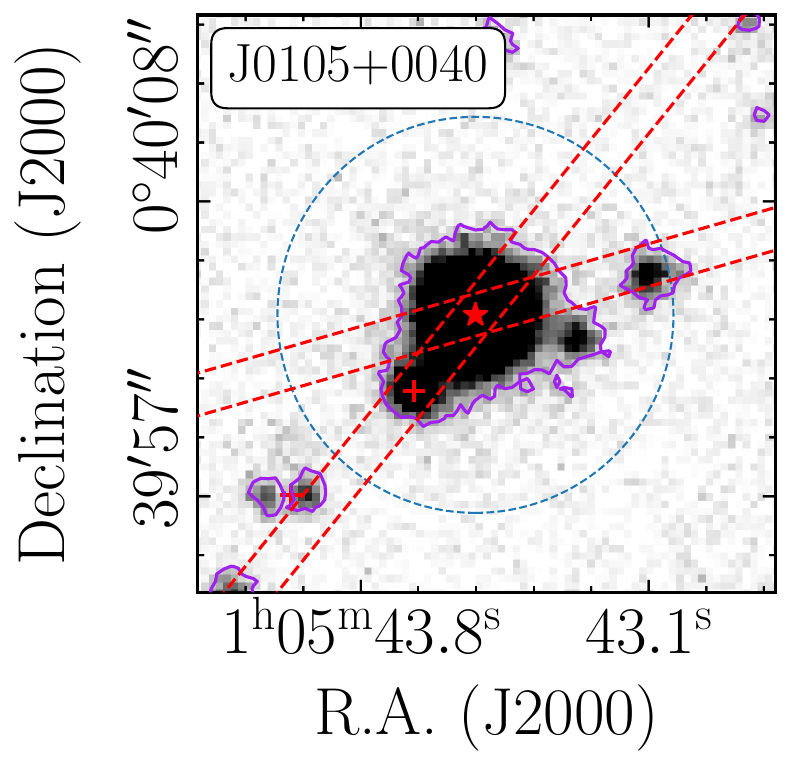}
  \end{subfigure}
  \newline
  
  \begin{subfigure}{0.195\textwidth}
    \centering\includegraphics[width=\textwidth]{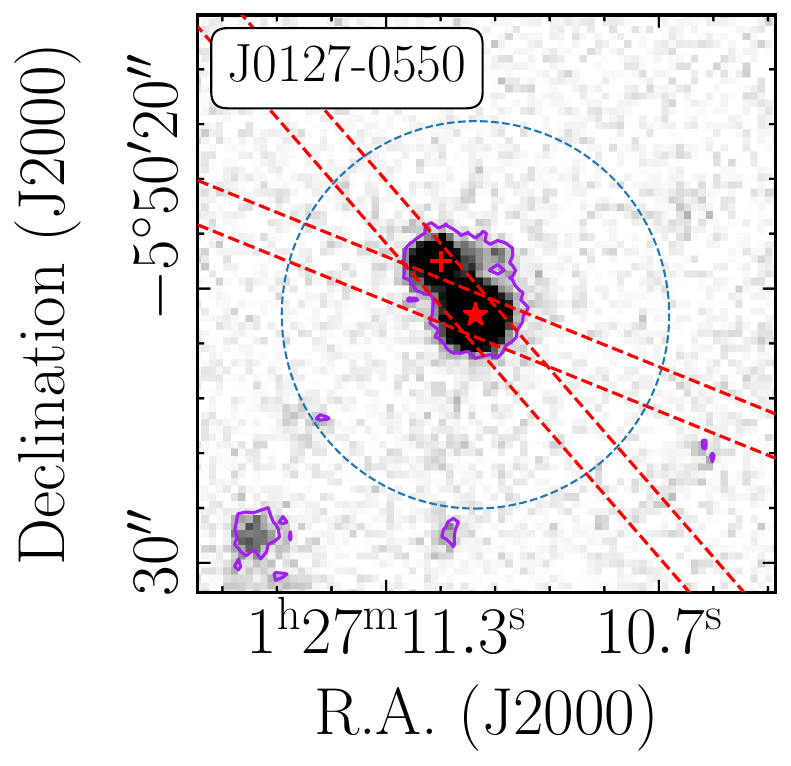}
  \end{subfigure}
  \begin{subfigure}{0.195\textwidth}
    \centering\includegraphics[width=\textwidth]{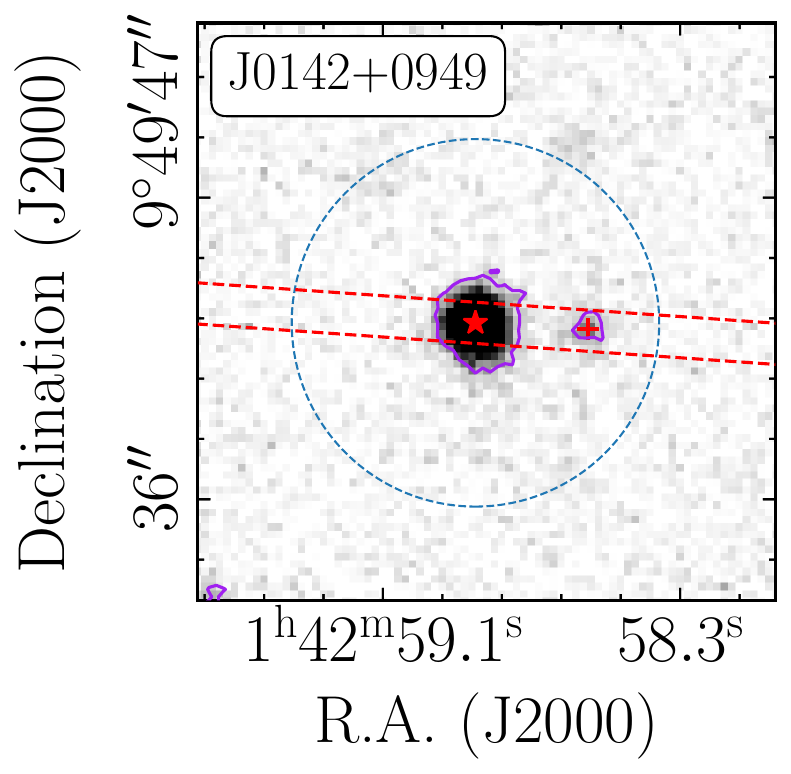}
  \end{subfigure}
  \begin{subfigure}{0.195\textwidth}
    \centering\includegraphics[width=\textwidth]{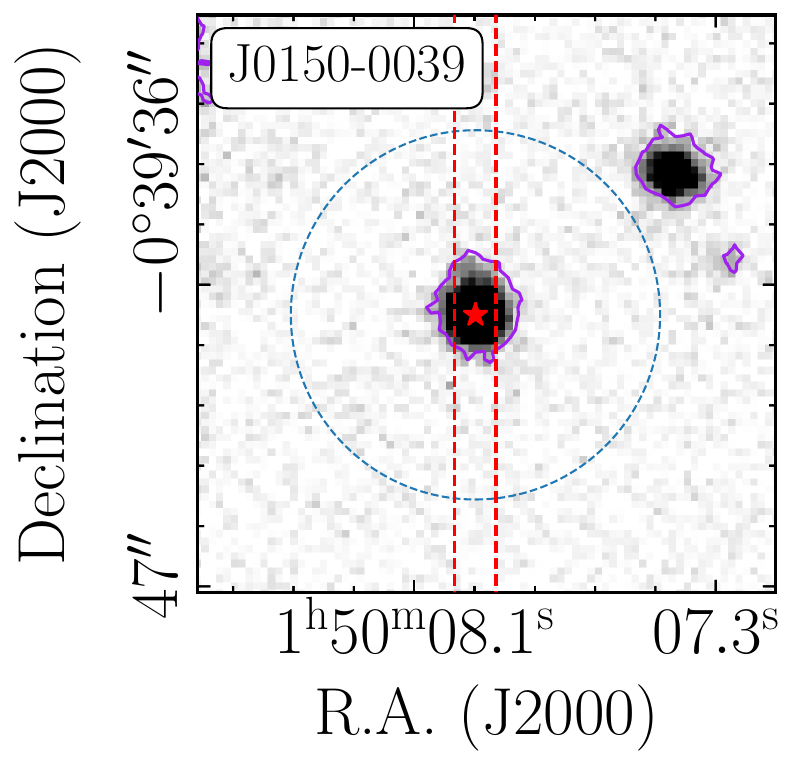}
  \end{subfigure}
  \begin{subfigure}{0.195\textwidth}
    \centering\includegraphics[width=\textwidth]{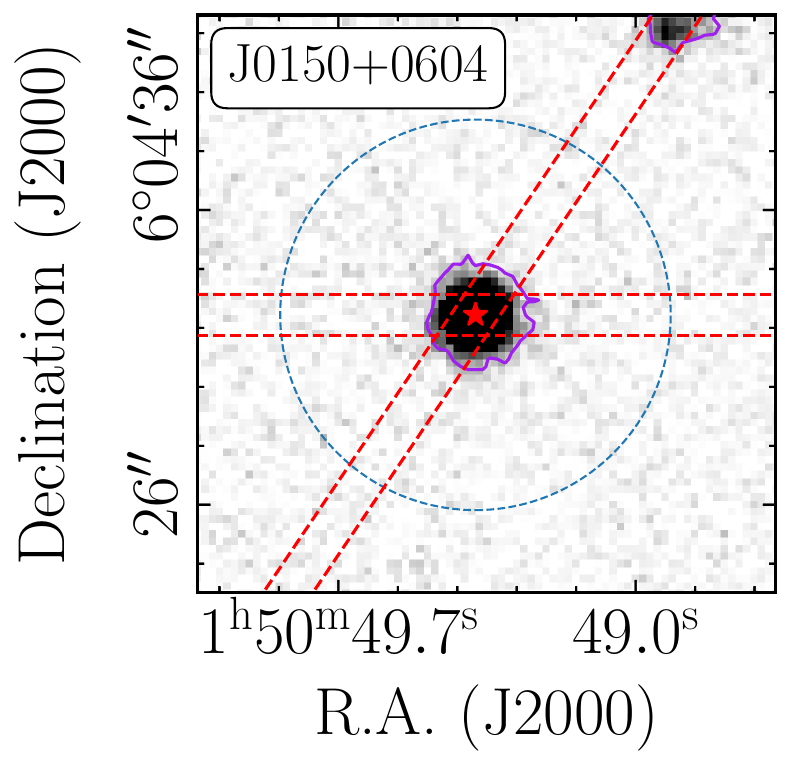}
  \end{subfigure}
  \begin{subfigure}{0.195\textwidth}
    \centering\includegraphics[width=\textwidth]{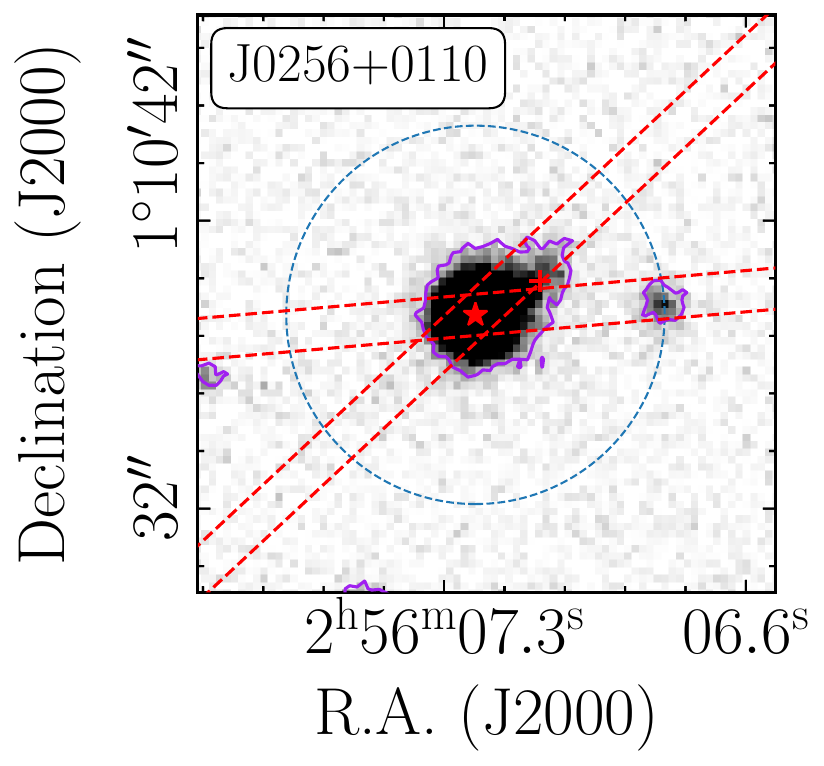}
  \end{subfigure}

  \begin{subfigure}{0.195\textwidth}
    \centering\includegraphics[width=\textwidth]{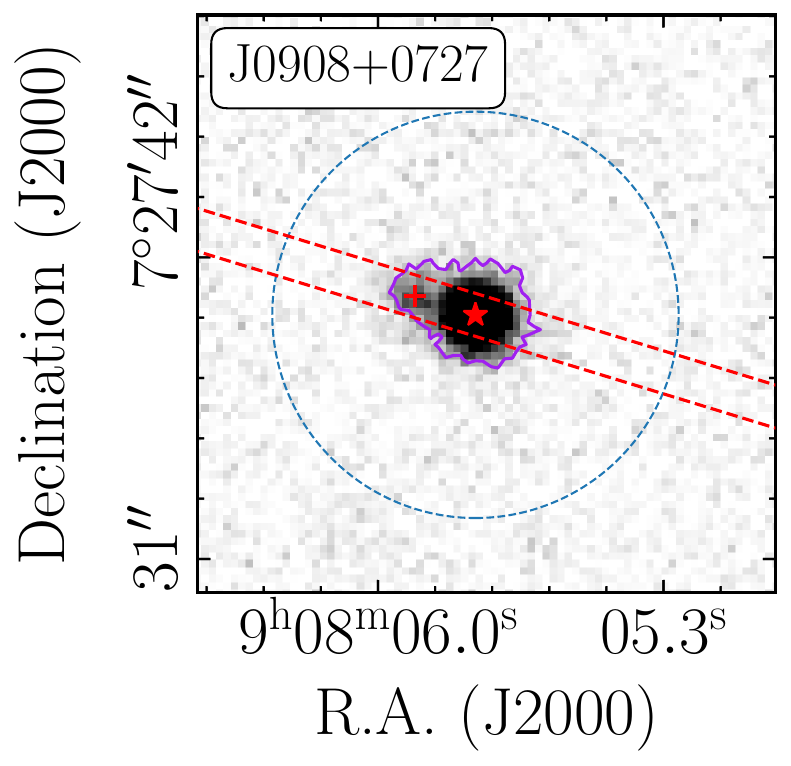}
  \end{subfigure}
  \begin{subfigure}{0.195\textwidth}
    \centering\includegraphics[width=\textwidth]{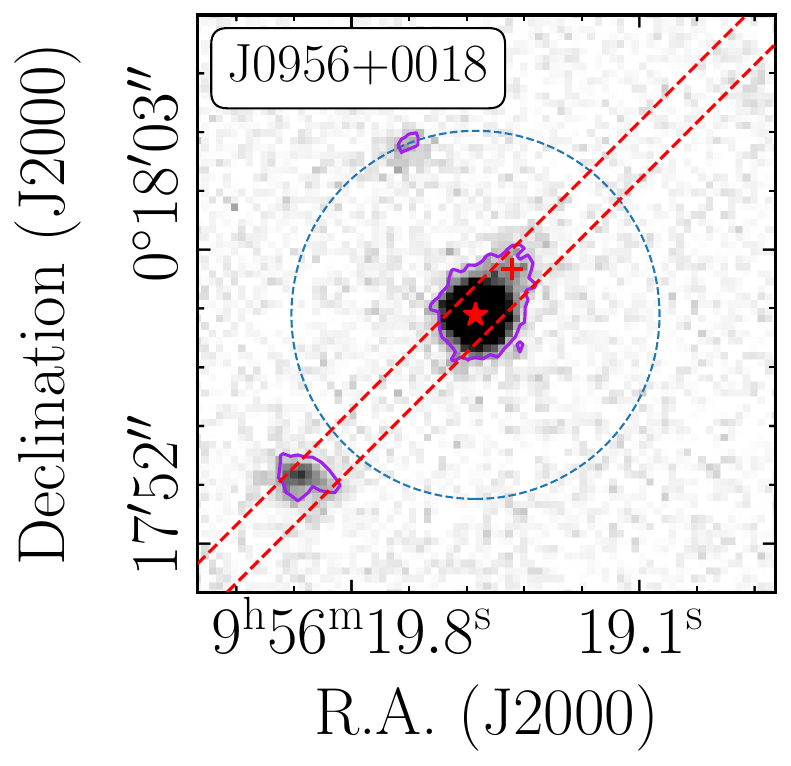}
  \end{subfigure}
  \begin{subfigure}{0.195\textwidth}
    \centering\includegraphics[width=\textwidth]{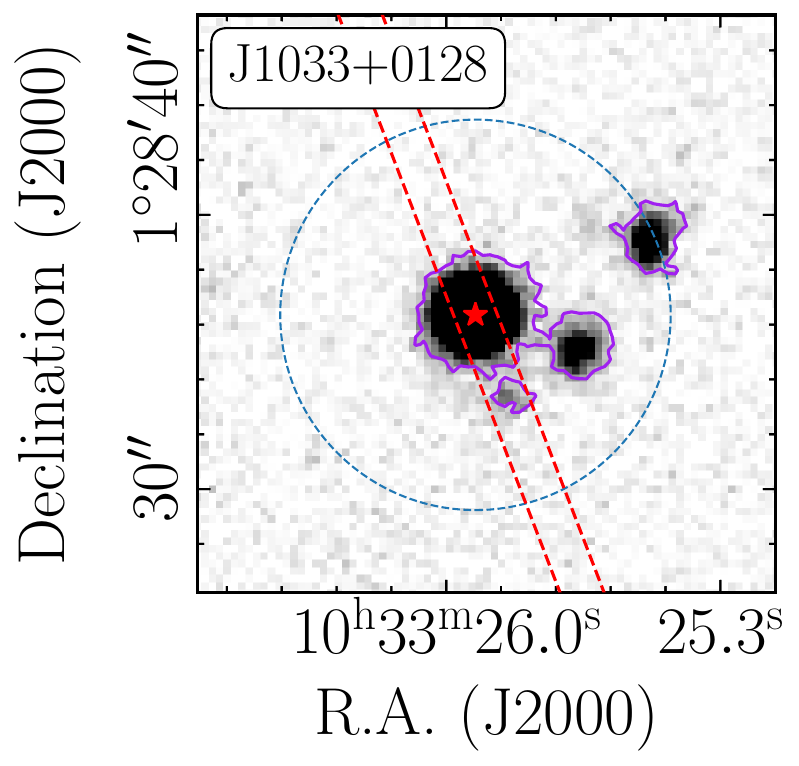}
  \end{subfigure}
  \begin{subfigure}{0.195\textwidth}
    \centering\includegraphics[width=\textwidth]{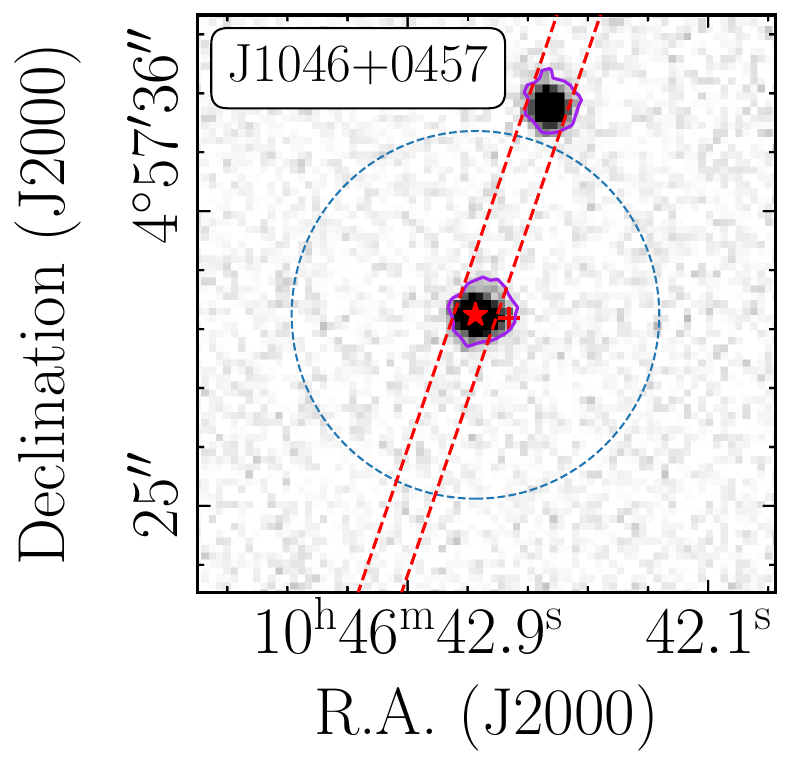}
  \end{subfigure}
  \begin{subfigure}{0.195\textwidth}
    \centering\includegraphics[width=\textwidth]{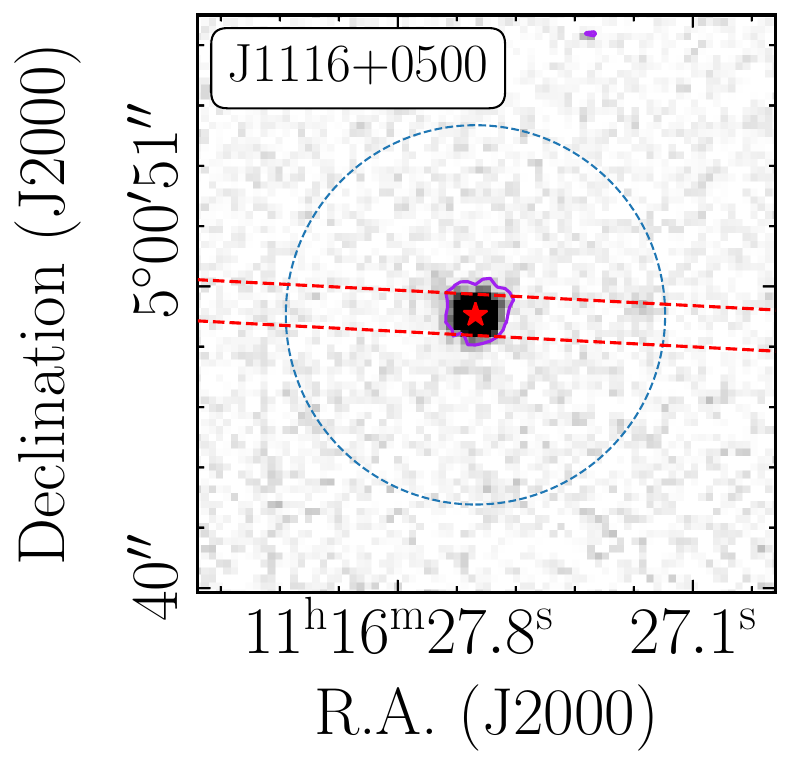}
  \end{subfigure}

  \begin{subfigure}{0.195\textwidth}
    \centering\includegraphics[width=\textwidth]{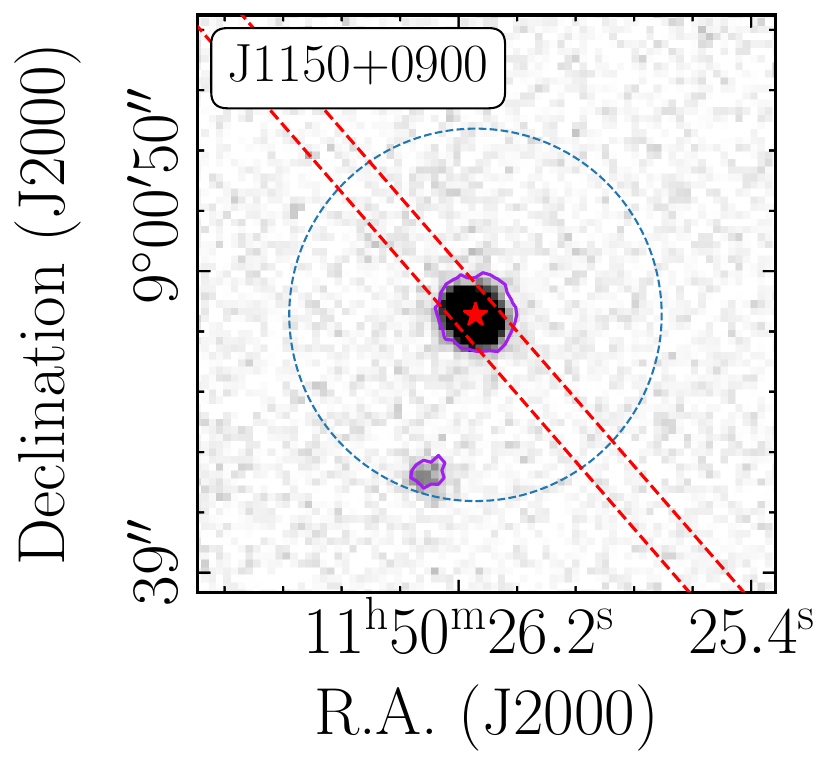}
  \end{subfigure}
  \begin{subfigure}{0.195\textwidth}
    \centering\includegraphics[width=\textwidth]{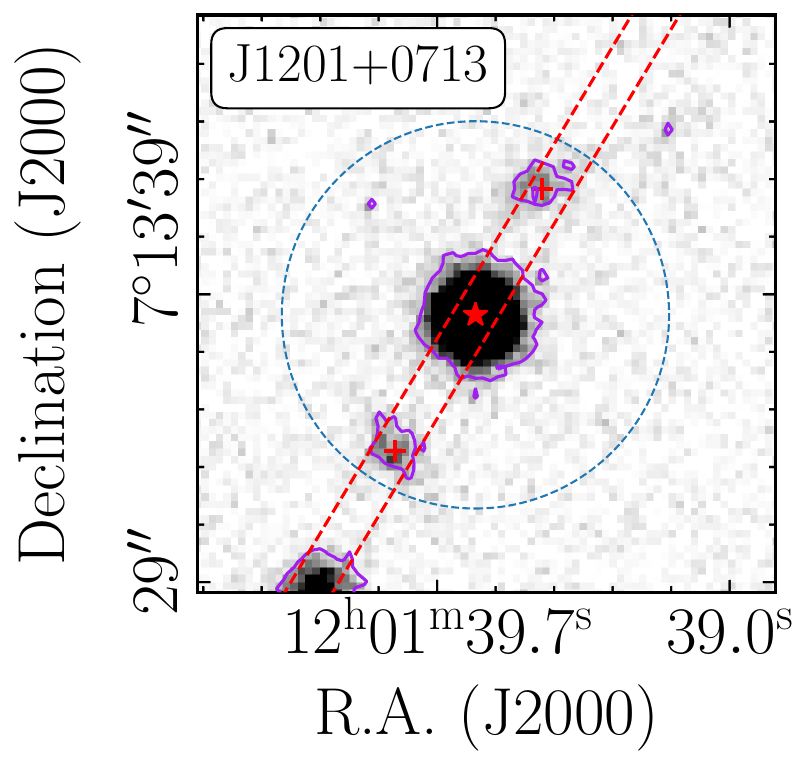}
  \end{subfigure}
  \begin{subfigure}{0.195\textwidth}
    \centering\includegraphics[width=\textwidth]{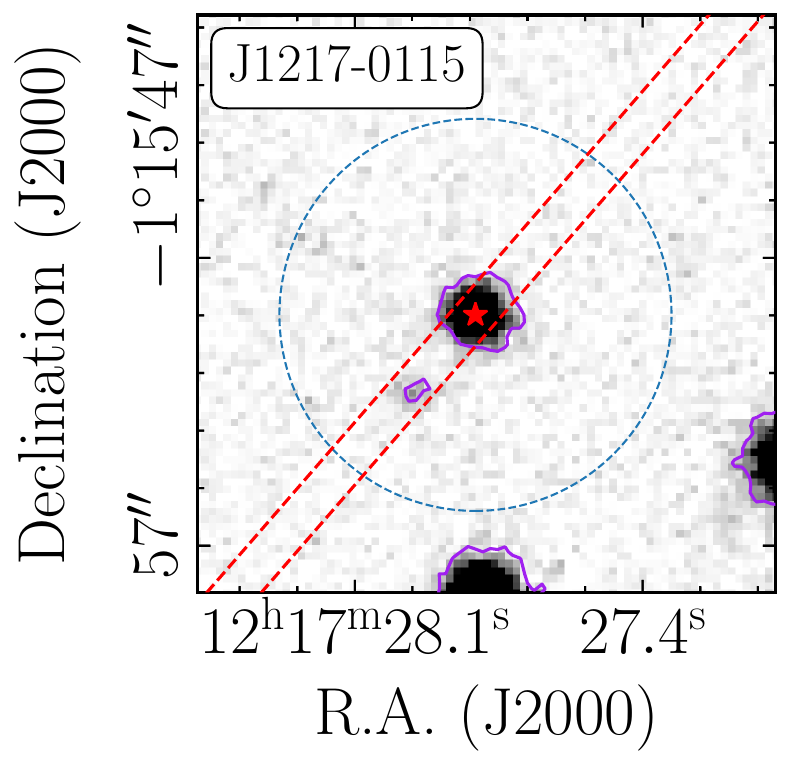}
  \end{subfigure}
  \begin{subfigure}{0.195\textwidth}
    \centering\includegraphics[width=\textwidth]{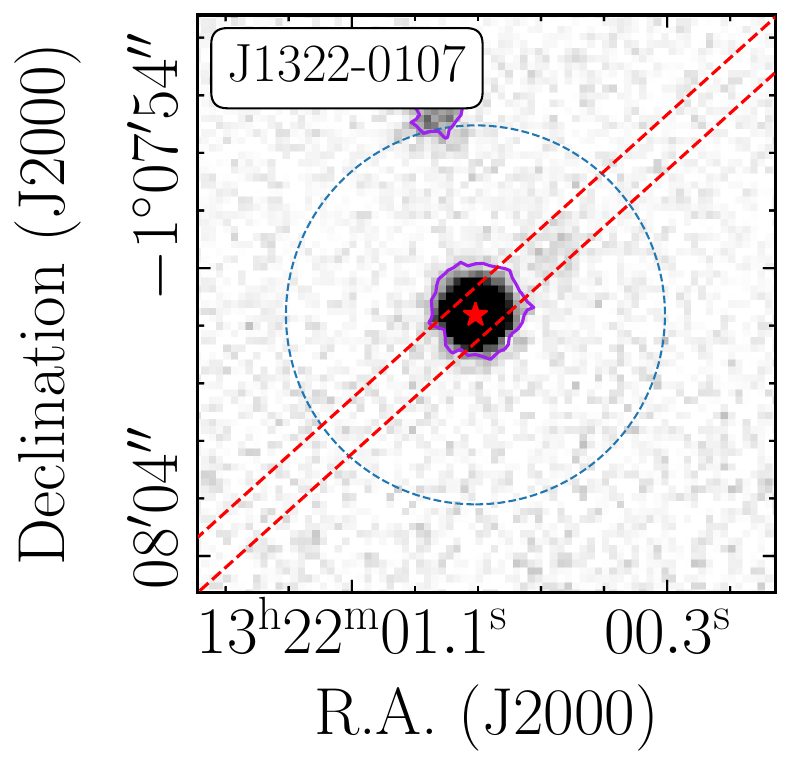}
  \end{subfigure}
   \begin{subfigure}{0.195\textwidth}
    \centering\includegraphics[width=\textwidth]{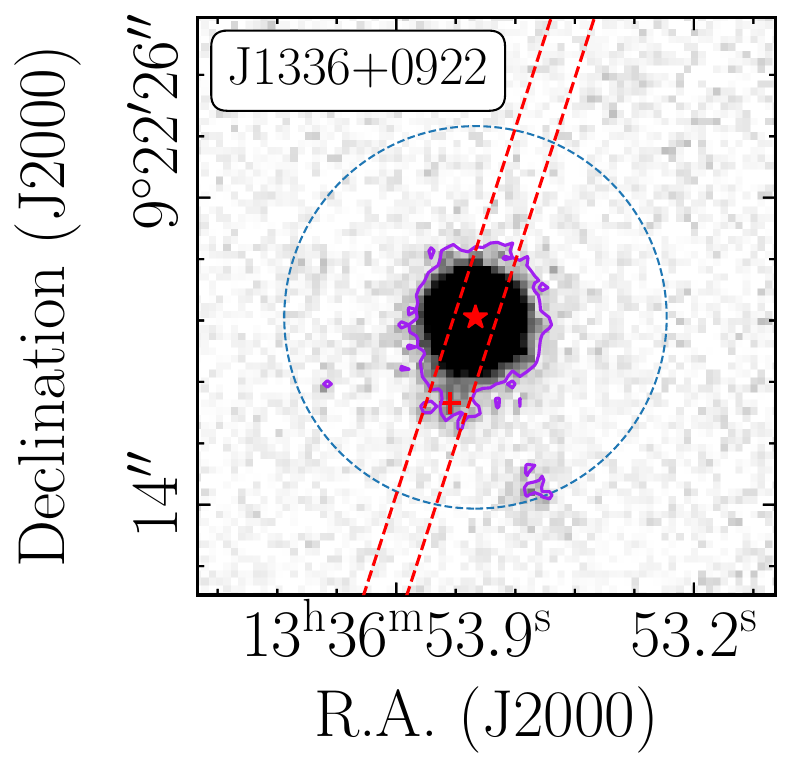}
  \end{subfigure}

  \begin{subfigure}{0.195\textwidth}
    \centering\includegraphics[width=\textwidth]{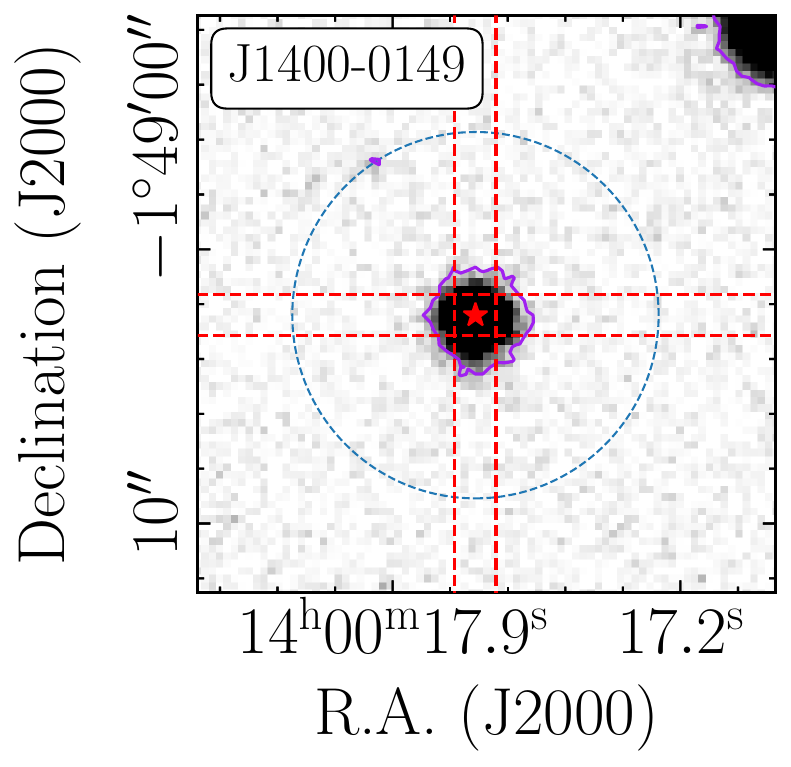}
  \end{subfigure}
  \begin{subfigure}{0.195\textwidth}
   \centering\includegraphics[width=\textwidth]{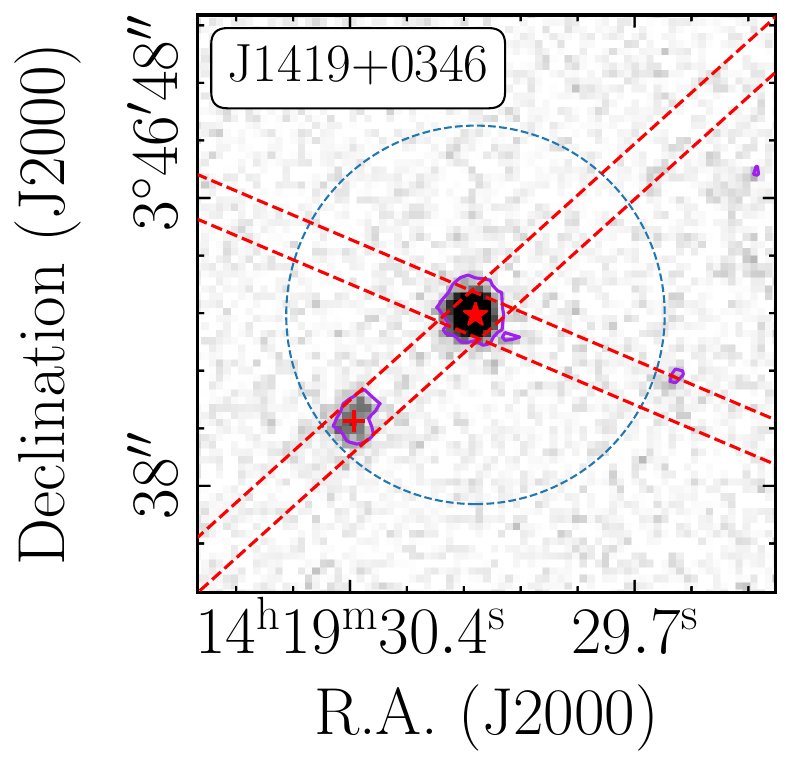}
  \end{subfigure}
  \begin{subfigure}{0.195\textwidth}
    \centering\includegraphics[width=\textwidth]{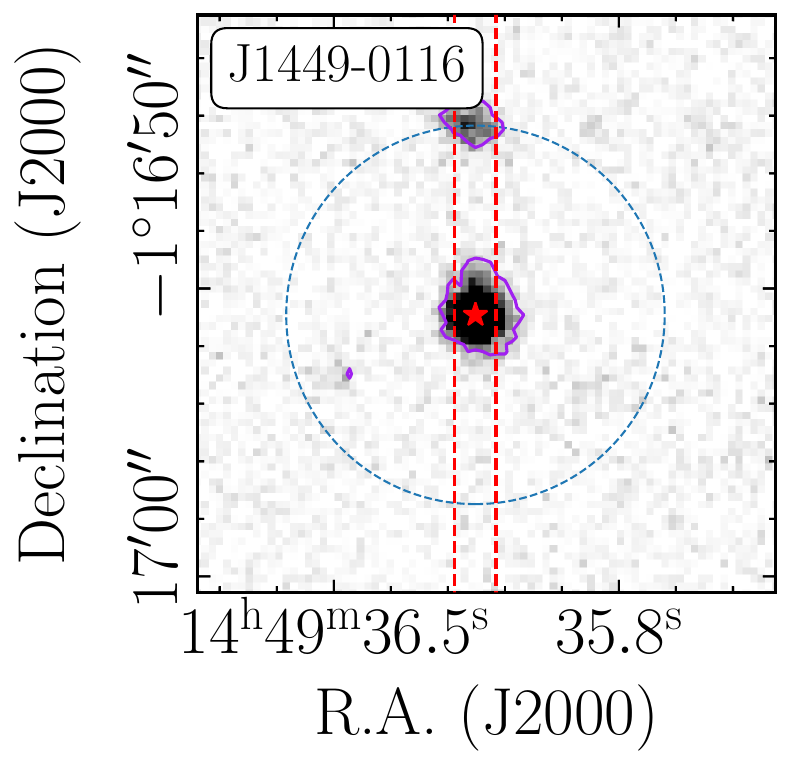}
  \end{subfigure}
  \begin{subfigure}{0.195\textwidth}
    \centering\includegraphics[width=\textwidth]{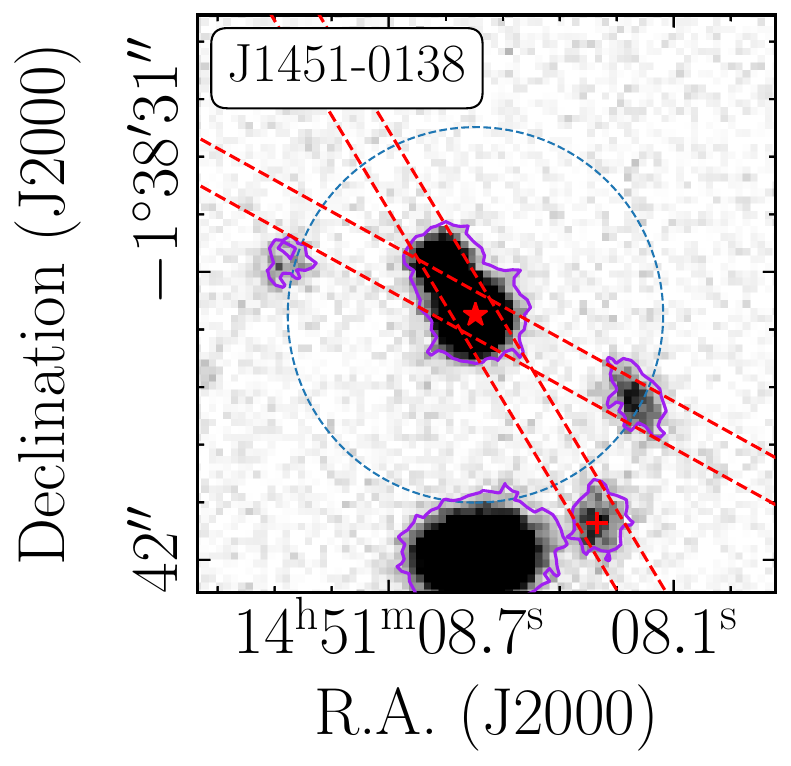}
  \end{subfigure}
  \begin{subfigure}{0.195\textwidth}
    \centering\includegraphics[width=\textwidth]{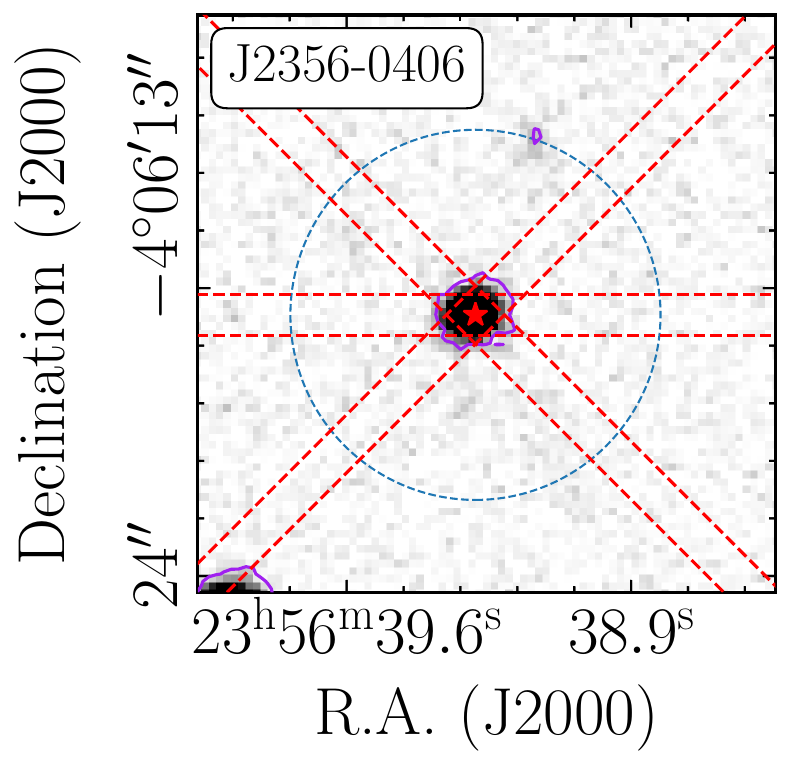}
  \end{subfigure}
  
  \caption{The DECaLS r-band images of the quasar fields with the quasar placed at the center and marked with a red `$\star$'. The slit configuration used is indicated with a pair of red dashed lines. The spectroscopically identified \usmg\ host galaxies are marked with a red `+'. The blue dashed circle in each panel corresponds to the circle of projected radius of 50 kpc at the absorption redshift. 
  }
  \label{fig:slit_configs}
  \end{figure*}


To identify the galaxy or the galaxy group giving rise to the \usmg\ absorption along the quasar line of sights, we first identify all the galaxies from the Dark Energy Spectroscopic Instrument Legacy Imaging Survey \citep[DESI-LIS,][typically complete upto $m_r \leqslant 23.6$]{Dey2019} within a projected distance of 100 kpc (at the photometric redshifts). Details of these galaxies are given in Table A2 of Appendix B. To optimize our spectroscopic identification process using long-slit spectroscopy, we mainly focus on the galaxies with photometric redshifts consistent within $1\sigma$ uncertainty with that of the \usmg\ systems and lying within 50~kpc from the quasar line-of-sight at the absorption redshift (as indicated by the dotted circles in Fig.~\ref{fig:slit_configs}). Next, we get the spectra of these candidate galaxies using the Southern African Large Telescope (SALT). In addition, whenever possible, we also observed potential galaxy candidates with impact parameters in the range 50$-$100 kpc.

The SALT spectroscopic observations were performed using the Robert Stobie Spectrograph \citep[RSS,][]{Burgh2003, Kobulnicky2003} in long-slit mode from June 2020 to February 2023 (Program IDs: 2020-1-SCI-010, 2020-2-SCI-019, 2021-1-SCI-006, 2021-2-SCI-012, 2022-1-SCI-016, 2022-2-SCI-015). The slit orientation (quantified through the position angle (PA) used) for each \usmg\ system is suitably chosen such that all the candidate galaxies and quasars are observed simultaneously. For all our observations, we use the PG0900 grating along with a long slit-width of width $1.5^{\prime\prime}$ and the grating angles are so chosen that the expected nebular lines (\OII, \OIII, and \textsc{H$\beta$}) from the \usmg\ host galaxies fall within the wavelength coverage of the spectrograph avoiding the CCD gaps. The details of the observations are provided in Table \ref{tab:observation_log}.  Column (2) provides the names of the quasars observed. The date of observations, total exposure time, the position angle (PA) from the north, the grating angle, and the wavelength range covered are provided in the next five subsequent columns in that order.

The raw CCD frames obtained from the observations are first processed with the SALT data reduction pipelines \citep{crawford2010}. Next, we use standard \texttt{PyRAF} \citep{pyraf} routines to obtain the wavelength and flux-calibrated spectra of the quasars as well as the candidate galaxies. In summary, the science frames were first flat-field corrected, cosmic ray zapped, and then wavelength calibrated against a standard lamp spectrum. Next, we corrected the extinction due to the Earth's atmosphere, and then the 1D spectra of the quasar and the galaxy were extracted. These 1D spectra were then flux-calibrated against standard stars observed with the same settings as the quasar. Finally, we apply the air to vacuum wavelength transformation and correct for the heliocentric velocity.

\section{Analysis and Results}
\label{sec:results}
\begin{table*}
    \centering
    \begin{tabular}{rcccccccc}
    \hline
    No. & Quasar        & Galaxy        & $z_{gal}$ & $D$   & $\log(M_\star / M_\odot)$ & $M_B$ & \OII\ Flux & SFR\\
        &    &     &           & (kpc) &           &   &  & ($M_\odot\, yr^{-1}$)\\
    (1) & (2) & (3) & (4) & (5) & (6) & (7) & (8) & (9)\\
    \hline
    \hline
    1 & J002022.66+000231.98   & J002022.65+000230.68 & 0.7684 & 9.7  &  -- & -- & 0.92$\pm$0.18 & -- (1.2)\\
      &                        & J002022.58+000227.38 & 0.7700 & 35.2 &  $9.33^{+0.31}_{-0.24}$ & $-$19.08 $\pm$ 0.11 & 0.71$\pm$0.23 & $1.04^{+0.88}_{-0.31}$ (1.0)\\
    2 & J002839.24+004103.05   & J002839.05+004104.58 & 0.6565 & 22.2 &  $10.70^{+0.12}_{-0.13}$ & $-$20.27 $\pm$ 0.09 & 0.94$\pm$0.27 & $11.56^{+10.80}_{-4.78}$ (0.3)\\
    3 & J003336.04+013851.06   & J003336.24+013844.75 & 0.7178 & 50.3 &  $11.76^{+0.04}_{-0.06}$ & $-$22.46 $\pm$ 0.07 & 1.88 $\pm$0.42 & $56.90^{+11.65}_{-16.15}$ (2.0)\\
    4 & J005554.25$-$010058.62 & J005554.28$-$010102.72 & 0.7150 & 29.8 & $11.49^{+0.04}_{-0.05}$ & $-$21.19 $\pm$ 0.07 & $<1.06$  & $4.72^{+5.31}_{-2.65}$ ($< 0.4$)\\
      &                        & J005554.42$-$010115.20 & 0.7161 & 120.7 & $10.03^{+0.18}_{-0.17}$ & $-$20.01 $\pm$ 0.10 & 0.80$\pm$0.19 & $3.47^{+3.66}_{-2.00}$ (0.5)\\
    5 & J010543.52+004003.86   & J010543.67+004001.10 & 0.6490 & 24.6 &  $10.35^{+0.11}_{-0.12}$ &  $-$20.90 $\pm$ 0.09 & 11.2$\pm$1.0 & $5.57^{+2.93}_{-2.04}$ (3.2)\\
      &                        & J010543.97+003957.28 & 0.6489 & 65.7 &  $9.54^{+0.14}_{-0.13}$ & $-$20.14 $\pm$ 0.12 & 0.81$\pm$0.21 & $3.04^{+2.19}_{-0.99}$ (0.5)\\
    6 & J012711.11$-$055020.95 & J012711.20$-$055019.01 & 0.6829 & 16.8 & $10.48^{+0.16}_{-0.17}$ & $-$21.35 $\pm$ 0.12 & 3.82$\pm$0.87 & $7.93^{+5.71}_{-2.96}$ (2.6)\\
      &                        & J012711.91$-$055016.45 & 0.6839 & 90.1 & $10.77^{+0.11}_{-0.18}$ & $-20.90\pm0.10$ & 1.73$\pm$0.23 & $9.91_{-2.94}^{+4.44}$ (1.1)\\
    7 & J014258.83+094942.43 & J014258.56+094942.20 & 0.7860 & 30.0 & $12.01^{+0.12}_{-0.14}$ & $-21.68\pm0.09$ & $< 1.10$ &  -- ($<0.4$)\\
    8 & J025607.25+011038.56 & J025607.10+011039.80 & 0.7244 & 18.4 & $11.04^{+0.14}_{-0.12}$ & $-21.26\pm0.10$ & $1.46\pm0.28$ & $36.18^{+44.52}_{-16.68}$ (1.5) \\
    9 & J090805.76+072739.90   & J090805.91+072740.58 & 0.6127 & 16.5 &  $10.73^{+0.13}_{-0.14}$ & $-$20.57 $\pm$ 0.10 & 2.53$\pm$0.33 & $15.61^{+12.78}_{-7.19}$ (1.3)\\
    10 & J095619.49+001800.34   & J095619.41+001802.00 & 0.7829 & 15.6 &  $10.72^{+0.14}_{-0.16}$ & $-$21.06 $\pm$ 0.13 & 6.89$\pm$0.52 & $14.42^{+20.12}_{-7.00}$ (16)\\
    11 & J104642.70+045731.96$^\star$   & J104642.62+045731.84 & 0.7848 & 8.9  &   -- & --  & 0.88$\pm$0.27 & -- (1.6)\\
    12 & J120139.57+071338.24   & J120139.41+071342.84 & 0.6854 & 36.3 &  $10.98^{+0.14}_{-0.11}$ & $-$21.43 $\pm$ 0.10 & 1.80$\pm$0.37 & $42.97^{+45.06}_{-24.86}$ (1.2)\\
       &                        & J120139.77+071333.28 & 0.6842 & 41.3 &  $11.42^{+0.13}_{-0.11}$ & $-$21.50 $\pm$ 0.09 & 2.91$\pm$0.50 & $82.69_{-24.86}^{+85.06}$ (1.7)\\
    13 & J133653.73+092221.23   & J133653.80+092217.96 & 0.7059 & 24.6 &  $10.47^{+0.16}_{-0.20}$ & $-$20.47 $\pm$ 0.14 & 3.05 $\pm$0.30 & $6.07^{+4.73}_{-2.10}$ (2.1)\\
    14 & J140017.69$-$014902.40$^\star$ & -- & 0.7933 & $\leqslant 8$ & -- & -- & -- & --\\ 
    15 & J141930.09+034643.73   & J141930.39+034639.86 & 0.7252 & 42.8 &  $10.97^{+0.16}_{-0.17}$ & $-$21.51 $\pm$ 0.12 & 8.0$\pm$1.0 & $23.29^{+17.46}_{-9.22}$ (2.7)\\
    16 & J144936.18$-$011650.46 & J144936.20$-$011643.58 & 0.6618 & 48.1 &  $10.74^{+.13}_{-0.08}$ & $-$21.05 $\pm$ 0.10 & 4.73$\pm$0.43 & $21.06^{+56.17}_{-10.87}$ (3.0)\\
    17 & J145108.53$-$013833.06 & J145108.24$-$013840.64 & 0.7414 & 64.1 &  $11.44^{+0.07}_{-0.10}$ & $-$21.94 $\pm$ 0.08 & 1.00$\pm$0.24 & $34.44^{+9.42}_{-9.25}$ (3.5)\\
    18 & J235639.31$-$040614.47$^\star$ & J235639.27$-$040413.80 & 0.7699 & 6.24 &  -- & -- & 5.49$\pm$0.43 & -- (3.9)\\
    
    \hline
    \end{tabular}
    \caption{Properties of the \usmg\ host galaxies. Columns (2) and (3) respectively indicate the quasars and corresponding \usmg\ host galaxies.  {Quasars marked with `$\star$' belong to the GOTOQs.} Columns (4) and (5) provide the \OII\ emission redshifts and the impact parameters of the host galaxies. Columns (6) and (7) provide the stellar masses and the rest frame absolute B band magnitude of the \usmg\ host galaxies, respectively. The \OII\ fluxes are measured in the units of $10^{-17}$ ergs cm$^{-2}$ s$^{-1}$ and are provided in Column (8). Column (9) corresponds to the star formation rates based on SED fitting analysis. The values in the parenthesis correspond to the star formation rates based on the \OII\ emission line luminosity. The typical errors associated are about one-third of the value.}
    \label{tab:spec_props}
\end{table*}

\subsection{Identification of the \usmg\ host galaxies}
\label{sec:gal_identify}
Spectroscopic observations of host-galaxy candidates were completed using SALT for 25 of the 27 \usmg\ systems (i.e., excluding the \usmg\ absorbers along J0010+0122 and J0141$-$0005) in our sample. The r-band images of the fields together with the slit orientations (parallel dashed lines in red) used, and the 50 kpc impact parameter at the redshift of the absorber (indicated by the dashed circle) are shown in Figure~\ref{fig:slit_configs}. The spectroscopically confirmed \usmg\ host galaxies are marked with a red `$+$', and the quasar locations are marked with a red `$\star$'. As can be seen from this figure and Table A2 of the Appendix (in the online material), there are ten cases where we do not find any galaxy with $r<23.6$ mag within an impact parameter of 50 kpc to the quasar sightline at the redshift of the \usmg\ system. These are, J0142$+$0949, J0150$-$0039, J0150$+$0604, J1046$+$0457, J1116$+$0500, J1217$-$0115, J1322$-$0107, J1400$-$0149, J1451$-$0138 and J2356$-$0406. The \usmg\ absorber towards J2356$-$0406 is a known GOTOQ \citep{Joshi2017}, and we obtained RSS spectra along three position angles to find the location of the host galaxy using triangulation (see Figure C1 of Appendix C). In the case of  J1046$+$0457 and J1400$-$0149, we detect \OII\ emission at the correct redshift in the spectrum of the quasar. We thus confirm the \usmg\ host galaxies to be GOTOQs at low impact parameters (i.e., $D< 10$ kpc). In the case of J1046$+$0457, we identify the location of the galaxy from the extension seen in the DECaLS images \citep[as discussed in][]{Guha2022b} and in the case of J1400$-$0149, we used the spectra obtained along two PAs.

{ In five cases (J0142$+$0949, J0150$-$0039, J1116$+$0500, J1217$-$0115, and J1322$-$0107), however,  we do have a candidate galaxy within 50 kpc if we slightly relax our candidate galaxy selection criteria. Here, we consider the galaxies having photo-z consistent within $\sim 2\sigma$ uncertainty and allow them to be fainter than the limiting magnitude of $m_r = 23.6$. The impact parameter of the candidate galaxies ranges from $D \sim 10$ to 50 kpc. In these cases, slit PA is chosen to simultaneously observe the quasar and the galaxy candidate.
In the case of J0142$+$0949, the targeted galaxy at $D \sim 30$ kpc ($z_{ph} \sim 0.9$) does not show any detectable nebular emission line in its spectrum. Still, we find a pair of absorption features consistent with the \CaII\ $\lambda\lambda\, 3935, 3970$ doublet at $z_{gal} = 0.786$ (see Figure B1 of Appendix B) and thus consistent with the redshift $z_{abs} = 0.7858$ of the \usmg\ system. Therefore, we consider this galaxy as the host galaxy of the \usmg\ absorber. 

In the case of J1451$-$0138, we find a few galaxies with a projected distance of less than 50 kpc from the quasar. However, none has a photometric redshift consistent with the \usmg\ absorption redshift. We covered three galaxies in our observations (using two slit orientations) plus other galaxies outside the 50~kpc distance. We confirm only one galaxy with correct spectroscopic redshift but at an impact parameter of 64.1 kpc ($\rm m_r = 21.59$). In the remaining three cases we do not detect \OII\ emission from the galaxy candidates (typical 3$\sigma$ limiting  \OII\ line flux of $7\times10^{-18}\, \rm{ergs\, cm^{-2}\, s^{-1}}$). In the case of J0150$+$0604, no faint candidate galaxies are found within 50 kpc. We do not have any indication of a faint galaxy coinciding with the quasar image (i.e., GOTOQs) either in the available photometry or in our spectroscopic data.} 

{In summary, out of the ten \usmg\ systems discussed above, we confirm three (J1046+0457, J1400$-$0149, and J2356$-$0406) of them to be GOTOQs. For J0142+0949 (even though the identified host galaxy has photo-z inconsistent within $1 \sigma$ uncertainty) and J1451$-$0138 (impact parameter is larger than 50 kpc), we spectroscopically confirm the \usmg\ host galaxies. In four (J0150$-$0039, J1116+0500, and J1217$-$0115, and J1322$-$0107) cases, we have candidate galaxies (either faint or consistent photo-z within $2\sigma$ uncertainty) without spectroscopic confirmation. For J0150+0604, we do not have any candidate galaxy present within 50 kpc.} 

In nine cases (J0028$+$0041, J0033$+$0138, J0055$-$0100, J0127$-$0550, J0908$+$0727, J1150$+$0900, J1336$+$0922, J1419+0346 and J1449$-$0116) we have only one galaxy with $r<23.6$ mag with a consistent photometric redshift within 50 kpc. We could get spectra of 8 candidate galaxies (apart from the case of J1150$+$0900 where our SALT observation was not scheduled) and confirm their redshifts using nebular emission lines to be consistent with the corresponding \usmg\ system. Therefore, our target completeness is 89\% for observations, and spectroscopic completeness is 100\% for cases with only one candidate galaxy within 50 kpc of the \usmg\ absorbers.

In the remaining six cases, there are two galaxies within 50 kpc with consistent photometric redshifts. In three cases (J0105$+$0040, J0256$+$0110, and J1201$+$0713), we could get spectra of both galaxies. We confirm both (one of) the candidate galaxies to have consistent spectroscopic redshifts in the case of J1201$+$0713 (J0105+0040 and J0256$+$0110).  In the case of J1033$+$0128, while we got the spectra of one of the candidates, we were unable to confirm the redshift using nebular emission lines. In the last two cases ({J0020+0002} and J0956+0018), we could observe only one of the candidate galaxies (with the closest impact parameter) each and confirm them to have consistent spectroscopic redshifts. Therefore, the overall target completeness is 75\% for observations, and spectroscopic completeness is 67\% for cases with one candidate galaxy (brighter than $23.6$ mag) within 50 kpc to the \usmg\ absorbers.

Among the 25 observed \usmg\ systems, we have successfully identified the \usmg\ host galaxies for 18 cases (16 based on \OII\ emission and two based on \CaII\ absorption). The quasars (that show \usmg\ absorption systems) and the associated \usmg\ host galaxies are indicated in columns (2) and  (3) of Table \ref{tab:spec_props}, respectively. Columns (4) and (5), respectively, provide the spectroscopic redshifts ($z_{gal}$) and the impact parameters (D). For four systems (J0020+0002, J0105+0040,
J0127$-$0550, and J1201+0713), we could identify another galaxy within 100 kpc having spectroscopic redshift consistent with \zabs\ of the \usmg\ system. For the case of J0055$-$0100, we find an additional galaxy having consistent spectroscopic redshift with the \usmg\ absorption at an impact parameter of 120 kpc.  Details of these additional galaxies are also provided in Table~\ref{tab:spec_props}. Therefore, we confirm that at least in 5 out of 18 cases, the \usmg\ absorber could be associated with more than one galaxy having correct spectroscopic redshifts and impact parameters less than 125~kpc.

Finally, for all the spectroscopically confirmed redshifts, we compare the photometric redshift and spectroscopic redshift (see Figure C2 of Appendix C). We find that the two match well within $2\sigma$ uncertainty.

\subsection{Properties of \usmg\ host galaxies}

Having identified the host galaxies of \usmg\ absorbers, we estimate their properties, such as the stellar mass (M$_*$), rest frame absolute B band magnitude ($M_B$), and the ongoing star formation rate (SFR) using the available photometric and spectroscopic data and the spectral energy distribution (SED) modeling. As in \citet{Guha2022a}, we use the freely accessible Python tool Bayesian Analysis of Galaxies for Physical Inference and Parameter EStimation \citep[BAGPIPES]{Carnall2018}. BAGPIPES takes into account \citet{Bruzual} stellar population models, which were built assuming the \citet{Kroupa} initial mass function (IMF) and most recently revised by \citet{Charlot2016} to incorporate the MILES stellar spectrum library and an updated stellar evolutionary track \citep{Marigo}. The interstellar dust is assumed to follow \citet{Calzetti1997} dust model. 

During the fit, we keep the redshifts of the \usmg\ host galaxies fixed to their spectroscopic redshifts. Although BAGPIPES can simultaneously fit photometric and spectroscopic data, we only use the DESI-LIS three broadband photometric measurements to fit the SED as the SNR of individual pixels for the continuum regions in the spectra of most of the \usmg\ host galaxies is quite poor (SNR $<3$). We also assume that all the stars in the galaxy have the same metallicity and use a flat prior in the range $0.01Z_\odot - 2.5Z_\odot$. We parameterize the star-formation histories with an exponential model \citep{Carnall2019}. We choose a uniform prior to the logarithm of the stellar mass in the range $0 \leqslant \log(M_\star / M_\odot) \leqslant 13$. However, note that the parametric models impose strong priors on physical parameters and may bias the inferred galaxy properties. Although the stellar mass is less sensitive to such effects (typical offset of 0.1 dex from the true value), the offset for SFR can be as large as 0.3 dex \citep{Carnall2019}. The derived parameters for each galaxy are listed in Table~\ref{tab:spec_props}. 

Note galaxy properties could not be measured for four \usmg\ systems as their host galaxy image is blended with the quasar image. For the same reason, we could measure the galaxy parameters only for 26 out of the 36 nearest host galaxies in our combined \usmg\ sample. To enable easy comparison, we have also measured all these parameters for host galaxies in the MAGIICAT sample using the same technique \citep[see][for details]{Guha2022a}. 

\subsubsection{\OII\ Luminosity}
\label{sec:oii_luminosity}

As mentioned earlier, except for two cases, we detected \OII\ nebular emission from \usmg\ host galaxies in our present sample. To obtain the spectroscopic redshifts, we fit the \OII\ emission line using a pair of Gaussian functions having the same redshifts and velocity widths with the centroids of the two Gaussian functions separated by the ratio of the rest wavelengths of the \OII\ $\lambda\lambda\, 3729,\, 3727$ lines. While fitting, the amplitude ratio of these two Gaussian functions is allowed to vary between 0.3 to 1.5 \citep{OsterbrockFerland2006}. In principle, this flux ratio can be used to constrain the electron density. However, due to the poor velocity resolution of the SALT spectra, the uncertainty in this ratio is large,  which prevents us from constraining the electron density accurately.

The \OII\ line fluxes (combined flux of the doublet) obtained from simple integration over the line profile are given in column (8) of Table \ref{tab:spec_props}. The \OII\ nebular line fluxes are then converted to the \OII\ line luminosities ($L_{[O\, \textsc{ii}]}$) based on their redshifts and the assumed background cosmology. Note that the measurements of \OII\ luminosity are affected by two factors -- the slit-loss and the dust attenuation. To correct for the slit-loss, we assume that the slit-loss is independent of the observed wavelength. Thus, we scale the observed spectra to match the synthetic spectra obtained from the SED fitting of available photometric points using the least square minimization method. We then multiply this scale factor with $L_{[O\, \textsc{ii}]}$ to correct for the slit-loss. However, to estimate the dust content in these galaxies, we would need either at least two \HI\ Balmer lines in emission or good-quality continuum spectra of these galaxies with sufficient SNR. As for all the cases, we have neither, so we do not correct for 
dust attenuation.

\begin{figure}
    \centering
    \includegraphics[width=0.49\textwidth]{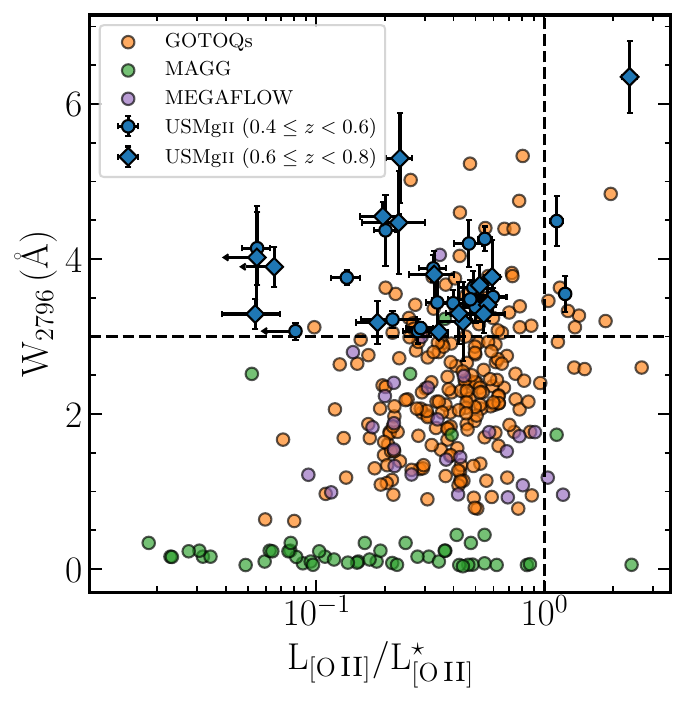}
    \caption{Scatter plot of $W_{2796}$ against the \OII\ line luminosity ($L_{[O\textsc{ii}]}$) of the associated \MgII\ host galaxies. To correct for redshift evolution, we have scaled the $L_{[O\textsc{ii}]}$ according to the characteristic \OII\ line luminosity ($L_{[O\textsc{ii}]}^\star$) at that redshift. The orange, green, purple, and blue points (diamond/circle) correspond to the \MgII\ host galaxies from the GOTOQs \citep{Joshi2017}, MAGG survey \citep{Dutta2020}, MEGAFLOW survey \citep{Schroetter2019}, and the \usmg\ host galaxies (high redshift / low redshift), respectively. The horizontal and vertical black dashed lines correspond to the $W_{2796}^{cut}$ of \usmg\ host galaxies and  $L_{[O\textsc{ii}]}^\star$ galaxies, respectively.}
    \label{fig:o2vsw}
\end{figure}

In Figure \ref{fig:o2vsw}, we show the scatter plot of $W_{2796}$ versus the $L_{[O\, \textsc{ii}]}$ of the associated \MgII\ host galaxies. To account for the redshift evolution, we have scaled the $L_{[O\, \textsc{ii}]}$ with respect to the characteristics \OII\ line luminosities \citep[$L_{[O\, \textsc{ii}]}^\star$,][]{Comparat2016} at the host galaxy redshifts. The orange, green, purple, and blue points correspond to the \MgII\ host galaxies from the GOTOQs \citep{Joshi2017}, MAGG survey \citep{Dutta2020}, MEGAFLOW survey \citep{Schroetter2019}, and the \usmg\ host galaxies, respectively. The $L_{[O\, \textsc{ii}]}$ for the \usmg\ host galaxies varies from $0.05L_{[O\, \textsc{ii}]}^\star$ to $2.37L_{[O\, \textsc{ii}]}^\star$ with a median value of $0.37L_{[O\, \textsc{ii}]}^\star$. Using the lowest \OII\ line luminosity as the detection threshold (i.e., $L_{min} = 0.05L^\star_{[O\, \textsc{ii}]}$) and the \OII\ line luminosity function \citep{Comparat2016} at $z\sim0.6$, we find that the total number of expected super-$L^{\star}_{[O\, \textsc{ii}]}$ galaxies for a random population of galaxies are only about 3\% implying only one super-$L^\star_{[O\, \textsc{ii}]}$ galaxies among the detected \usmg\ host galaxies. The expected median line luminosity for the random population of galaxies (at $z \sim 0.6$) is $\sim 0.1L^\star_{[O\, \textsc{ii}]}$. For our \usmg\ sample, however, we have three super-$L^{\star}_{[O\, \textsc{ii}]}$ galaxies and the median line luminosity ($0.37L_{[O\, \textsc{ii}]}^\star$) is also significantly high, even without correcting for dust extinction.

Out of all the GOTOQs, only  $\sim 5\%$ are super-$L^\star_{[O\, \textsc{ii}]}$ galaxies. If we restrict ourselves to GOTOQs that produce  \usmg\ absorption, we find $\sim 15\%$ of them are super-$L^\star_{[O\, \textsc{ii}]}$. Like the \usmg\ galaxies, the \OII\ luminosities of GOTOQs should  be taken as lower limits as they are not corrected for fiber loss and dust extinction. For the MEGAFLOW sample, out of 26 \MgII\ host galaxies, only two ($\sim 8\%$) are super-$L_{[O\, \textsc{ii}]}^\star$ galaxies. Among the 53 galaxies associated with \MgII\ absorption in the MAGG sample, only two ($\sim 4\%$) are super-$L_{[O\, \textsc{ii}]}^\star$ galaxies. In the above two samples, there are only three \usmg\ systems, and none of them are found to be a super-$L_{[O\, \textsc{ii}]}^\star$ galaxy.  Therefore, when we combine all the \usmg\ absorbers in different samples, we find $\sim12\pm4$\% of the host galaxies have  super-$L_{[O\, \textsc{ii}]}^\star$. This, together with the high mean luminosity found above, confirms the early finding of \citet{Guha2022a} that \usmg\ absorbers are preferentially hosted by galaxies having higher-$L_{[O\, \textsc{ii}]}$ compared to field galaxies.

A Spearman rank correlation analysis between $W_{2796}$ and $L_{[O\, \textsc{ii}]} / L_{[O\textsc{ii}]}^\star$ yields a significant correlation between these two quantities ($r_S = 0.28$ and $p$-value $\sim 10^{-7}$), which is also reported in literature \citep{Menard2011, Joshi2018}. However, this correlation is driven by weak \MgII\ absorbers. The correlation gets weaker as we restrict ourselves to the stronger \MgII\ absorbers \citep[see][for details]{Guha2022a}. For example, for $W_{2796} > 1$\AA, Spearman rank correlation analysis gives $r_S = 0.16$ and $p$-value = 0.008, while for $W_{2796} > 2$\AA, we get $r_S = 0.05$ and $p$-value = 0.52. It is also clear from Figure~\ref{fig:o2vsw} that there is no clear trend among the \usmg\ absorbers between $W_{2796}$ and $L_{[O\, \textsc{ii}]}$.

Next, we look for any redshift evolution in the \OII\ luminosity of \usmg\ host galaxies over the redshift range $0.4 \leqslant z \leqslant 0.8$. We do not find any redshift evolution in the \OII\ luminosities of the \usmg\ host galaxies with a Spearman correlation coefficient $r_s = 0.07$ with $p$ value of 0.68. This is contrary to the mild evolution shown by $L_{[O\, \textsc{ii}]}^\star$ over the same redshift range \citep{Comparat2016}. It will be useful to confirm this with better-quality spectra after applying appropriate dust corrections.

\subsubsection{Star formation rates}
Next, we consider the current star formation rates (SFR) of the \usmg\ host galaxies. We measure the star formation rates using two different methods. The first method is based on the \OII\ nebular line luminosity \citep[][given inside the brackets in the last column of Table~\ref{tab:spec_props}]{Kennicat1998}, and the second method is based on the SED fitting (last column of Table~\ref{tab:spec_props}). The SFR obtained using the \OII\ emission line has systematic uncertainties that depend on variation in reddening, chemical abundance (stellar mass through the mass metallicity relations), and ionization \citep{Moustakas2006, Gilbank2010}.
The inferred SFR using \OII\ nebular emission is less than measured using SED fitting  and as in \citet{Guha2022a}, for the reasons mentioned above,
we use the SFR measurements based on SED fitting in our analysis.

\begin{figure}
    \centering
    \includegraphics[width=0.49\textwidth]{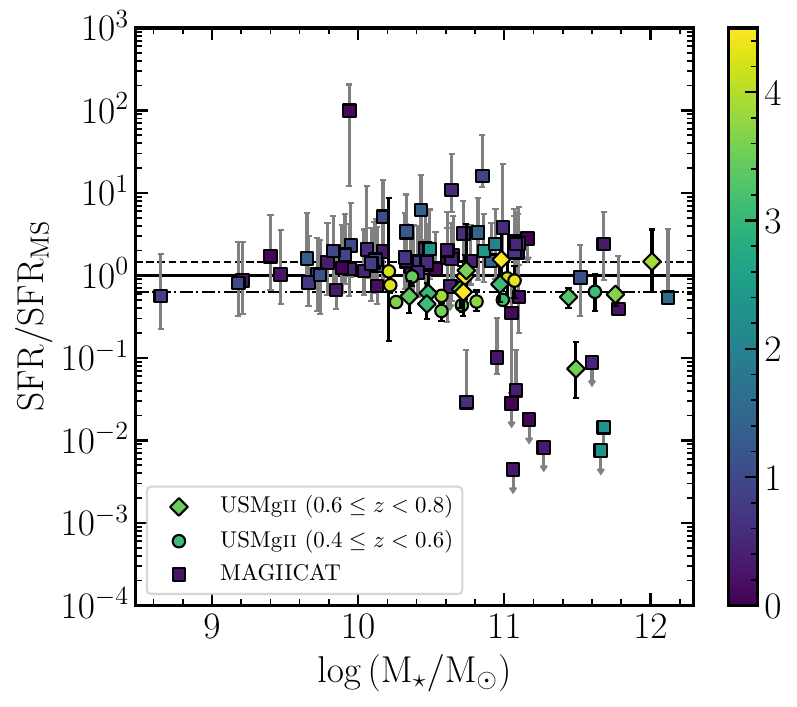}
    \caption{Scatter plot of the ongoing star formation rates  (SFR) scaled by the star formation rate of a main sequence galaxy  (SFR$_{\rm MS}$) of the same mass at the same redshift, color-coded as $W_{2796}$ against the stellar masses of the \MgII\ absorbers. The diamond, circle, and square markers correspond to the high redshift \usmg\ host galaxies, low redshift \usmg\ host galaxies, and the \Magiicat\ host galaxies, respectively. The solid black line corresponds to the main sequence SFR. The dotted and the dash-dotted line correspond to the median SFR of \Magiicat\ and \usmg\ host galaxies, respectively.}
    \label{fig:mass_vs_sfr}
\end{figure}

We find a mild redshift evolution of the SFR of the \usmg\ host galaxies, obtained using SED fitting, over the redshift range  $0.4 \leqslant z \leqslant 0.8$. The Spearman correlation coefficient between the SFR and $z$, $r_s =  0.61$, with $p$ value being $\sim$0.001. { Since our host galaxy sample is magnitude limited some of the redshift evolution may be biased by this.}
In Figure \ref{fig:mass_vs_sfr}, we show the scatter plot between the ongoing star formation rates obtained from the SED fitting scaled by the star formation rate of a main sequence galaxy(SFR$_{\rm MS}$) of the same mass at the same redshift \citep{Speagle_2014}, against their stellar masses. We scale the SFR with the main sequence SFR to account for redshift evolution.  
The points are color-coded by $W_{2796}$. The solid black line corresponds to the main sequence SFR. The dotted and the dash-dotted line correspond to the median SFR of \Magiicat\ and \usmg\ host galaxies, respectively. The new data added from this work confirms that the \usmg\ host galaxies are not starburst galaxies and have star formation rates slightly lower than that expected for the main-sequence galaxies \citep{Guha2022a}. However, note that the star formation rates are obtained from SED fittings based on the multiband galaxy photometries without the galaxy spectra can have large offsets (up to a factor 2) and can at best be main sequence galaxies.


\subsection{$\rm W_{2796}$ versus impact parameter}
\label{sec:w_vs_d}
\begin{figure*}
    \includegraphics[width=\textwidth]{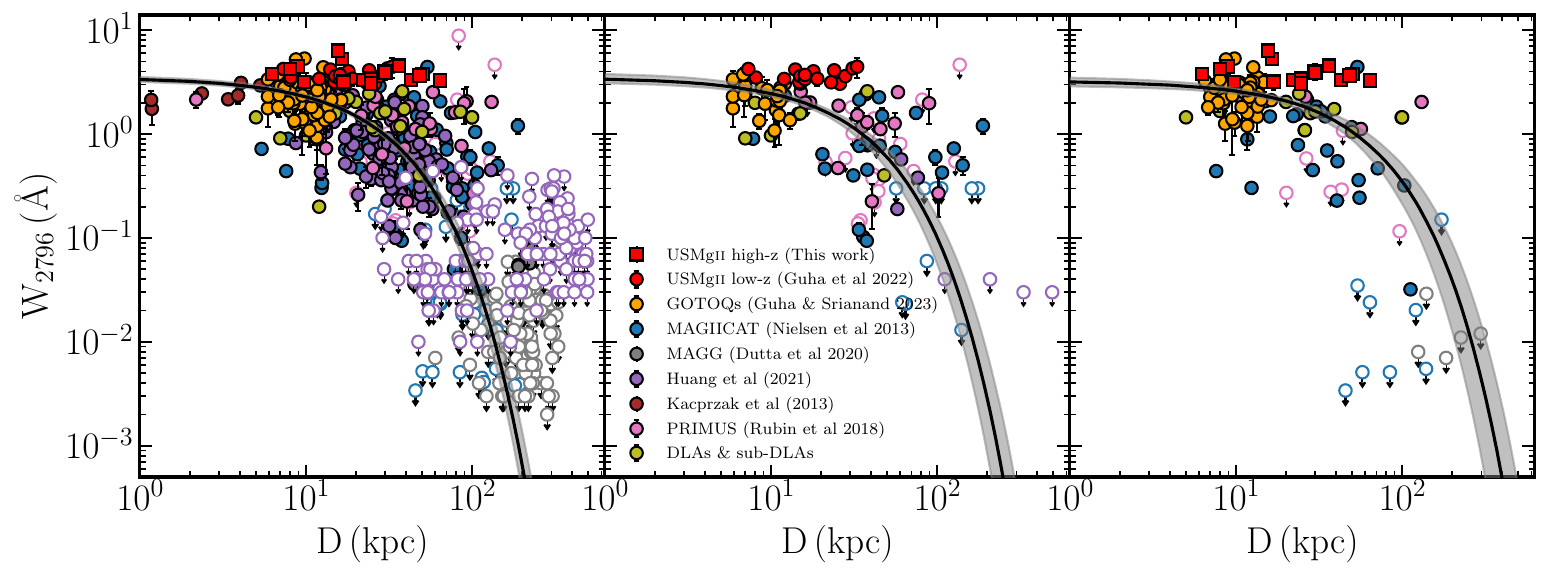}
    \caption{The \emph{left panel} shows the impact parameter (D) versus the $\rm{W_{2796}}$ anti-correlation over the full redshift range for the isolated galaxies. The red squares correspond to the high-redshift \usmg\ absorption systems. The red (circle), orange, blue, violet, gray, brown, green, and pink points are taken from the low redshift \usmg\ survey \citep{Guha2022a}, GOTOQs \citep{Guha2022b},  MAGIICAT survey \citep{Nielsen_2013}, \citet{Huang2021}, MAGG survey \citep{Dutta2020}, \citet{Kacprzak_2013}, DLAs and sub-DLAs, and \citet{Rubin2018} respectively. The solid black line corresponds to the best-fit log-linear model, and the shaded region corresponds to the $1\sigma$ errors associated with it. The \emph{middle } and \emph{right panel} show the same, but only for the redshift ranges, $0.4 \leqslant z_{abs} < 0.6$, and $0.6 \leqslant z_{abs} \leqslant 0.9$ respectively.}
    \label{fig:w_vs_d}
\end{figure*}

In this section, we revisit the well-known anti-correlation between $W_{2796}$ and impact parameter \citep{bergeron1991, steidel1995, Chen_2010, Nielsen_2013} for the \MgII\ absorbers. In particular, we are interested in (i) understanding this relation for \usmg\ absorbers by combining the present data with those from \citet[][]{Guha2022a} and (ii) how the global fit for the \MgII\ systems gets modified by the inclusion of  \usmg\ absorbers that mostly populate the low D - high $W_{2796}$ region of the $\rm W_{2796}-D$ plane.  For consistency and simplicity, we consider the host galaxy as the one with the smallest impact parameter in any of the samples used here when there is more than one galaxy identified at the correct redshift.

In the left panel of Figure \ref{fig:w_vs_d}, we show the points from our \usmg\ sample, GOTOQs from \citet{Guha2022b}, and additional measurements from different samples from the literature. The red points (circles for galaxies from \citet{Guha2022a} and squares for galaxies from this work) correspond to the \usmg\ host galaxies from our observations. The orange, blue, violet, gray, brown, green, and pink circular points correspond to the isolated galaxies in \citet{Guha2022b},  MAGIICAT survey \citep{Nielsen_2013}, \citet{Huang2021}, MAGG survey \citep{Dutta2020}, \citet{Kacprzak_2013}, DLAs and sub-DLAs \citep{Rahmani2016}, and \citet{Rubin2018} respectively. For DLAs and sub-DLAs, we obtained the $W_{2796}$ from measurements available in the literature and used the impact parameters quoted in \citet{Rahmani2016}. It is good to recollect that the $W_{2796}$  measurements reported in  \citet{Rubin2018} are towards background galaxies, unlike all other measurements towards quasars.
We found some of the \MgII\ systems in the literature sample are observed in two different surveys, and we made sure that the individual points plotted in Figure~\ref{fig:w_vs_d} and used for our statistical analysis correspond to a unique \MgII\ system by removing the duplicates. 

\subsubsection{Relationship for the \usmg\ systems}
\begin{figure}
    \centering
    \includegraphics[width=0.49\textwidth]{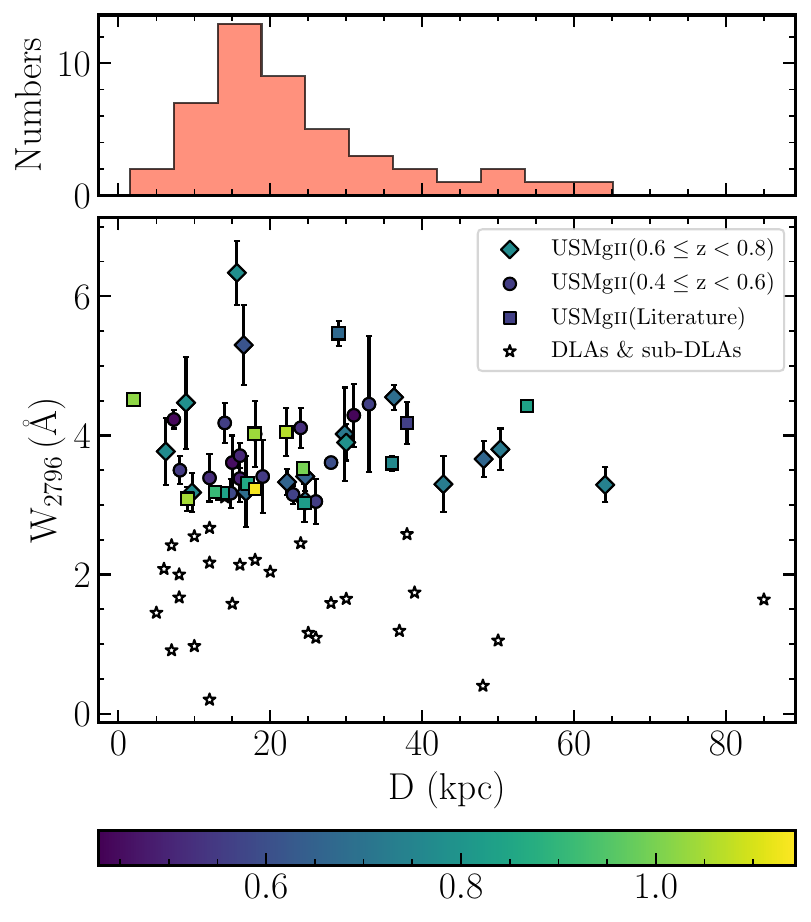}
    \caption{$W_{2796}$ vs the impact parameters ($D$) for the \usmg\ host galaxies color coded by their absorption redshifts. The diamonds (circles) indicate our measurements for $0.4\lesssim z \lesssim 0.6$ ($0.6 \lesssim z\lesssim 0.9$) \usmg\ sample. Squares show the data from the existing literature. The open stars are for the DLAs and sub-DLAs in the redshift range $0.4 \lesssim z \lesssim 1.0$ \citep{Rahmani2016}. The top panels show the distribution of impact parameters for the \usmg\ absorbers.
    }
    \label{fig:usmg_only}
\end{figure}

\begin{figure}
    \centering
    \includegraphics[width=0.475\textwidth]{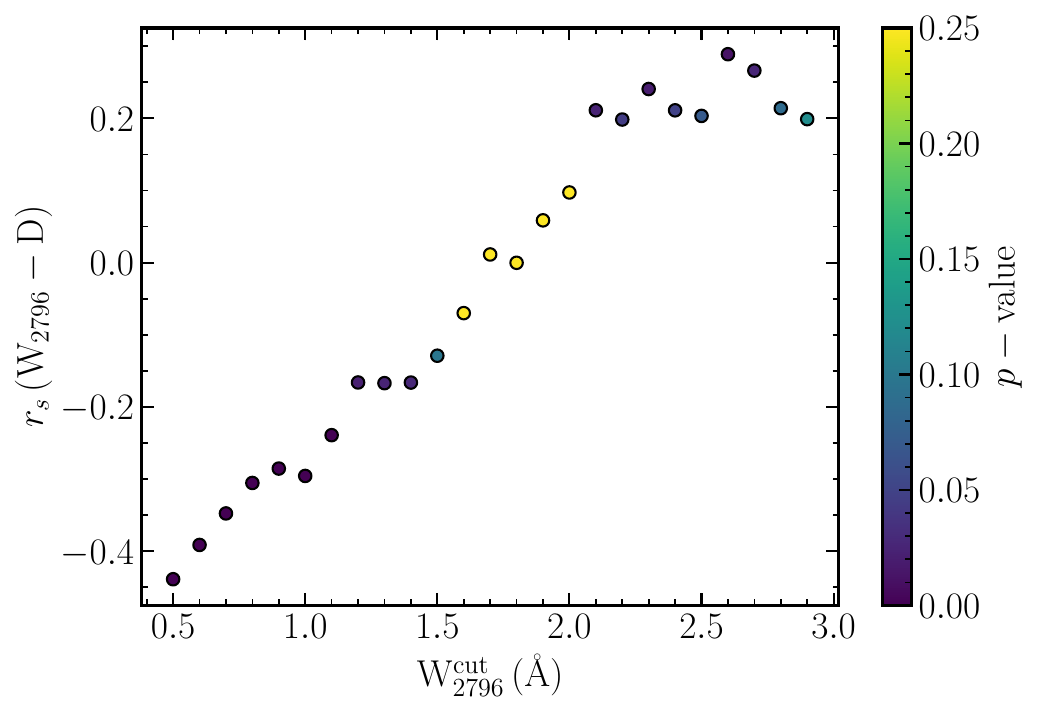}
    \caption{The Sprearmann rank correlation coefficient between the $W_{2796}$ and D for a given $W_{2796}^{cut}$ color-coded according to the $p$ values with the null hypothesis that these two datasets are uncorrelated. For a small $W_{2796}^{cut}$, there is a significant anti-correlation between $W_{2796}$ and D. However, as $W_{2796}^{cut}$ increases, the significance of the anti-correlation keeps on decreasing and completely vanishes for $W_{2796}^{cut}$ of 1.5 \AA .}
    \label{fig:w_corr}
\end{figure}

First, we study the $W_{2796}-D$ distribution of \usmg\ systems
(see Figure~\ref{fig:usmg_only}). 
The top part of Figure~\ref{fig:usmg_only} shows the distribution of impact parameters for \usmg\ systems. The impact parameter of the \usmg\ host galaxies in the present sample varies from 6.24 to 64.1 kpc with a median of 23 kpc. Points in this plot are color-coded by their absorption redshifts. The red squares (circles) correspond to our measurements for high-$z$ (low-$z$)  \usmg\ systems, whereas those from the  literature \citep[from][]{Dutta2020, Bouche2007, Schroetter2015, Nestor2011} are shown with circles of different colors. For the combined \usmg\ sample, the impact parameter ranges from 6.24 kpc to 79 kpc with a median of 19 kpc. As noted by \citet{Guha2022a}, the \usmg\ systems do not follow the $W_{2796}$-D anti-correlation. The Spearman correlation test yields no significant correlation ($r_S = 0.12$, $p$ = 0.40) between these two quantities.

It is well known that the $W_{2796}-D$ relationship has a large scatter, and absorbers of a given rest equivalent width can originate from a wide range of impact parameters. Since our sample is based on large $W_{2796}$ cutoff (i.e $\mathrm{W_{2796}^{cut}} = 3$\AA), it is important to check how  the imposed $\mathrm{W_{2796}^{cut}}$ influences the measured correlation. To explore this, we measure the Spearman rank correlation coefficient $r_s$ between $W_{2796}$ and $D$ for different values of  $W_{2796}^{\rm{cut}}$ for the full sample. We notice that the strength of the anti-correlation between the $W_{2796}$ and $D$ becomes statistically insignificant if we confine ourselves to the stronger $W_{2796}$, i.e., $W_{2796}^{\rm{cut}}\ge 1.5$ \AA (see Figure \ref{fig:w_corr}).  For higher values of $W_{2796}^{\rm{cut}}$, the data suggests a positive correlation coefficient but with less statistical significance. This demonstrates that when host galaxies are studied for an absorber-centric sample, the derived relationship between $W_{2796}$ and $D$ will be very sensitive to the $W_{2796}$ cut-off used. Recently \citet{DeFelippis2021} discussed the $W_{2796}$ vs. $D$ relationship in their cosmological hydrodynamical simulations (see their figure 4). While high $W_{2796}$ absorption are not well represented in their simulations, there is also a lack of correlation between $W_{2796}$ and D for strong \MgII\ absorber (defined as $W_{2796}>0.5$\AA) along selected sightlines. On the other hand, for sightlines generated around halos, the anti-correlation is seen with a large scatter. Our findings align with these results even though the simulated $W_{2796}$ are much weaker than that of \usmg\ systems.

It is known that galaxy properties like $M_*$, $M_B$, and the star formation rate may introduce scatter in the $W_{2796}$ vs. D relationship. In the following sections, we will explore whether the large scatter in D is related to the scatter in the galaxy properties. Alternatively, the lack of correlation between $W_{2796}$ and D could be related to a good fraction of \usmg\ absorbers not originating from the CGM around a single galaxy but contributed by several galaxies. Below we will discuss this possibility as well.

\begin{figure}
    \centering
    \includegraphics[width=0.475\textwidth]{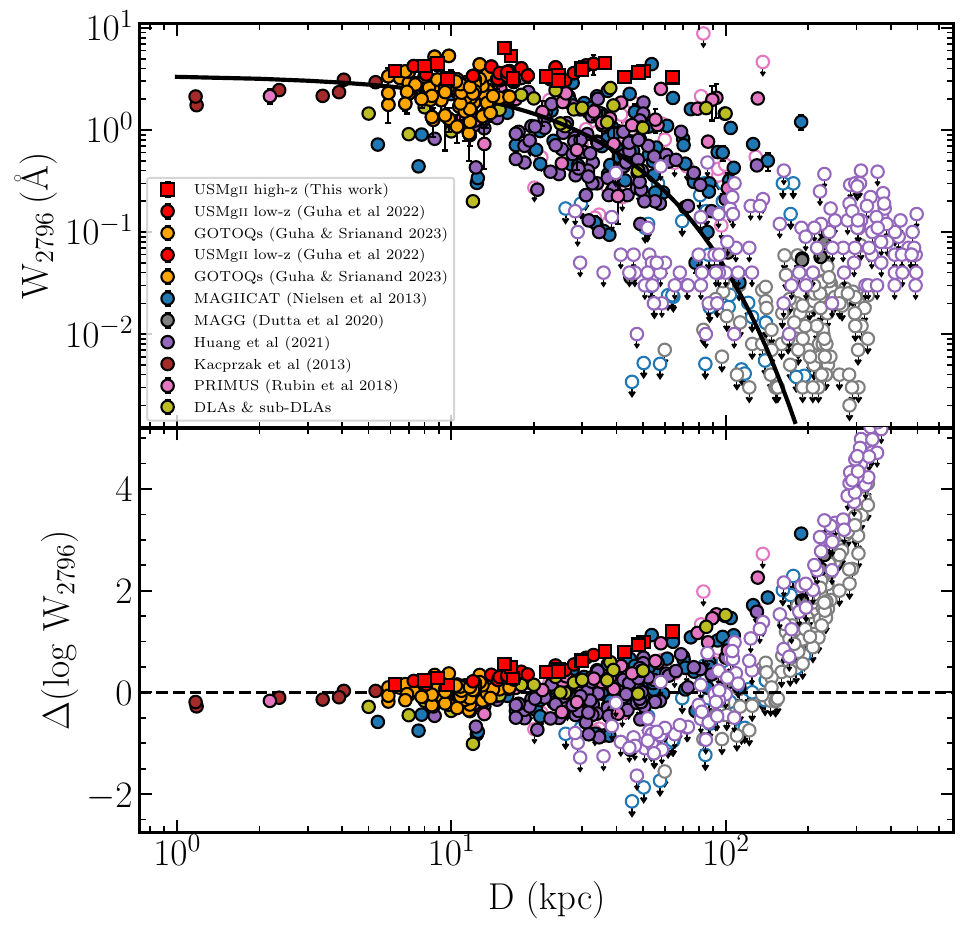}
    \caption{The scatter in the $\rm W_{2796}-D$ anti-correlations. The top panel shows various \MgII\ absorption systems in the $\rm W_{2796}-D$ space from this work and the literature. The solid black line represents the best-fit log-linear curve. This plot is the same as the left panel of Figure \ref{fig:w_vs_d}. The bottom panel shows the difference between the observed $\rm \log\, (W_{2796})$ and the best-fit model for that impact parameter.}
    \label{fig:w_vs_d_scatter}
\end{figure}

\subsubsection{Redshift evolution}
Here, we study the possible redshift evolution of the $W_{2796}-D$ relationship for the full \MgII\ absorber population. The solid black line in Figure~\ref{fig:w_vs_d} corresponds to the best-fit log-linear model, and the shaded region corresponds to the $1\sigma$ error associated with it. The maximum likelihood fitting prescription we use, including the upper limits, is provided in \citet{Guha2022a}. The middle and right panels show the same, but for the redshift ranges, $0.4 \leqslant z_{abs} < 0.6$, and $0.6 \leqslant z_{abs} \leqslant 0.9$ respectively. The best-fitted parameters for the full sample and systems in two redshift ranges are given in Table \ref{tab:best_fits}. The quoted values of $\sigma$ in Table~\ref{tab:best_fits} are larger than most of the literature values. In the bottom panel of Figure~\ref{fig:w_vs_d_scatter}, we show the residual of log of equivalent width (i.e., observed log~$W_{2796}$ minus the predicted value from our best fit) as a function of impact parameter. It is evident that \usmg\ points form the upper envelope of the scatter, and the scatter increases as we go towards higher D.

We attribute the large scatter around the best-fit relation to the lack of correlation between $W_{2796}$ and $D$ found for the \usmg\ systems and the large scatter in $W_{2796}$ at the low-D found for GOTOQs. As described in \citet{Guha2022b} the inclusion of data at low impact parameters has increased the value of $\beta$ compared to what has been reported in the literature. For the combined sample, our best fit parameters suggest $W_{2796}$ (D=0)$\sim$3.5\AA\ and characteristic impact parameter scale (i.e $-1/(2.303 \times \alpha)$)  is $\sim$ 23 kpc. 

\begin{figure}
    \centering
    \includegraphics[width=0.475\textwidth]{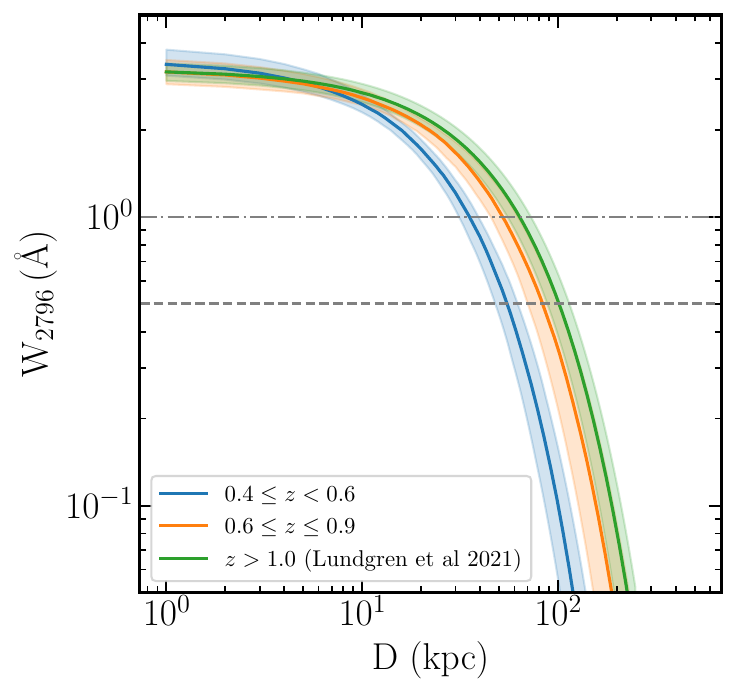}
    \caption{Redshift evolution of the $\rm W_{2796}-D$ anti-correlation. The blue and orange lines correspond to the best fit $\rm W_{2796}-D$ curve for the redshift range $0.4 \leqslant z < 0.6$ and $0.6 \leqslant z \leqslant 0.8$, respectively. The green curve is taken from \citet{Lundgren2021} and corresponds to the best fit $\rm W_{2796}-D$ anti-correlation for the redshift range $z \geqslant 1$. The grey dot-dashed and dashed horizontal lines correspond to $\rm W_{2796} = 1$\AA\ and $\rm W_{2796} = 0.5$\AA\ respectively.}
    \label{fig:w_d_evolve}
\end{figure}

Neither $\alpha$ nor $\beta$ shows any significant change between the two distinct redshift ranges considered here. In addition, the KS-test confirms the $W_{2796}$ distribution in these two redshift bins is not significantly different (p-value of 0.06). To avoid any bias introduced by differences in the $W_{2796}$ distribution, we considered several sub-sample by matching the distribution of $W_{2796}$ and confirming the lack of any significant redshift evolution in both $\alpha$ and $\beta$. For ease of viewing, in Figure~\ref{fig:w_d_evolve}, we plot the best-fit relations and associated errors for the two redshift ranges. As mentioned, W$_{2796}$ (D=0) does not evolve with redshift.  Even though there is no significant difference between the best-fitted values of $\alpha$, the plot suggests a slight excess in  W$_{2796}$ at large D for the high-$z$ sub-sample.

\citet{Lundgren2021} have reported $\alpha = -0.008\pm0.001$ and $\beta = 0.51\pm0.03$  for $z>1$ using their data combined with the available data from the literature \citep[from][]{Bouche2012nosfr, Lundgren2012, Lundgren2021, Schroetter2019}.  The $\beta$ value they obtained is consistent with what we report in Table~\ref{tab:best_fits}.  The $\alpha$ value quoted by \citet{Lundgren2021} is slightly low but consistent within 2.3$\sigma$ with our measurements in two low-$z$ bins.  By comparing their values of $\alpha$ and $\beta$ with those of \citet{Nielsen_2013} (i.e., $\alpha = -0.015\pm0.002$ and $\beta = 0.27\pm0.11$) \citet{Lundgren2021} suggested a possible evolution in the gas distribution around galaxies at $z\sim 0.4$ and $z\sim1.5$. To ensure that our result is not dominated by any bias arising from differences in the $W_{2796}$ distribution, we have measured $\alpha$ and $\beta$ of 500 random sub-samples from our sample that have the same number of absorbers having the distribution of  $W_{2796}$ similar to that of the sample of \citet{Lundgren2021}. Using the log-linear fits of these realizations, we  obtain $\alpha=-0.011\pm0.001$ and $\beta=0.48^{+0.02}_{-0.01}$.  This again confirms the lack of strong redshift evolution in $\alpha$ and $\beta$. In Figure~\ref{fig:w_d_evolve}, we also show the best-fit relationship for the high-$z$ sample of \citet{Lundgren2021}. As before, we notice that at large impact parameters, the best fits suggest higher $W_{2796}$ at high-$z$. This figure also suggests no significant redshift evolution in the $\rm W_{2796}$ vs. $D$ relationship at $D<20$ kpc.

\citet{Dutta2020} have obtained $\beta$ = $-0.05^{+0.42}_{-0.38}$ and $\alpha=-0.010\pm0.003$ for closest galaxies at $z\sim0.8-1.5$ to the line of sight to 28 background quasars. Their $\alpha$ values are consistent with our values. However, the inferred   $W_{2796}$(D=0) value is lower than ours, albeit with large errors.  This is related to the poor representation of galaxies with low impact parameters (i.e., $D<20$ kpc) in their sample. 

\begin{table}
    \centering
    \begin{tabular}{lccc}
        \hline
         Redshift & $\alpha$ & $\beta$ & $\sigma$ \\
         \hline
         $0 < z < 1.5$ & $-0.019\pm0.002$ & $0.540\pm0.028$ & $1.09\pm0.05$ \\
         $0.4 \leqslant z < 0.6$ & $-0.015\pm0.003$ & $0.538\pm0.045$ & $1.13\pm0.10$ \\
         $0.6 \leqslant z \leqslant 0.9$ &  $-0.010\pm0.002$ & $0.514\pm0.044$ & $1.29\pm0.10$ \\
         \hline
    \end{tabular}
    \caption{Best fitting parameters for the log-linear characterization of the $W_{2796}$ vs $D$ anti-correlations for different redshift ranges.}
    \label{tab:best_fits}
\end{table}

How the $W_{2796}-D$ relationship evolves with $z$ has an important consequence on the redshift evolution of the number of \MgII\ absorbers per unit redshift path length (i.e., $dN/dz$). For absorbers with $W_{2796}>1$\AA, $dN/dz$ = 0.16, 0.20 and 0.31 for $z$ = 0.5, 0.7 and 1.3 respectively \citep{Zhu2013}.  When we consider $W_{2796}>0.5$\AA, the corresponding values are 0.41, 0.47, and 0.62. Thus systems with the largest equivalent width tend to show a slightly more rapid increase with redshift compared to those with the lowest equivalent width. In a simple model where all known galaxies have a spherical gaseous halo of the same radius and gas covering factor $f_c$, the average number of \MgII\ absorbers per unit redshift interval is given by $\langle dN/dz \rangle = f_c \pi D_{max}^2\int_{L_{min}}^\infty \phi(L, z) dL \times dl / dz$. Here, $D_{max}$  is the maximum impact parameter for a given value of $W_{2796}^{cut}$ for which $dN/dz$ is computed. $\phi(L, z)$ is the galaxy luminosity function at redshift $z$, and $dl/dz$ is the differential comoving path length per unit redshift interval.  

From \citet{Faber2007} we find the average number density of galaxies (with $-30\le M_B \le -10$ and $M^*_B \sim -21.3$) per unit physical volume is $\sim$0.90 and $\sim$1.74 Mpc$^{-3}$ at $z$ = 0.5 and 1.1 respectively. Thus a factor of 2 change in $dN/dz$ over the redshift range $0.5\le z \le 1.1$ can purely originate from the redshift evolution of the galaxy luminosity function. From Figure~\ref{fig:w_d_evolve}, if the covering fraction is independent of $D$, then the number density of low $W_{2796}$ is expected to show stronger redshift evolution. This is contrary to the observed results quoted above. Thus one requires the covering fraction to change with D. This expectation is consistent with the finding of \citet{Lan2020}. Their results suggest that over the redshift range $0.5 \leqslant z \leqslant 1.1$, the covering fraction of star-forming galaxies ($M_\star \sim 10^{10}\, M_\odot$) associated with strong \MgII\ absorption ($W_{2796} \ge 1$\AA) increases from 0.19 to 0.41, implying an increase of the covering fraction by a factor of $\sim 2$. The same for weak absorbers ($W_{2796} < 1$\AA) changes from 0.15 to 0.24, implying an increase by a factor of $\sim 1.5$.

\begin{figure*}
    \begin{subfigure}{0.49\textwidth}
    \centering\includegraphics[width=\textwidth]{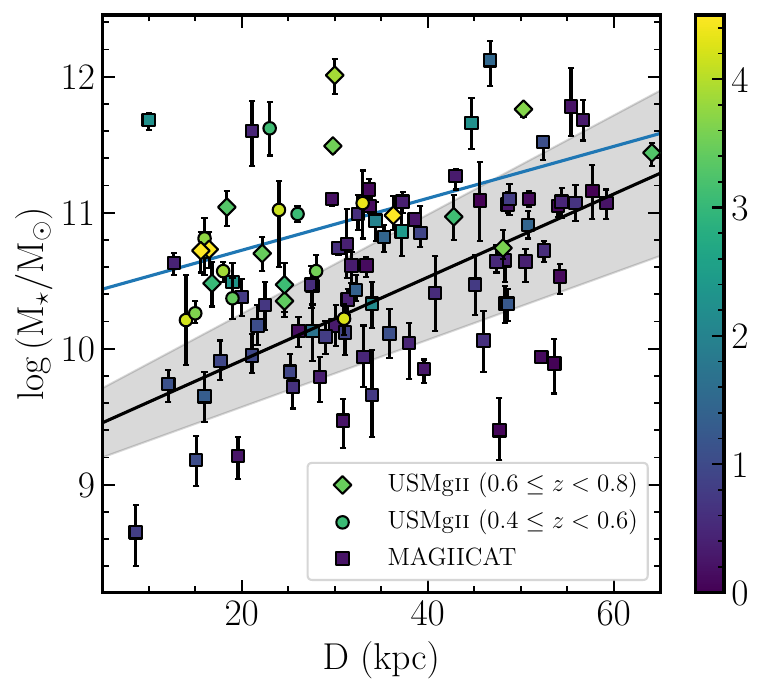}
  \end{subfigure}
  \begin{subfigure}{0.49\textwidth}
    \centering\includegraphics[width=\textwidth]{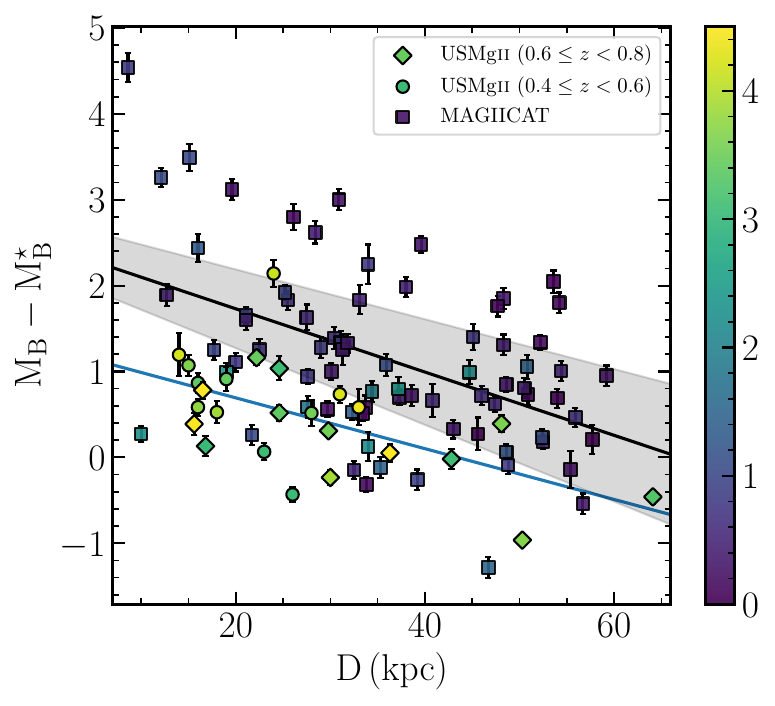}
  \end{subfigure}
  \caption{Left panel: Stellar masses of the \MgII\ host galaxies against their impact parameters, color-coded as $W_{2796}$.  Right Panel: Rest frame absolute B-band magnitudes of the \MgII\ host galaxies relative to the characteristic B-band magnitude at the same redshift against their impact parameters. For both panels, the diamond, circle, and square markers correspond to the high redshift \usmg\ host galaxies, low redshift \usmg\ host galaxies, and the \Magiicat\ host galaxies. The solid black lines correspond to the linear regression fit to the \Magiicat\ host galaxies, and the grey regions correspond to the 1$\sigma$ uncertainty to the fit. The solid blue line corresponds to the linear fit, but only for the \usmg\ host galaxies from our survey.}
    \label{fig:d_vs_mass_mag}
\end{figure*}

\subsubsection{Dependence on stellar mass}
\label{sec:stellarmass}

\citet{churchill2013} have noted that for a given $W_{2796}$, the host galaxies at high impact parameters tend to have larger halo masses. This section investigates the presence of such a trend in our \usmg\ sample. In column 6 of Table \ref{tab:spec_props}, we provide the present sample's stellar masses of the \usmg\ host galaxies.  The stellar mass of the \usmg\ host galaxies at $z \sim 0.7$ ranges from $10.35 \leqslant \log(M_\star / M_\odot) \leqslant 12.01$ with a median value of 10.86. In the combined sample of \usmg\ host galaxies, the stellar mass ranges from $10.21 \leqslant \log(M_\star / M_\odot) \leqslant 12.01$ with the median of $\log(M_\star / M_\odot) = 10.73$. We find a mild evolution in the stellar masses of the \usmg\ host galaxies over the redshift range $0.4 \leqslant z \leqslant 0.8$ (Spearman correlation coefficient, $r_s = 0.48$ with $p$ value 0.01). {As mentioned before, some redshift evolution may be influenced by the fact that we are using a magnitude-limited sample.}
For the MAGIICAT host galaxies, the stellar masses are found to be in the range $8.65 \leqslant \log(M_\star / M_\odot) \leqslant 12.12$ with a median value of 10.51. Even though the two samples overlap in $M_*$ ranges, the \usmg\ host galaxies tend to have higher $M_*$ values as suggested by the median values.

In the left panel of Figure \ref{fig:d_vs_mass_mag}, we show the scatter plot of $M_\star$ versus $D$ for host galaxies of the general \MgII\ system population from the MAGIICAT sample and \usmg\ systems in our sample color-coded according to the $W_{2796}$. For the MAGIICAT host galaxies, we also find a correlation between $M_\star$ and $D$, which can be characterized by a linear fit of the form $\log(M_\star / M_\odot) = (0.036\pm0.006)D\,(\text{kpc}) + (9.301 \pm 0.225)$ \citep[see][]{Guha2022a}. The solid black straight line corresponds to this fit, and the grey region to the 1$\sigma$ uncertainty to the fit. This aligns  with the findings of \citet{Churchill_2013, Rubin2018, Huang2021}, who reported that massive galaxies produce stronger absorption for a similar impact parameter.
Considering the \usmg\ host galaxies alone, we find a possible mild correlation between $M_\star$ and $D$. A Spearman rank correlation analysis returns a correlation coefficient of $r_S = 0.5$ and a $ p$-value of 0.009. A linear fit to the data for the \usmg\ galaxies gives  $\log(M_\star / M_\odot) = (0.019\pm0.007)D\,(\text{kpc})  + (10.341 \pm 0.203)$. The solid blue line in the left panel of Figure \ref{fig:d_vs_mass_mag} corresponds to this fit. For a given impact parameter,  the stellar mass of host galaxies of \usmg\ absorbers is systematically higher than the low equivalent width systems. For example, at median D$\sim$23 kpc (or D$\sim$0 kpc)  of the \usmg\ sample, the host galaxies of \usmg\ systems are a factor 4 (or$\sim$ 10) times higher than absorbers seen in the MAGIICAT sample.

In the MAGIICAT sample, when we consider all cases with $W_{2796}$ measurements and not upper limits,  we find that the correlation is stronger between $W_{2796}$ and $\rm D +0.16 (\log (M_\star/M_\odot)- 10.51)$ (i.e.,  $r_s = -0.47$ with a p-value of $\sim 10^{-4}$) compared to the correlation between $\rm W_{2796}$ and $D$ (i.e.,  $r_s = -0.37$ with a p-value of $\sim 10^{-3}$). We repeated the  same exercise for \usmg\ absorbers alone. We find for  $W_{2796}-D$ relation the correlation coefficient is  $r_s$ = $-0.113$ with p-value = 0.59. After minimizing  the mass dependent scatter, (i.e $\rm W_{2796}$  vs $\rm D +  0.017(log (M_*/M_\odot) - 10.73)$) correlation coefficient is $r_s$ = $-0.112$ with p-value = 0.590. 

We quantify this further using the relationship between $W_{2796}$, D and $M_*$ found by \citet{Huang2021}. For the median stellar mass (i.e. log $(M_* / M_\odot)$ = 10.73) and impact parameter (D$\sim$24 kpc), their best-fitted relationship predicts $W_{2796}\sim1.15$\AA. On the contrary, to have $W_{2796}>3$\AA\, D should be less than 9.5 kpc for the host galaxy with $M_*$ close to the median value of the \usmg\ host galaxies. {\it While \usmg\ absorbers are massive for a given impact parameter (as found for the general population of \MgII\ absorbers), they do not follow the best-fit relationship between $W_{2796}$, $D$ and $M_\star$ obtained for the general \MgII\ population. This could be related to any prevailing physical conditions specific to \usmg\ absorbers or some of the high impact parameter \usmg\ absorbers originating from more than one galaxy.}

\subsubsection{Dependence on rest frame B-band absolute magnitude}
\label{sec:b_band_mag}
Next, we consider the rest frame absolute B band magnitudes of the \usmg\ and the host galaxies in the MAGIICAT sample. The rest-frame absolute B band magnitudes ($M_B$) are obtained from the synthetic SED fitted spectra of the galaxies using the method described in \citet{Guha2022a}. The rest-frame B-band magnitude for the \usmg\ host galaxies at $z \sim 0.7$ ranges from $-22.46 \leqslant M_B \leqslant -20.27 $ with a median value of $-21.23$. Upon combining with our $z \sim 0.5$ \usmg\ host galaxies \citep{Guha2022a}, the rest frame B-band magnitude ranges from $-22.46 \leqslant M_B \leqslant -19.05$ with a median value of $-20.85$. For the MAGIICAT galaxies, $M_B$ ranges $-22.56 \leqslant M_B \leqslant -16.53$ with a median value of $-20.08$. Column (7) of Table \ref{tab:spec_props} provides the $M_B$ of the \usmg\ host galaxies identified in this work. Note, as in the case of $M_\star$, $M_B$ measurements are possible only for 26 out of 36 host galaxies as there is the contamination of quasar light in the remaining cases.

In the right panel of Figure \ref{fig:d_vs_mass_mag}, we show the scatter plot of 
$M_B$ relative to the B-band characteristic magnitude \citep[$M_B^\star$,][]{Faber2007} at the galaxy redshift versus $D$ color-coded according to the $W_{2796}$. For the MAGIICAT host galaxies, just like for the stellar mass, we also find a correlation between host galaxy B band absolute magnitude and $D$, which can be characterized by a linear fit of the form $M_B - M_B^\star = (-0.037\pm0.008)D\,(\text{kpc})  + (2.445\pm0.302)$. The solid black straight line corresponds to this fit, and the gray region associated 1$\sigma$ uncertainty to the fit. It is evident from the figure that compared to the MAGIICAT host galaxies, for a given $D$, the \usmg\ host galaxies tend to be brighter. This aligns with the findings of \citet{Chen_2010, Huang2021}, who reported that brighter galaxies produce stronger absorption for a given impact parameter. Like the MAGIICAT host galaxies, we find that $M_B - M_B^\star$ of \usmg\ host galaxies are also anti-correlated with $D$. A Spearman rank correlation analysis provides $r_S = -0.58$ with $p$ value 0.002. A linear fit to the data for the \usmg\ galaxies alone gives  $M_B - M_B^\star = (-0.028\pm0.009)D \,(\text{kpc}) + (1.234 \pm 0.252)$.  The solid blue line in the left panel of Figure \ref{fig:d_vs_mass_mag} corresponds to this fit.  At the median impact parameter for the \usmg\ host galaxies (i.e., D=23 kpc), the host galaxies of the \usmg\ absorbers are $\sim$1 mag brighter than that of the host galaxies in the MAGIICAT sample.

Just like the stellar mass, for the MAGIICAT sample, we find that the anti-correlation is stronger for the quantity $\rm D\, (kpc) -0.153(M_B + 20.08)$ and $\rm W_{2796}$ compared to simple $\rm W_{2796}-D$ anti-correlation. The correlation coefficient increases from $r_s = -0.37$ (p-value $\sim 10^{-3}$) to $r_s = -0.49$ (p-value $\sim 10^{-5}$). However, such a linear combination does not significantly alter the correlation coefficients for the \usmg\ host galaxies. If we use the relationship between $W_{2796}$, D and $M_B$ found by \citet{Huang2021} the expected $W_{2796}$ is 1.17\AA\ when we use the median D and $M_B$.  {\it Thus, even though the host galaxies of \usmg\ are bright at a given $D$ compare to the host galaxies of normal \MgII\ absorbers they do not follow the relationship between $W_{2796}$, D and $M_B$ obtained for the general \MgII\ host galaxies.}

\subsubsection {$W_{2796}$ versus normalized impact parameter}

\begin{figure}
    \centering
    \includegraphics[viewport=11 10 350 500, clip=true,width=0.475\textwidth]{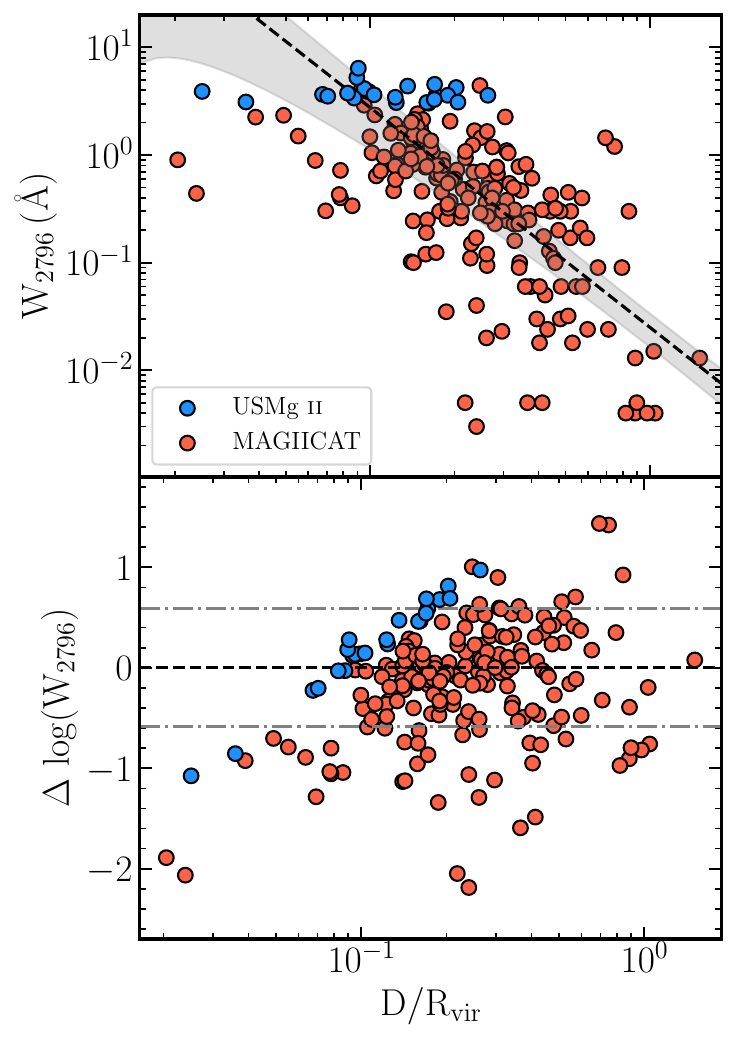}
    \caption{Top panel: Scatter plot of $\rm W_{2796}$ against the normalized impact parameter. Blue points correspond to the \usmg\ absorbers from our survey and the red points correspond to the \Magiicat\ survey. The black dashed line shows the power law fit from \citet{Churchill_2013} with the shaded region corresponding to the 1$\sigma$ uncertainty. Bottom panel: The difference between the observed $\rm \log\, (W_{2796})$ and the best-fit model for that normalized impact parameter.}
    \label{fig:w_vs_d_scaled}
\end{figure}

\citet{Churchill_2013} have suggested the scatter in the $\rm W_{2796}-D$ plane is substantially reduced if one uses the normalized impact parameter (i.e. D scaled by the virial radius, $R_{vir}$) instead of D. In Figure \ref{fig:w_vs_d_scaled}, we plot the $W_{2796}$ versus the normalized impact parameter  for both the MAGIICAT host galaxies (squares) and the \usmg\ host galaxies (circle) from our sample. The halo masses for the MAGIICAT host galaxies are obtained from abundance matching techniques \citep{Nielsen_2013b}. In contrast, the halo masses for the \usmg\ galaxies are measured using the stellar mass to halo relations (see Section \ref{sec:stellarmass} and \ref{sec:escape_vel} for details). \citet{Churchill_2013} found a strong anti-correlation between two quantities best characterized by a power law with a power-law index of $\sim -2$ (shown in the Figure with an orange dashed line), albeit a large scatter. When we considered only the \usmg\ host galaxies, we found no significant anti-correlation (i.e., $r_s = -0.17$ and p-value = 0.59). As the $W_{2796}$ range probed by the \usmg\ sample is narrow, we expect the scatter in the normalized impact parameter to be lower than that of D if the scaling works as in the case of MAGIICAT host galaxies.  In the bottom panel of Figure~\ref{fig:w_vs_d_scaled}, we show the residual of log~$W_{2796}$ as a function of the normalized impact parameter.  The deviation of \usmg\ point from the best fit is consistent with the scatter seen for the general population of \MgII\ absorbers.

\subsubsection{Comparison of \usmg\ and DLAs:}
A large fraction of these \usmg\ absorbers is expected to be DLAs and sub-DLAs \citep{rao2006, Nestor_2007}. It is also well known that for a given $z$, a strong correlation exists between $N$(\HI) and $W_{2796}$. Using the latest relationship between the two quantities found by \citet{LanFukugita2017} for the \usmg\ systems, we find log~$N$(H~{\sc i})$\ge20.20$ at $z\sim0.7$. Therefore, it is most likely that the \usmg\ absorbers in our sample are either sub-DLAs or DLAs. It is also well known that DLAs and sub-DLAs show a significant anti-correlation between $N$(H~{\sc i}) and $D$ \citep[see for example,][]{rao2011, Rahmani2016, kulkarni2022}.

Information on the impact parameter for these low-$z$ DLA and the sub-DLA host galaxies are obtained from \citet{Rahmani2016}, who compiled all the available DLAs and sub-DLAs at low-$z$ from the literature. We obtained $W_{2796}$ for each DLA and sub-DLA from the original references. The $W_{2796}$ associated with the DLAs and the sub-DLAs varies from 0.2\AA\ to 3.11\AA\ with a median value of 1.67\AA. In the bottom panel of Figure~\ref{fig:usmg_only}, we plot the $W_{2796}-D$ relationship for DLAs and sub-DLAs at $0.4 \lesssim z \lesssim 1.0$. It is evident from Figure~\ref{fig:usmg_only} that DLAs and \usmg\ systems probe similar impact parameter ranges. The KS-test suggests no statistically significant difference between the two impact parameter distributions with a p-value of 0.32. It is clear from Figure~\ref{fig:usmg_only} that there is no correlation between $W_{2796}$ and $D$ even in the case of low-$z$ DLAs and sub-DLAs. The rank correlation test provides the correlation coefficient of {$r_s = -0.17$} and p-value = 0.38. This is consistent with our finding that the anti-correlation weakens when we consider strong Mg~{\sc ii} absorbers. Interestingly, for the same DLA sample, the anti-correlation between $N$(\HI) and D is also not statistically significant (with $r_s$ = $-0.31$ and p-value of 0.11 for the Spearman rank correlation test).
 
At high-$z$, some of the DLAs selected based on C~{\sc i} absorption and extremely strong DLAs tend to have $W_{2796}>3$\AA\citep[see  figure 3 of ][]{Ranjan2022, Zou2018}. Thus lack of \usmg\ systems among the low-$z$ DLAs mainly identified through \MgII\ based selection (where \usmg\ systems are a very rare population) is not very surprising. However, the lack of correlation between $W_{2796}$ and $D$ is consistent with our finding in Figure~\ref{fig:w_corr}. {\it Discussions presented here clearly demonstrate how system selection influences the derived $w_{2796}-D$ relationship.}

\section{Discussion}
\label{sec:discuss}

\subsection{Isolated galaxy vs. a group of galaxies}

There are indications that some of the strong \MgII\ absorption systems are associated with multiple host galaxies in literature \citep[e.g.,][]{Whiting2006, Gauthier2013}. \citet{Gauthier2013} have identified a $z \sim 0.5$ galaxy group dominated by passive galaxies associated with a \usmg\ absorption system and argued that such strong absorption are driven by cool intragroup gas rather than star-formation-driven outflows. \citet{Nielsen2022} have identified a compact galaxy group associated with a DLA, which is also a \usmg\ system at $z \sim 2.4$, using KCWI and argued the gas in absorption is related to star-formation driven outflows, accretion from the IGM, and the tidal streams due to the galaxy-galaxy interactions. Some of the deviations shown by \usmg\ absorbers with respect to the relationships found for the general population of \MgII\ absorbers could be explained if a good fraction of \usmg\ absorbers is not originating from the CGM of individual galaxies.

Ideally, one requires IFU spectroscopic observations to identify the galaxies contributing to the \usmg\ absorbers.  In recent IFU studies, the only \usmg\ system present in the MAGG survey \citep{Dutta2020} and 3 in the sample of \citet{Schroetter2019} are found to be associated with isolated galaxies (no other host galaxy found within 100 kpc and velocity separations within 500 \kms). In \citet{Guha2022a}, 33 percent of \usmg\ absorbers (i.e., 7 out of 21) have only one host galaxy (with $m_r<23.6$ mag) within an impact parameter of 100 kpc. In Table A2 of the Appendix, we list all the galaxies within a projected distance of 100 kpc at the redshift of the \usmg\ absorption with $m_r < 23.6$ (typical completeness of the DESI-LIS r-band images) studied in this work. As mentioned before, this sets the completeness of our survey up to $\sim 0.3L_\star$ galaxies. Using photometric redshifts, we have found that up to a maximum impact parameter of 100 kpc, for four systems (i.e., along the line of sight to J0055-0100, J1336+0922, J1419+0346, and J2356-0406), there are no galaxies other than the confirmed \usmg\ host seen in the DESI-LIS images. This gives about $22^{+23}_{-13}$\% (95\% confidence Wilson score) of the \usmg\ absorption systems originating from isolated galaxies. Therefore, in the combined sample, we have at least 29\% of the \usmg\ systems associated with isolated host galaxies. 

On the other hand, in the present sample, 5 out of 18 ($28^{+23}_{-16}$\%) of the \usmg\ absorbers are confirmed to be associated with multiple galaxies. In the combined sample, this  fraction of \usmg\ absorbers confirmed to be associated with more than one host galaxy is $\ge$21\%. Among the rest of the six \usmg\ systems in the present sample, other than the confirmed \usmg\ host galaxies, there is at least one galaxy with consistent photo-$z$ within 100 kpc with $m_r < 24$. However, we do not have the spectroscopic redshift of these galaxies, so they may or may not be related to the \usmg\ absorption.

\begin{table}
    \centering
    \begin{tabular}{lcccc}
        \hline
         Properties &  $D$ value & $p$ value & Median  & Median \\
                    &            &           & (Isolated) & (Group) \\
         \hline
         \hline
         $W_{2796}$ &  0.31      &    0.62   &  3.44  &  3.71 \\
         $W_{2600}$ &  0.30      &    0.72   &  2.22  &  2.60 \\
         $W_{2796} / W_{2803}$ &  0.47 & 0.17 & 1.07  &  1.11\\
         $W_{2852} / W_{2796}$ &  0.41 & 0.33 & 0.30 &  0.23  \\
         D         &   0.38   &  0.38 &  15  &  27.2\\
         $\rm{\log\, (M_\star / M_\odot)}$ & 0.20      & 1.0        &  10.47  & 10.48  \\
         \hline
    \end{tabular}
    \caption{Comparison of properties of  \usmg\ absorbers associated with single and multiple host galaxies.}
    \label{tab:compare_isolated_group}
\end{table}

Next, we ask whether there is any difference between the \MgII\ absorbers properties identified with a single galaxy or multiple galaxies in our sample. Table~\ref{tab:compare_isolated_group} gives the median values of different measured quantities of the two sub-samples and KS-statistics results (D and p-value). Median values of observed equivalent widths and equivalent width ratios are consistent within 20\% uncertainty. This is also reflected by the large $p$-values found for the KS test. We also find the median value of the inferred stellar mass ($M_\star$) of the two samples are almost same. The most significant difference we notice (i.e., by 80\%) is for the median value of the impact parameter. The median impact parameter is higher for the nearest galaxy  of the \usmg\ absorber for which more than one host galaxy is identified. Interestingly the difference is not statistically significant due to the small number of systems involved. 

\subsection{Line of sight velocity of the absorbing gas}
\label{sec:escape_vel}

\begin{figure}
    \includegraphics[width=0.47\textwidth]{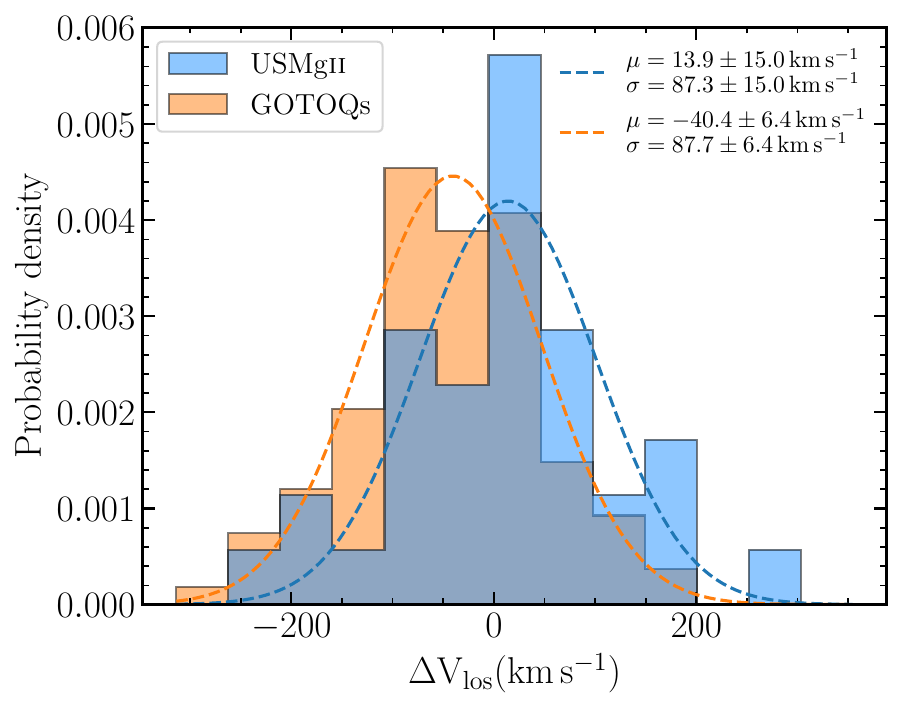}
    \caption{Distribution of the rest frame line of sight velocity offset of the \MgII\ absorbing gas with respect to the \MgII\ host galaxies. The blue histogram corresponds to the \usmg\ host galaxies, whereas the orange histogram corresponds to the GOTOQs. The blue and orange dashed lines correspond to the Gaussian fits to the blue and orange histograms, respectively.}
    \label{fig:veloffset}
\end{figure}

\begin{figure}
    \includegraphics[width=0.5\textwidth]{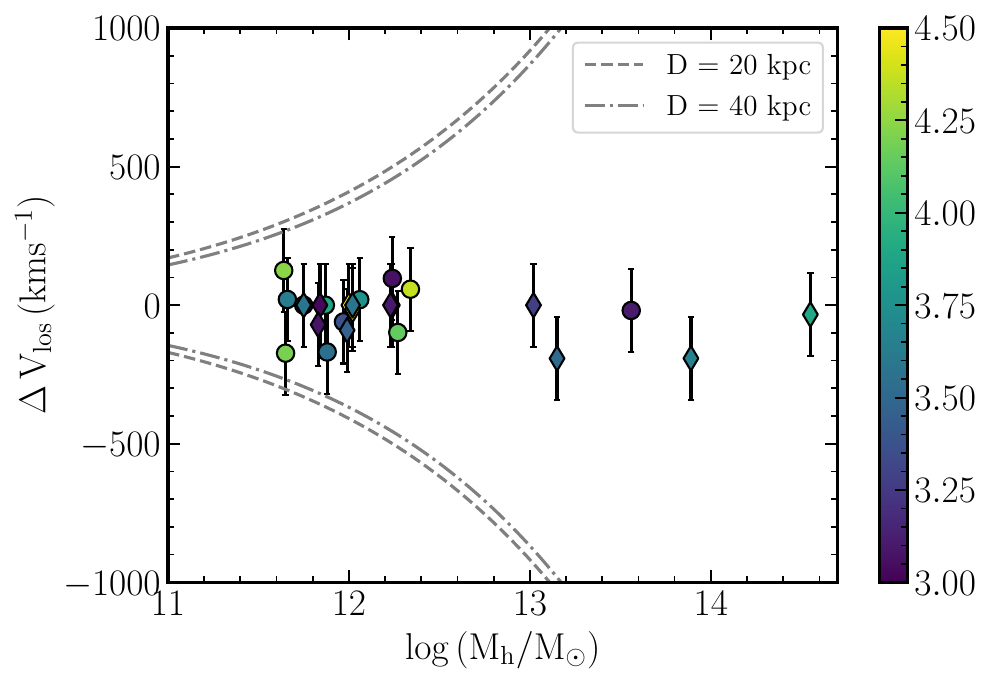}
    \caption{Line of sight velocity offset of the \usmg\ absorbing gas with respect to the \usmg\ host galaxies is plotted against the halo masses of the \usmg\ host galaxies and color-coded according to the $W_{2796}$respectively correspond to the low and high redshift \usmg\ absorbers from our sample. Circular and diamond points  The dashed and dot-dashed lines show the escape velocity for a given halo mass at the impact parameters of 20 kpc and 40 kpc, respectively.}
    \label{fig:escape_vel}
\end{figure}

Next, we discuss the rest-frame line of sight velocity offset of the  absorbing gas with respect to the  host galaxies and the fate of the absorbing gas: whether it can escape the host galaxy or is bound to it. In Figure \ref{fig:veloffset}, we show the rest-frame line of sight velocity offset of the \MgII\ absorbing gas with respect to the \MgII\ host galaxies. The rest-frame velocity offsets are calculated using $\Delta V_{los} = c (z_{abs} - z_{em}) / (1 + z_{em})$ where $z_{abs}$ and $z_{em}$ are absorption and emission redshifts obtained from the \MgII\ absorption towards the quasar line of sights and \OII\ nebular emissions of the \MgII\ host galaxies respectively. $c$ denotes the speed of light. The typical error in this offset measurement is 100 \kms. The blue histogram corresponds to the \usmg\ host galaxies, whereas the orange histogram corresponds to the GOTOQs. The blue and orange dashed lines correspond to the Gaussian fit of the blue and orange histograms. The line of sight velocity offset for the \usmg\ absorbers is centered around $\mu \sim 14\, kms^{-1}$  with a standard deviations of about $\sigma \sim 87\, kms^{-1}$, whereas for the GOTOQs the center is at $\mu \sim -40\, kms^{-1}$ with a standard deviation of $\sigma \sim 88\, kms^{-1}$. The most important thing to notice is that the width of the $\Delta V_{los}$ distribution is very similar for the \usmg\ host galaxies and the GOTOQs. Considering the isolated host galaxies in our sample (discussed above), we find $\mu$ = 48\kms\ and $\sigma$ =78\kms. The same is for the nearest galaxy where the absorber can be associated with more than one galaxy, and we find $\mu$ = 17 \kms and $\sigma$ = 147 \kms. The $\sigma$ being large in the case of absorbers associated with multiple galaxies found here is consistent with the results of \citep{Huang2021}.

Next, we investigate whether or not the absorbing gas is gravitationally bound to the \usmg\ host galaxies. In Figure \ref{fig:escape_vel}, we plot the line of sight velocity offset of the \usmg\ absorbing gas with respect to the \usmg\ host galaxies against the halo masses of the \usmg\ host galaxies color-coded according to the $W_{2796}$. The dashed and the dot-dashed lines show the escape velocity for a given halo mass at the impact parameters of 20 kpc and 40 kpc calculated using an NFW profile \citep{Navarro1997}, respectively. The halo masses are obtained from the stellar mass to halo mass relations at the host galaxy redshifts \citep{Girelli2020}. Note that this relation has a typical uncertainty of 0.2 dex. As seen from Figure \ref{fig:escape_vel}, the line of sight velocity offset of absorbing gas is insufficient to escape the halo's gravitational potential. Even if we correct for the three-dimensional velocity components by multiplying $\Delta V_{los}$ with a factor of $\sqrt{3}$, all the points are still consistent within 1$\sigma$ with the absorbing gas to be bound to the host galaxies.

Given the large velocity spread in absorption, it is difficult to imagine that all the gas responsible for the \usmg\ absorption will be confined to the host galaxy. For a simple saturated absorption, we expect the \usmg\ absorption to spread over more than 300 \kms. However, from Figure 10 of \citet{Ranjan2022}, we can see that the velocity spread can be anywhere between 300 to 700 \kms\ when $W_{2796}>3$\AA. Therefore, high-resolution spectroscopic data are required to directly quantify the fraction of absorbing gas that can be bound to the host galaxy.

\subsection{ The nature of \usmg\ host galaxies}
\label{sec:compare_usmg}

Our results indicate that for a given impact parameter \usmg\ galaxies tend to be more massive and brighter in the B-band compared to the host galaxies of absorbers with low $W_{2796}$. It is also known that $W_{2796}$ is primarily driven by the total number of such clumps along the line of sight \citep{Petitjean1990, Churchill2020} and velocity dispersion between them rather than the \MgII\ column density. The requirement of large velocity widths for the \usmg\ can come from (i) large-scale bi-conical outflow; (ii) gaseous in the CGM under the influence of the large gravitational potential of a massive galaxy; or (iii) as discussed above, large velocity widths can originate from gas in the interacting system of galaxies. 

Recent high-resolution TNG50 cosmological simulations by \citet{DeFelippis2021} also found that more massive halos produce stronger and broader \MgII\ absorption. However, the \usmg\ absorbers are extremely rare in these simulations.  High mass halos producing stronger \MgII\ absorption is also suggested by various authors in the literature \citep[e.g.,][]{Lan2014, Dutta2020, Anand2021} while studying relatively low $W_{2796}$ systems compared to \usmg\ systems. However, we find that the \usmg\ systems do not follow the best fit relationship between $W_{2796}$, D and $M_\star$ and/or $M_B$ \citep[see for example,][]{Huang2021} followed by low $W_{2796}$ absorbers. We require high spectral resolution data to address the origin of this difference. Also, knowing what fraction of the \usmg\ absorbers originate from more than one host galaxy (using IFU spectroscopy) will help understand this issue.

Our results also indicate that the star formation rate in the \usmg\ host galaxies is slightly lower than that of the main sequence galaxies at the same redshift having the same stellar mass. Even when one accounts for the possible bias in the SFR estimations it is clear that \usmg\ host galaxies will, at best, follow the main sequence. Similar trend is found for the massive \MgII\ host galaxies ($\rm M_\star \geqslant 10^{10} \,M_\odot$) in the MEGAFLOW survey \citep{Schroetter2019}. \citet{Rhodin2018} also found that the massive galaxies producing DLAs and sub-DLAs at $z \sim 0.7$ also have significantly smaller ongoing star formation rates compared to main-sequence galaxies of the same mass. \citet{Anand2021} found that the covering fractions for strong \MgII\ absorption ($\rm W_{2796} \geqslant 2$ \AA\ ) within $D\sim 30$ kpc is similar for both the star-forming and the passive galaxies. This could be related to the fact that these galaxies have been through a phase of rapid star formation and consequently became massive. The star-formation-driven outflows enrich the circumgalactic medium while the star-formation-driven feedback starts quenching the galaxies, and hence the ongoing SFR drops below the main sequence. Another contributing factor to the suppression of the star formation rates could be galaxy-galaxy interactions. If a significant fraction of the \usmg\ host galaxies live in a group environment, the motion of these galaxies through the intra-group medium may cause the gas in the galactic disks to be stripped away, shock heated, and turbulent which henceforth further reduced star formation activity.

\section{Summary \& Conclusions}
\label{sec:summary}

In this work, we extend our previous study \citep{Guha2022a} of \usmg\ absorbers to a slightly higher redshift range (i.e., $0.6\le z\le0.8$). From an input sample of 151 \usmg\ systems observed in the SDSS-DR12 \citep{Zhu2013} in the above-mentioned redshift range, we identified 27 secured \usmg\ systems. We carried out long-slit observations with SALT to identify the host galaxies. This, combined with the sample presented in \citet{Guha2022a}, constitutes  the most extensive sample of \usmg\ systems studied to date. Our main findings are summarized below.

\begin{enumerate}
\item [1.] Among the 27 \usmg\ systems in the present sample, SALT observations could be completed for 25 systems. Based on the \OII\ emission or the \CaII\ absorption present in the spectra, we have successfully identified the host galaxies for the 18 \usmg\ systems. For 5 of the \usmg\ systems, we have identified more than one host galaxy associated with the \usmg\ absorption. The impact parameters of the spectroscopically confirmed \usmg\ host galaxies range from 6.24 kpc to 120.7 kpc (see Table \ref{tab:spec_props}).

\item[2.] Inclusion of \usmg\ data points to the measurements available in the literature increases the scatter in the well-known anti-correlation between $W_{2796}$ and D (see Figure \ref{fig:w_vs_d}). When we consider only the \usmg\ data points, we do not find any significant correlation between $W_{2796}$ and D (see Figure \ref{fig:usmg_only}). We find that the best-fit value of $W_{2796}(D=0)$ is higher than the measurement previously quoted in the literature. We also find that the significance of anti-correlation between $W_{2796}$ and D depends on how the sample is defined. If one defines a sample with a higher $W_{2796}$ threshold there is no trend evident between $W_{2796}$ and D (see Figure~\ref{fig:w_corr}) . Similarly we notice that  low-$z$ DLAs do not show any anti-correlation between  $W_{2796}$ and D.

\item[3.] 
We find for a given impact parameter, the \usmg\ host galaxies are brighter and more massive compared to the low equivalent width \MgII\ absorbers. Such a mass dependence is also seen among the non-\usmg\ absorbers. However, we find the \usmg\ absorbers not to follow the relationship between $W_{2796}$, D and $M_\star$(or $M_B$) found for the general population of \MgII\ absorbers (see Figure \ref{fig:d_vs_mass_mag}).

\item[4.] Although \usmg\ absorbers do not follow the canonical $W_{2796}-D$ anti-correlations, they seem to follow the $W_{2796}-D/R_{vir}$ anti-correlation. The scatter seen for the \usmg\ around the global best-fit solution is similar to the scatter seen for the low equivalent width \MgII\ absorbers. This once again indicates more gaseous content in massive galaxies (see Figure \ref{fig:w_vs_d_scaled}).

\item[5.] We find that parameters $\alpha$ and $\beta$ that fit the $W_{2796}-D$ anti-correlation does not evolve significantly with redshift over the redshift range $0 < z \lesssim 1.5$ (see Figure \ref{fig:w_vs_d}). However, even small differences in $\alpha$ can lead to large differences in $W_{2796}$ at high D. For $D>20$ kpc, the fits are consistent with the $W_{2796}$ at higher redshift being higher for a given impact parameter.  Using a simple model, we argue that major evolution in $dn/dz$ of the \MgII\ absorbers can come from the evolution of the galaxy luminosity function. The differential evolution seen between high and low equivalent width systems will require a covering factor as a function of D to evolve with time, as found by \citet{Lan2014}.

\item[6.] Majority of the \usmg\ host galaxies are detected based on their \OII\ emissions, which indicates that they are all star-forming galaxies. However, in the SFR-$M_\star$ plane, \usmg\ host galaxies fall slightly below the main sequence galaxies. We, therefore, infer that the \usmg\ host galaxies are likely to be post-starburst galaxies transitioning slowly from the main sequence to the quenched galaxies (see Figure \ref{fig:mass_vs_sfr}).

\item[7.] Based on the line of sight velocity differences between the absorbing gas and the host galaxies, we find that the overall \usmg\ absorbing gas is bound to the host galaxies. However, we expect the velocity width of the \MgII\ absorption to be very large. Most likely, some of the absorbing gas components may have excess velocities outside the escape velocity predicted based on the $M_\star$-$M_h$ relationship. The spectral resolution of our spectra does not allow us to quantify the fraction of gas in each \usmg\ absorber that is bound to the galaxy. For this, we need higher-resolution spectra (see Figure \ref{fig:escape_vel}).

\item[8.] In our combined \usmg\ sample, at least 28\% of the absorbers are associated with a single isolated galaxy. No other galaxy with correct spectroscopic or photometric redshift having $m_r<23.6$ is found. Similarly, in the $\sim$21\% case, we have identified more than two host galaxies within an impact parameter of 100 kpc. In the remaining 50\% cases, our spectroscopic observations are not available for potential candidate host galaxies at $50\le D[kpc]\le 100$. Completing spectroscopic observations of some of these galaxies will be important to quantify the origin of large velocity fields in \usmg\ absorbers.

\end{enumerate}
\section*{Data Availability}
Data used in this work are obtained using SALT. Raw data will become available for public use 1.5 years after the observing date at https://ssda.saao.ac.za/.

\section*{Acknowledgement}
\label{sec:acknowledge}

This project makes use of the following softwares : NumPy \citep{numpy2020}, SciPy \citep{scipy2020}, Matplotlib \citep{matplotlib2007}, AstroPy \citep{astropy:2013, astropy:2018}, and Ultranest \citep{Buchner2021}.

All the new observations reported in this paper were obtained with the Southern African Large Telescope (SALT).

PPJ thanks Camille No\^us (Laboratoire Cogitamus) for inappreciable and often unnoticed discussions,  advice, and support. PPJ is partly supported by the Agence Nationale de la Recherche under contract HZ-3D-MAP, ANR-22-CE31-0009.

This paper makes use of SDSS observational data. Funding for the Sloan Digital Sky Survey IV has been provided by the Alfred P. Sloan Foundation, the U.S. Department of Energy Office of Science, and the Participating Institutions. SDSS-IV acknowledges support and resources from the Center for High-Performance Computing  at the University of Utah. The SDSS website is www.sdss.org. SDSS-IV is managed by the Astrophysical Research Consortium for the Participating Institutions of the SDSS Collaboration, including  the Brazilian Participation Group, the Carnegie Institution for Science, Carnegie Mellon University, Center for Astrophysics | Harvard \& Smithsonian, the Chilean Participation 
Group, the French Participation Group, Instituto de Astrof\'isica de 
Canarias, The Johns Hopkins University, Kavli Institute for the 
Physics and Mathematics of the Universe (IPMU) / University of Tokyo, the Korean Participation Group, Lawrence Berkeley National Laboratory, Leibniz Institut f\"ur Astrophysik Potsdam (AIP),  Max-Planck-Institut 
f\"ur Astronomie (MPIA Heidelberg), Max-Planck-Institut f\"ur Astrophysik (MPA Garching), Max-Planck-Institut f\"ur Extraterrestrische Physik (MPE), National Astronomical Observatories of China, New Mexico State University, 
New York University, University of Notre Dame, Observat\'ario Nacional / MCTI, The Ohio State University, Pennsylvania State University, Shanghai 
Astronomical Observatory, United Kingdom Participation Group, 
niversidad Nacional Aut\'onoma de M\'exico, University of Arizona, University of Colorado Boulder, University of Oxford, University of Portsmouth, University of Utah, University of Virginia, University of Washington, University of Wisconsin, Vanderbilt University, and Yale University.

The DESI Legacy Imaging Surveys consist of three individual and complementary projects: the Dark Energy Camera Legacy Survey (DECaLS), the Beijing-Arizona Sky Survey (BASS), and the Mayall z-band Legacy Survey (MzLS). DECaLS, BASS, and MzLS together include data obtained, respectively, at the Blanco telescope, Cerro Tololo Inter-American Observatory, NSF’s NOIRLab; the Bok telescope, Steward Observatory, University of Arizona; and the Mayall telescope, Kitt Peak National Observatory, NOIRLab. NOIRLab is operated by the Association of Universities for Research in Astronomy (AURA) under a cooperative agreement with the National Science Foundation. Pipeline processing and analyses of the data were supported by NOIRLab and the Lawrence Berkeley National Laboratory (LBNL). Legacy Surveys also uses data products from the Near-Earth Object Wide-field Infrared Survey Explorer (NEOWISE), a project of the Jet Propulsion Laboratory/California Institute of Technology, funded by the National Aeronautics and Space Administration. Legacy Surveys was supported by: the Director, Office of Science, Office of High Energy Physics of the U.S. Department of Energy; the National Energy Research Scientific Computing Center, a DOE Office of Science User Facility; the U.S. National Science Foundation, Division of Astronomical Sciences; the National Astronomical Observatories of China, the Chinese Academy of Sciences and the Chinese National Natural Science Foundation. LBNL is managed by the Regents of the University of California under contract to the U.S. Department of Energy. The complete acknowledgments can be found at https://www.legacysurvey.org/acknowledgment/. The Photometric Redshifts for the Legacy Surveys (PRLS) catalog used in this paper was produced thanks to funding from the U.S. Department of Energy Office of Science, Office of High Energy Physics via grant DE-SC0007914.



\bibliographystyle{mnras}
\bibliography{example} 




\appendix
\section{BAL quasars as false \usmg\ absorption systems in the SDSS absorbers catalog}
\label{sec:bal_quasars}

Table \ref{tab:zhu_catalog} provides the details of all the 151 \usmg\ absorption systems in the redshift range $0.6 \geqslant z_{abs} \geqslant 0.8$ that is accessible to the SALT (declination $\delta \leqslant 10^\circ$) from SDSS-DR12 \FeII\ / \MgII\ metal absorber catalog \citep{Zhu2013}. For each of the listed absorption systems, we explicitly mention the reasoning for whether the system is included in our sample or not.

\begin{table*}
\caption{Details of all the 151 \usmg\ absorbers from \citet{Zhu2013} in the redshift range $0.6 \leqslant z_{abs} \leqslant 0.8$ with declination, $\delta \leqslant 10^\circ$.}
\label{tab:zhu_catalog}
\begin{tabular}{llcccccc}
\hline

No. & Quasar & Plate & Fiber   & MJD & $z_{qso}$ & $z_{abs}$ & Comment\\
\hline

1 &  J001030.81+012203.43  &  4298  &  322  &  55511  &  2.197  &  0.6431  &  Secure \usmg\ identification. \\
2 &  J000931.24+031420.14  &  4297  &  182  &  55806  &  2.323  &  0.7752  &  False identification of \usmg\ absorption.\\
3 &  J001156.33-013701.47  &  4365  &  874  &  55539  &  2.305  &  0.7837  &  BAL quasar\\
4 &  J001227.00+092701.75  &  4536  &  650  &  55857  &  2.453  &  0.7365  &  False identification of \usmg\ absorption.\\
5 &  J001232.06+052657.94  &  4416  &  320  &  55828  &  2.621  &  0.7952  &  BAL quasar\\
6 &  J001436.53-090622.34  &  7169  &  890  &  56628  &  2.518  &  0.7283  &  BAL quasar\\
7 &  J001602.40-001225.10  &  4218  &  147  &  55479  &  2.080  &  0.6241  &  False identification of \usmg\ absorption.\\
8 &  J002022.66+000231.98  &  4219  &  734  &  55480  &  2.749  &  0.7699  &  Secure \usmg\ system\\
9 &  J002410.02+064620.69  &  4417  &  796  &  55829  &  2.915  &  0.7606  &  Blended lines detected as \usmg \\
  &                        &  4537  &  24   &  55806  &  2.915  &  0.7605  &  \\
10 &  J002509.22-071616.78  &  7150  &  40   &  56597  &  2.130  &  0.6932  &  BAL quasar\\
11 &  J002529.79+001339.69  &  4220  &  618  &  55447  &  2.085  &  0.6691  &  BAL quasar\\
12 &  J002839.24+004103.05  &  3586  &  564  &  55181  &  2.493  &  0.6565  &  Secure \usmg\ system\\
   &                        &  4220  &  846  &  55447  &  2.493  &  0.6567  &  \\
13 &  J003336.04+013851.06  &  4302  &  150  &  55531  &  2.658  &  0.7187  &  Secure \usmg\ system\\
14 &  J003749.07-065428.97  &  7152  &  990  &  56660  &  2.331  &  0.7931  &  BAL quasar\\
15 &  J003904.01+023313.28  &  4304  &  681  &  55506  &  2.304  &  0.7917  &  BAL quasar\\
16 &  J004045.20+040646.21  &  4303  &  978  &  55508  &  2.143  &  0.6931  &  BAL quasar\\
17 &  J004444.06+001303.59  &  4222  &  972  &  55444  &  2.280  &  0.7791  &  BAL quasar\\
18 &  J004618.61-000936.78  &  4222  &  10   &  55444  &  2.439  &  0.6256  &  Blended lines detected as \usmg\ \\
19 &  J005521.73-002817.19  &  4224  &  186  &  55481  &  2.254  &  0.7351  &  BAL quasar\\
20 &  J005551.22-002837.17  &  4224  &  149  &  55481  &  2.428  &  0.6302  &  BAL quasar\\
21 &  J005554.25-010058.62  &  4224  &  170  &  55481  &  2.363  &  0.7150  &  Secure \usmg\ system\\
22 &  J005958.70+095338.68  &  4547  &  726  &  55893  &  0.693  &  0.6844  &  BAL quasar\\
23 &  J010543.52+004003.86  &  4225  &  964  &  55455  &  1.078  &  0.6489  &  Secure \usmg\ system\\
   &                        &  3735  &  821  &  55209  &  1.078  &  0.6489  &  \\
24 &  J010915.69+055138.52  &  4423  &  270  &  55889  &  2.324  &  0.6270  &  Poor SNR. False identification of \usmg\ .\\
25 &  J011448.97+042018.91  &  4312  &  846  &  55511  &  2.157  &  0.7131  &  BAL quasar\\
26 &  J011722.00-012734.86  &  4353  &  752  &  55532  &  2.179  &  0.7182  &  BAL quasar\\
27 &  J011903.30-021746.91  &  4353  &  859  &  55532  &  2.319  &  0.7909  &  BAL quasar\\
28 &  J012424.98-010327.66  &  4228  &  96   &  55484  &  2.148  &  0.7034  &  BAL quasar\\
29 &  J012711.11-055020.95  &  7047  &  207  &  56572  &  2.137  &  0.6837  &  Secure \usmg\ system\\
30 &  J013311.86-054258.13  &  7048  &  214  &  56575  &  2.139  &  0.6266  &  False identification of \usmg \\
31 &  J013537.43-003853.06  &  4230  &  184  &  55483  &  2.276  &  0.6233  &  BAL quasar\\
32 &  J014115.32-000500.98  &  3639  &  773  &  55205  &  2.130  &  0.6105  &  Secure \usmg\ system\\
33 &  J014118.88-021214.49  &  4350  &  659  &  55556  &  2.268  &  0.7721  &  BAL quasar\\
34 &  J014118.88-021214.49  &  4350  &  659  &  55556  &  2.268  &  0.7443  &  BAL quasar\\
35 &  J014258.83+094942.43  &  4545  &  360  &  55567  &  0.981  &  0.7858  &  Secure \usmg\ system\\
36 &  J015007.91-003937.09  &  4232  &  194  &  55447  &  2.730  &  0.7747  &  Secure \usmg\ system\\
37 &  J015049.39+060432.42  &  4403  &  274  &  55536  &  2.676  &  0.6707  &  Secure \usmg\ system\\
38 &  J015522.03-034740.11  &  7052  &  630  &  56539  &  2.433  &  0.6128  &  False identification of \usmg\ \\
39 &  J020100.99+020659.50  &  4271  &  196  &  55507  &  2.149  &  0.6720  &  BAL quasar\\
40 &  J020733.90-040847.28  &  7238  &  869  &  56660  &  2.157  &  0.6195  &  BAL quasar\\
41 &  J020812.49-004928.84  &  4235  &  299  &  55451  &  2.220  &  0.7475  &  BAL quasar\\
42 &  J020949.08-054824.53  &  4398  &  962  &  55946  &  0.738  &  0.7310  &  BAL quasar\\
43 &  J022419.07+022813.62  &  4265  &  740  &  55505  &  2.317  &  0.7225  &  Poor SNR. False identification of \usmg\ \\
44 &  J022445.16+001028.34  &  6780  &  536  &  56267  &  2.161  &  0.7124  &  BAL quasar\\
45 &  J022545.65+005149.96  &  3647  &  638  &  55241  &  2.525  &  0.7437  &  BAL quasar\\
   &                        &  3647  &  610  &  55476  &  2.525  &  0.7436  &  \\
   &                        &  3615  &  634  &  55179  &  2.525  &  0.7445  &  \\
   &                        &  3647  &  638  &  55241  &  2.525  &  0.7437  &  \\
46 &  J022812.19+002657.82  &  3647  &  722  &  55476  &  2.585  &  0.6402  &  BAL quasar\\
47 &  J023252.80-001351.15  &  3615  &  40   &  56544  &  2.025  &  0.6497  &  BAL quasar\\
   &                        &  3615  &  30   &  55179  &  2.025  &  0.6499  &  \\
48 &  J024004.18+004500.68  &  4239  &  958  &  55458  &  2.233  &  0.6887  &  Blended lines detected as \usmg\ \\
   &                        &  4240  &  635  &  55455  &  2.233  &  0.6888  &  \\

\hline
\end{tabular}
\label{tab:usmg_all}
\end{table*}

\begin{table*}
\ContinuedFloat
\caption{Continued.}
\begin{tabular}{lccccccc}
\hline

   &                        &  3651  &  556  &  55247  &  2.233  &  0.6888  &  \\
   &                        &  3650  &  754  &  55244  &  2.233  &  0.6890  &  \\
49 &  J024046.76-062829.65  &  7055  &  474  &  56576  &  2.579  &  0.6852  &  BAL quasar\\
50 &  J025607.25+011038.56  &  4242  &  784  &  55476  &  1.349  &  0.7256  &  Secure \usmg\ system\\
51 &  J081214.92+013608.90  &  4787  &  404  &  55863  &  2.177  &  0.7250  &  BAL quasar\\
52 &  J081829.68+070655.74  &  4841  &  634  &  55895  &  2.215  &  0.6037  &  False identification of \usmg\ \\
53 &  J081831.10+012608.74  &  4792  &  450  &  55925  &  2.618  &  0.7622  &  Poor SNR. False identification of \usmg\ .\\
54 &  J084502.73+081214.26  &  5285  &  952  &  55946  &  2.365  &  0.7034  &  BAL quasar\\
55 &  J085731.54+041951.40  &  3814  &  880  &  55535  &  2.445  &  0.7862  &  Poor SNR. False \usmg\ identification. \\
56 &  J085932.56+064254.29  &  4868  &  608  &  55895  &  2.034  &  0.6422  &  BAL quasar\\
57 &  J090805.76+072739.90  &  5299  &  215  &  55927  &  2.414  &  0.6122  &  Secure \usmg\ system\\
58 &  J092550.26+041503.41  &  4796  &  376  &  55924  &  2.381  &  0.6833  &  BAL quasar\\
59 &  J092550.26+041503.41  &  4796  &  376  &  55924  &  2.381  &  0.6724  &  BAL quasar\\
60 &  J093415.29+090059.49  &  5314  &  650  &  55952  &  2.268  &  0.7259  &  Poor SNR. False \usmg\ identification.\\
61 &  J093955.67-013513.90  &  3782  &  554  &  55244  &  2.315  &  0.7364  &  BAL quasar\\
62 &  J094202.95+082643.74  &  5314  &  8    &  55952  &  2.228  &  0.7467  &  Poor SNR. False \usmg\ identification.\\
63 &  J094548.46+023108.29  &  4736  &  973  &  55631  &  2.631  &  0.7495  &  Blended lines detected as \usmg\ \\
64 &  J094852.19+011506.73  &  4743  &  348  &  55645  &  2.314  &  0.6891  &  Poor SNR. False \usmg\ identification.\\
65 &  J095619.49+001800.34  &  3828  &  774  &  55539  &  2.172  &  0.7821  &  Secure \usmg\ system\\
66 &  J095745.89+062811.70  &  4874  &  360  &  55673  &  2.308  &  0.7359  &  BAL quasar\\
67 &  J100029.85+095434.30  &  5327  &  250  &  55979  &  2.320  &  0.7925  &  BAL quasar\\
68 &  J100154.84-005500.81  &  3783  &  750  &  55246  &  2.455  &  0.6444  &  Blended lines detected as \usmg\ \\
69 &  J100339.82+082236.33  &  5329  &  312  &  55946  &  2.121  &  0.6706  &  BAL quasar\\
70 &  J100513.61+004028.31  &  3829  &  802  &  55300  &  2.568  &  0.7952  &  BAL quasar\\
71 &  J100716.69+030438.75  &  4738  &  646  &  55650  &  2.127  &  0.6109  &  BAL quasar\\
72 &  J102219.71-020735.81  &  3770  &  620  &  55234  &  2.162  &  0.6681  &  BAL quasar\\
73 &  J103325.92+012836.35  &  4734  &  442  &  55646  &  2.180  &  0.6709  &  Secure \usmg\ system\\
74 &  J103400.72+032557.35  &  4772  &  130  &  55654  &  2.370  &  0.7070  &  BAL quasar\\
75 &  J103433.72-011732.74  &  3785  &  918  &  55241  &  2.615  &  0.7831  &  BAL quasar\\
   &                        &  3785  &  914  &  55273  &  2.615  &  0.7834  &  \\
76 &  J104642.70+045731.96  &  4773  &  998  &  55648  &  2.542  &  0.7849  &  Secure \usmg\ system\\
77 &  J110650.53+061049.86  &  4855  &  200  &  55926  &  2.251  &  0.7553  &  BAL quasar\\
78 &  J110728.16+051016.40  &  4770  &  600  &  55928  &  2.570  &  0.7131  &  BAL quasar\\
79 &  J111108.20+095513.98  &  5361  &  4    &  55973  &  2.272  &  0.7160  &  BAL quasar\\
80 &  J111535.12-011442.50  &  3838  &  364  &  55588  &  2.291  &  0.7843  &  BAL quasar\\
81 &  J111627.65+050049.96  &  4769  &  526  &  55931  &  2.571  &  0.7208  &  Secure \usmg\ system\\
   &                        &  4770  &  990  &  55928  &  2.571  &  0.7207  &  \\
82 &  J114747.38+092108.76  &  5380  &  102  &  55980  &  2.176  &  0.7091  &  BAL quasar\\
83 &  J114842.43-023948.03  &  3790  &  260  &  55208  &  2.197  &  0.7308  &  BAL quasar\\
84 &  J115026.11+090048.40  &  5382  &  714  &  55982  &  2.492  &  0.7569  &  Secure \usmg\ system\\
85 &  J115911.34+025831.12  &  4747  &  364  &  55652  &  2.250  &  0.7104  &  Poor SNR. False \usmg\ identification.\\
86 &  J120131.70+054123.44  &  4831  &  467  &  55679  &  2.231  &  0.7346  &  BAL quasar\\
87 &  J120139.57+071338.24  &  5389  &  296  &  55953  &  1.205  &  0.6842  &  Secure \usmg\ system\\
88 &  J120326.31+052445.77  &  4831  &  399  &  55679  &  2.118  &  0.7149  &  BAL quasar\\
89 &  J120403.00+064829.56  &  5389  &  166  &  55953  &  2.263  &  0.7625  &  BAL quasar\\
90 &  J121212.09+083216.44  &  5393  &  764  &  55946  &  2.119  &  0.6721  &  BAL quasar\\
91 &  J121727.80-011548.57  &  3777  &  758  &  55210  &  2.624  &  0.6642  &  Secure \usmg system\\
92 &  J123858.06+064042.34  &  5407  &  366  &  55926  &  2.250  &  0.7450  &  BAL quasar\\
93 &  J123917.42-013132.42  &  3793  &  524  &  55214  &  2.310  &  0.7910  &  BAL quasar\\
94 &  J124602.10+075425.69  &  5407  &  39   &  55926  &  2.139  &  0.7233  &  BAL quasar\\
95 &  J125046.21+024414.37  &  4756  &  885  &  55631  &  2.332  &  0.7798  &  BAL quasar\\
96 &  J131216.58-013145.00  &  4053  &  742  &  55591  &  2.163  &  0.7089  &  BAL quasar\\
97 &  J132200.79-010755.70  &  4050  &  451  &  55599  &  2.160  &  0.7226  &  Secure \usmg\ system\\
98 &  J132216.24+052446.33  &  4839  &  442  &  55703  &  2.045  &  0.6439  &  BAL quasar\\
   &                        &  4761  &  794  &  55633  &  2.045  &  0.6441  &  BAL quasar\\
99 &  J132251.16+004901.21  &  4050  &  586  &  55599  &  2.397  &  0.6627  &  \\
100 &  J133653.73+092221.23  &  5437  &  508  &  55973  &  2.531  &  0.7059  &  Secure \usmg\ system\\
101 &  J134857.12+032234.35  &  4785  &  260  &  55659  &  2.363  &  0.7737  &  BAL quasar\\
102 &  J135009.61+092417.23  &  5442  &  786  &  55978  &  2.415  &  0.6799  &  BAL quasar\\

\hline
\end{tabular}
\end{table*}

\begin{table*}
\ContinuedFloat
\caption{Continued.}
\begin{tabular}{lccccccc}
\hline

103 &  J140017.69-014902.40  &  4041  &  964  &  55361  &  2.555  &  0.7928  &  Secure \usmg\ system\\
104 &  J141930.09+034643.73  &  4782  &  100  &  55654  &  2.316  &  0.7250  &  Secure \usmg\ system\\
105 &  J142140.27-020239.03  &  4032  &  736  &  55333  &  2.084  &  0.6707  &  BAL quasar\\
106 &  J142218.92-025250.56  &  4032  &  254  &  55333  &  2.380  &  0.6976  &  False identification of \usmg\ absorption. \\
107 &  J142452.06+025007.78  &  4027  &  828  &  55629  &  2.287  &  0.6095  &  BAL quasar\\
108 &  J142735.32-003936.43  &  4028  &  131  &  55621  &  2.650  &  0.7883  &  BAL quasar\\
109 &  J143023.18+084219.53  &  5462  &  62   &  55978  &  2.396  &  0.7192  &  Poor SNR. False identification of \usmg\ \\
110 &  J143158.00+075448.29  &  5467  &  456  &  55973  &  2.522  &  0.7453  &  Poor SNR. False identification of \usmg\ \\
111 &  J143223.09-000116.43  &  4025  &  613  &  55350  &  2.474  &  0.6455  &  Blended lines detected as \usmg\ \\
112 &  J143550.09+084854.09  &  5467  &  776  &  55973  &  2.777  &  0.7890  &  False identification of \usmg\ \\
113 &  J143830.68-014807.43  &  4026  &  696  &  55325  &  2.220  &  0.7408  &  BAL quasar\\
114 &  J144936.18-011650.46  &  4023  &  758  &  55328  &  0.772  &  0.6621  &  Secure \usmg\ system\\
115 &  J145108.53-013833.06  &  4023  &  814  &  55328  &  2.390  &  0.7408  &  Secure \usmg\ system\\
116 &  J150341.63+063456.57  &  4856  &  318  &  55712  &  2.463  &  0.6494  &  BAL quasar\\
117 &  J151841.30+094936.26  &  5489  &  210  &  55990  &  0.666  &  0.6418  &  BAL quasar\\
118 &  J152218.01+051007.01  &  4803  &  558  &  55734  &  2.454  &  0.6744  &  BAL quasar\\
119 &  J152316.86+091030.01  &  5490  &  862  &  56003  &  2.378  &  0.7573  &  False identification of \usmg\ \\
120 &  J153635.03+060412.68  &  4885  &  62   &  55735  &  2.478  &  0.6126  &  BAL quasar\\
121 &  J154000.99+070854.50  &  5211  &  324  &  56002  &  2.122  &  0.6731  &  BAL quasar\\
122 &  J154143.28+021536.37  &  4054  &  977  &  55358  &  2.267  &  0.7847  &  BAL quasar\\
123 &  J155625.89+045248.80  &  4808  &  509  &  55705  &  2.700  &  0.7174  &  BAL quasar\\
124 &  J160759.22+054038.63  &  4895  &  378  &  55708  &  2.446  &  0.7966  &  BAL quasar\\
125 &  J212521.07+001054.41  &  4193  &  778  &  55476  &  2.150  &  0.6385  &  BAL quasar\\
126 &  J214015.57+012024.17  &  5145  &  617  &  55835  &  2.267  &  0.7750  &  BAL quasar\\
127 &  J215002.02-010938.48  &  4197  &  402  &  55479  &  2.426  &  0.6496  &  Poor SNR. Blended lines detected as \usmg .\\
128 &  J215217.10+042603.13  &  4096  &  450  &  55501  &  2.596  &  0.6854  &  False identification of \usmg \\
129 &  J215422.57-003253.53  &  4197  &  154  &  55479  &  2.833  &  0.7280  &  False identification of \usmg \\
130 &  J221820.92+044944.33  &  4319  &  646  &  55507  &  2.193  &  0.6772  &  False identification of \usmg \\
131 &  J222312.24+094106.93  &  5052  &  154  &  55857  &  2.609  &  0.6639  &  False identification of \usmg \\
132 &  J222519.53+004523.99  &  4202  &  718  &  55445  &  2.132  &  0.6835  &  BAL quasar\\
133 &  J222635.48+093117.81  &  5053  &  452  &  56213  &  2.364  &  0.7261  &  BAL quasar\\
134 &  J224028.14-003813.17  &  4204  &  232  &  55470  &  0.658  &  0.6520  &  BAL quasar\\
135 &  J224101.17+075224.97  &  5059  &  537  &  56190  &  2.176  &  0.7040  &  BAL quasar\\
136 &  J225211.14+061927.72  &  4412  &  886  &  55912  &  2.224  &  0.7195  &  BAL quasar\\
137 &  J225251.12+000501.20  &  4206  &  738  &  55471  &  2.130  &  0.6322  &  BAL quasar\\
138 &  J225750.19+032623.61  &  4293  &  124  &  55509  &  2.563  &  0.6714  &  False identification of \usmg \\
139 &  J230027.63-022610.26  &  4362  &  280  &  55828  &  2.163  &  0.6246  &  BAL quasar\\
140 &  J230143.68-003909.52  &  4208  &  484  &  55476  &  2.276  &  0.6663  &  False identification of \usmg \\
141 &  J231014.92+070746.11  &  6168  &  350  &  56187  &  2.388  &  0.6741  &  False identification of \usmg \\
142 &  J231155.29+090153.81  &  6163  &  496  &  56219  &  2.222  &  0.6973  &  False identification of \usmg \\
143 &  J231355.83-022039.10  &  4360  &  360  &  55539  &  2.275  &  0.7483  &  BAL quasar\\
144 &  J232107.95-002756.62  &  4210  &  196  &  55444  &  2.285  &  0.6494  &  False identification of \usmg \\
145 &  J232117.95+035840.72  &  4285  &  734  &  55881  &  2.119  &  0.6089  &  BAL quasar\\
146 &  J232339.29+045126.75  &  4285  &  870  &  55881  &  2.891  &  0.7550  &  Blended lines detected as \usmg \\
147 &  J233739.99-002701.16  &  4213  &  396  &  55449  &  2.510  &  0.6502  &  BAL quasar\\
148 &  J234603.48-000258.01  &  4213  &  11   &  55449  &  2.081  &  0.6666  &  BAL quasar\\
149 &  J234626.48+052737.48  &  4406  &  854  &  55858  &  2.221  &  0.7471  &  BAL quasar\\
150 &  J235242.02-103837.72  &  7166  &  298  &  56602  &  2.919  &  0.7784  &  BAL quasar\\
151 &  J235639.31-040614.47  &  7034  &  506  &  56564  &  2.880  &  0.7709  &  Secure \usmg\ system\\

\hline
\end{tabular}
\end{table*}

\begin{table*}
\caption{Details of galaxies around the \usmg\ absorption systems in our sample with D $<$ 100 kpc with $m_r < 24$ }
    \centering
    \begin{tabular}{cccccccc}
    \hline
No. & Quasar Name & \zabs & Galaxy coordinates & $m_r$ & D(kpc) & Photo-z  & Comments \\
\hline

1 & J0020+0002 & 0.7699 & J002022.50+000235.52 & 23.19 & 31.5 & 0.619 $\pm$ 0.233 & Consistent\\
  &            &        & J002022.65+000230.68 & 23.39 & 9.7  & 1.152 $\pm$ 0.180 & Spectroscopically confirmed\\
  
2 & J0028+0041 & 0.6565 & J002839.78+004057.62 & 23.78 & 68.0 & 1.063 $\pm$ 0.148 & Inconsistent\\
  &            &        & J002838.78+004100.76 & 21.92 & 50.1 & 0.410 $\pm$ 0.110 & Inconsistent\\
  &            &        & J002839.05+004104.58 & 22.73 & 22.2 & 0.765 $\pm$ 0.128 & Spectroscopically confirmed\\
  &            &        & J002839.13+004049.56 & 23.55 & 94.5 & 0.812 $\pm$ 0.410 & Consistent\\
  &            &        & J002839.53+004115.93 & 23.39 & 94.7 & 0.599 $\pm$ 0.228 & Consistent\\

3 & J0033+0138 & 0.7187 & J003336.24+013844.75 & 20.96 & 50.3 & 0.784 $\pm$ 0.057 & Spectroscopically confirmed\\
  &            &        & J003335.83+013838.37 & 22.83 & 94.5 & 0.818 $\pm$ 0.057 & Inconsistent\\ 
  &            &        & J003335.90+013900.31 & 23.82 & 68.6 & 1.129 $\pm$ 0.653 & Consistent\\
  &            &        & J003335.90+013847.27 & 22.01 & 31.4 & 0.427 $\pm$ 0.077 & Inconsistent\\
  &            &        & J003336.11+013843.10 & 23.86 & 57.9 & 1.079 $\pm$ 0.216 & Inconsistent\\

4 & J0055-0100 & 0.7150 & J005554.95-010052.02 & 23.46 & 89.7 & 1.020 $\pm$ 0.242 & Inconsistent\\ 
  &            &        & J005554.28-010102.72 & 22.24 & 29.8 & 0.697 $\pm$ 0.032 & Spectroscopically confirmed\\ 
  &            &        & J005554.44-010056.53 & 23.96 & 25.7 & 0.962 $\pm$ 0.211 & Inconsistent\\ 

5 & J0105+0040 & 0.6489 & J010542.82+004011.05 & 23.78 & 87.9 & 0.981 $\pm$ 0.116 & Inconsistent\\
  &            &        & J010543.05+004016.30 & 23.52 & 99.0 & 0.635 $\pm$ 0.205 & Consistent\\
  &            &        & J010543.09+004005.21 & 22.27 & 44.8 & 0.578 $\pm$ 0.219 & Consistent\\
  &            &        & J010543.27+004002.93 & 22.45 & 25.9 & 1.033 $\pm$ 0.287 & Inconsistent\\
  &            &        & J010543.48+004013.75 & 23.04 & 68.6 & 0.926 $\pm$ 0.091 & Inconsistent\\
  &            &        & J010543.67+004001.10 & 21.79 & 24.6 & 0.658 $\pm$ 0.154 & Spectroscopically confirmed\\
  &            &        & J010543.97+003957.28 & 22.45 & 65.7 & 0.541 $\pm$ 0.251 & Spectroscopically confirmed\\
  &            &        & J010544.11+003953.44 & 22.52 & 94.7 & 0.627 $\pm$ 0.172 & Consistent\\

6 & J0127-0550 & 0.6837 & J012711.91-055016.45 & 22.17 & 90.1 & 0.733 $\pm$ 0.092 & Spectroscopically confirmed\\
  &            &        & J012711.20-055019.01 & 21.52 & 16.8 & 0.678 $\pm$ 0.073 & Spectroscopically confirmed\\ 
  &            &        & J012711.65-055029.19 & 22.48 & 81.6 & 0.647 $\pm$ 0.115 & Consistent\\ 

7 & J0142+0949 & 0.7858 & J014258.56+094942.20 & 22.52 & 30.0 & 0.905 $\pm$ 0.057 & Spectroscopically confirmed\\ 
  &            &        & J014258.58+094949.01 & 23.51 & 56.5 & 0.807 $\pm$ 0.103 &  Consistent\\ 
  &            &        & J014258.60+094936.54 & 23.85 & 50.7 & 0.886 $\pm$ 0.128 &  Consistent\\ 
  &            &        & J014259.09+094947.92 & 22.66 & 50.1 & 0.669 $\pm$ 0.470 &  Consistent\\

8 & J0150-0039 & 0.7747 & J015007.31-003935.11 & 23.95 & 68.5 & 1.024 $\pm$ 0.404 & Consistent \\
  &            &        & J015007.44-003931.92 & 21.66 & 65.5 & 0.694 $\pm$ 0.032 & Inconsistent\\
  &            &        & J015007.92-003935.52 & 23.71 & 11.7 & 0.916 $\pm$ 0.201 & Consistent\\
  &            &        & J015008.43-003931.82 & 23.41 & 70.0 & 1.057 $\pm$ 0.676 & Consistent\\
  &            &        & J015008.60-003929.18 & 23.68 & 96.5 & 1.134 $\pm$ 0.153 & Inconsistent\\
  
9 & J0150+0604 & 0.6707 & J015048.84+060443.27 & 22.67 & 95.4 & 0.783 $\pm$ 0.073 & Inconsistent\\ 
   &            &        & J015048.92+060442.74 & 22.33 & 87.4 & 0.563 $\pm$ 0.256 & Consistent\\

10 & J0256+0110 & 0.7256 & J025608.12+011034.20 & 23.98 & 99.3 & 0.967 $\pm$ 0.124 & Inconsistent\\
   &            &        & J025606.80+011038.87 & 22.95 & 48.8 & 0.680 $\pm$ 0.057 & Consistent\\
   &            &        & J025607.10+011039.80 & 22.18 & 18.4 & 0.727 $\pm$ 0.051 & Spectroscopically confirmed\\
   &            &        & J025607.52+011027.60 & 22.43 & 84.7 & 0.707 $\pm$ 0.032 & Consistent\\
   &            &        & J025607.91+011036.25 & 23.83 & 73.1 & 1.231 $\pm$ 0.602 & Consistent\\

11 & J0908+0727 & 0.6122 & J090805.91+072740.58 & 22.12 & 16.5 & 0.568 $\pm$ 0.068 & Spectroscopically confirmed\\ 
   &            &        & J090806.07+072732.04 & 23.57 & 61.6 & 0.658 $\pm$ 0.237 & Consistent\\ 

12 & J0956+0018 & 0.7821 & J095619.41+001802.00 & 22.50 & 15.6 & 0.619 $\pm$ 0.096 & Spectroscopically confirmed\\ 
   &            &        & J095619.67+001806.24 & 22.98 & 48.0 & 0.709 $\pm$ 0.170 & Consistent\\ 
   &            &        & J095619.92+001754.41 & 21.87 & 65.0 & 0.310 $\pm$ 0.146 & Inconsistent\\ 

13 & J1033+0128 & 0.6709 & J103325.50+012839.08 & 21.63 & 48.5 & 1.030 $\pm$ 0.294 & Inconsistent\\
   &            &        & J103325.65+012839.82 & 23.80 & 38.0 & 0.937 $\pm$ 0.157 & Inconsistent\\
   &            &        & J103325.67+012835.11 & 21.79 & 27.7 & 0.635 $\pm$ 0.092 & Consistent\\
   &            &        & J103325.85+012833.25 & 23.14 & 23.3 & 0.619 $\pm$ 0.198 & Consistent\\

14 & J1046+0457 & 0.7848  & J104642.79+045719.64 & 22.93 & 92.6 & 0.607 $\pm$ 0.314 & Consistent\\ 
   &            &         & J104642.52+045739.54 & 21.54 & 60.0 & 0.501 $\pm$ 0.278 & Consistent\\

15 & J1116+0500 & 0.7208 & J111627.24+050050.21 & 23.74 & 43.6 & 0.687 $\pm$ 0.321 & Consistent\\
   &            &        & J111627.78+050050.44 & 23.84 & 14.6 & 0.928 $\pm$ 0.160 & Inconsistent\\

\hline
\end{tabular}
\end{table*}

\begin{table*}
\ContinuedFloat
\caption{Continued.}
\begin{tabular}{lccccccc}
\hline
No. & Quasar Name & \zabs & Galaxy coordinates & $m_r$ & D(kpc) & Photo-z  & Comments \\
\hline

16 & J1150+0900 & 0.7569 & J115026.23+090042.52 & 22.93 & 45.2 & 1.124 $\pm$ 0.498 & Consistent\\

17 & J1201+0713 & 0.6842 & J120139.96+071328.30 & 20.86 & 81.4 & 0.467 $\pm$ 0.054 & Inconsistent\\
   &            &        & J120139.11+071344.91 & 23.19 & 67.7 & 0.813 $\pm$ 0.184 & Consistent\\
   &            &        & J120139.41+071342.84 & 21.67 & 36.6 & 0.653 $\pm$ 0.089 & Spectroscopically confirmed\\
   &            &        & J120139.77+071333.28 & 21.75 & 41.3 & 0.697 $\pm$ 0.049 & Spectroscopically confirmed\\

18 & J1217-0115 & 0.6642 & J121727.02-011553.88 & 19.87 & 90.2 & 0.308 $\pm$ 0.035 & Inconsistent\\
   &            &        & J121727.21-011556.15 & 23.65 & 81.2 & 1.033 $\pm$ 0.431 & Consistent\\
   &            &        & J121727.78-011558.88 & 19.42 & 72.1 & 0.070 $\pm$ 0.069 & Inconsistent\\
   &            &        & J121727.96-011551.50 & 22.97 & 26.1 & 0.800 $\pm$ 0.108 & Inconsistent\\
   &            &        & J121728.19-011545.93 & 23.97 & 44.9 & 1.110 $\pm$ 0.351 & Inconsistent\\
   &            &        & J121728.60-011550.74 & 23.33 & 85.0 & 0.945 $\pm$ 0.120 & Inconsistent\\

19 & J1322-0107 & 0.7226 & J132200.59-010753.28 & 22.92 & 28.2 & 0.526 $\pm$ 0.144 & Inconsistent \\
   &            &        & J132200.90-010748.64  & 22.14 & 52.3 & 0.266 $\pm$ 0.227 & Inconsistent \\

20 & J1336+0922 & 0.7059 & J133653.38+092229.76 & 23.54 & 71.6 & 0.960 $\pm$ 0.132 & Inconsistent\\
   &            &        & J133653.59+092215.03 & 22.71 & 46.9 & 0.828 $\pm$ 0.097 & Inconsistent\\
   &            &        & J133653.80+092217.96 & 22.69 & 24.6 & 0.701 $\pm$ 0.128 & Spectroscopically confirmed\\

21 & J1400-0149 & 0.7928 & J140016.86-014859.22 & 23.74 & 96.3 & 0.849 $\pm$ 0.474 & Consistent\\
   &            &        & J140017.98-014853.62 & 23.95 & 73.2 & 0.842 $\pm$ 0.348 & Consistent\\

22 & J1419+0346 & 0.7250 & J141929.42+034645.82 & 23.63 & 73.7 & 1.020 $\pm$ 0.160 & Inconsistent\\
   &            &        & J141929.46+034648.31 & 23.54 & 75.7 & 0.930 $\pm$ 0.158 & Inconsistent\\
   &            &        & J141929.62+034641.15 & 23.37 & 54.2 & 0.894 $\pm$ 0.082 & Inconsistent\\
   &            &        & J141930.39+034639.86 & 21.82 & 42.8 & 0.767 $\pm$ 0.073 & Spectroscopically confirmed\\

23 & J1449-0116 & 0.6621 & J144935.43-011652.92 & 23.88 & 80.7 & 1.116 $\pm$ 0.230 & Inconsistent\\
   &            &        & J144935.61-011643.87 & 23.88 & 75.9 & 0.981 $\pm$ 0.115 & Inconsistent\\
   &            &        & J144935.69-011702.53 & 23.45 & 99.1 & 0.845 $\pm$ 0.202 & Consistent\\
   &            &        & J144935.92-011637.35 & 22.08 & 95.7 & 0.408 $\pm$ 0.098 & Inconsistent\\
   &            &        & J144936.20-011643.58 & 21.85 & 48.1 & 0.559 $\pm$ 0.086 & Spectroscopically confirmed\\
   &            &        & J144936.52-011652.65 & 23.64 & 37.9 & 0.615 $\pm$ 0.227 & Consistent\\
   &            &        & J144937.02-011646.41 & 22.79 & 92.1 & 0.539 $\pm$ 0.258 & Consistent\\

24 & J1451-0138 & 0.7408 & J145108.15-013836.17 & 21.92 & 47.9 & 0.480 $\pm$ 0.063 & Inconsistent\\
   &            &        & J145108.24-013840.64 & 21.59 & 64.1 & 0.811 $\pm$ 0.058 & Spectroscopically confirmed\\
   &            &        & J145108.52-013841.80 & 19.30 & 63.8 & 0.191 $\pm$ 0.035 & Inconsistent\\
   &            &        & J145108.62-013831.19 & 20.76 & 16.7 & 0.248 $\pm$ 0.164 & Inconsistent\\
   &            &        & J145109.00-013831.22 & 22.62 & 53.2 & 0.558 $\pm$ 0.088 & Inconsistent\\
   &            &        & J145109.35-013837.29 & 22.71 & 95.0 & 0.574 $\pm$ 0.235 & Consistent\\

25 & J2356-0406 & 0.7059 & J235639.14-040618.27 & 23.52 & 33.9 & 1.123 $\pm$ 0.319 & Inconsistent\\
   &            &        & J235639.17-040608.35 & 23.43 & 47.9 & 1.058 $\pm$ 0.186 & Inconsistent\\
   
\hline
\end{tabular}
\label{tab:completeness}
\end{table*}

\newpage
\section{Detection of \usmg\ host galaxies based on the \CaII\ absorption}

\begin{figure*}
  \begin{subfigure}{0.475\textwidth}
    \centering\includegraphics[width=\textwidth]{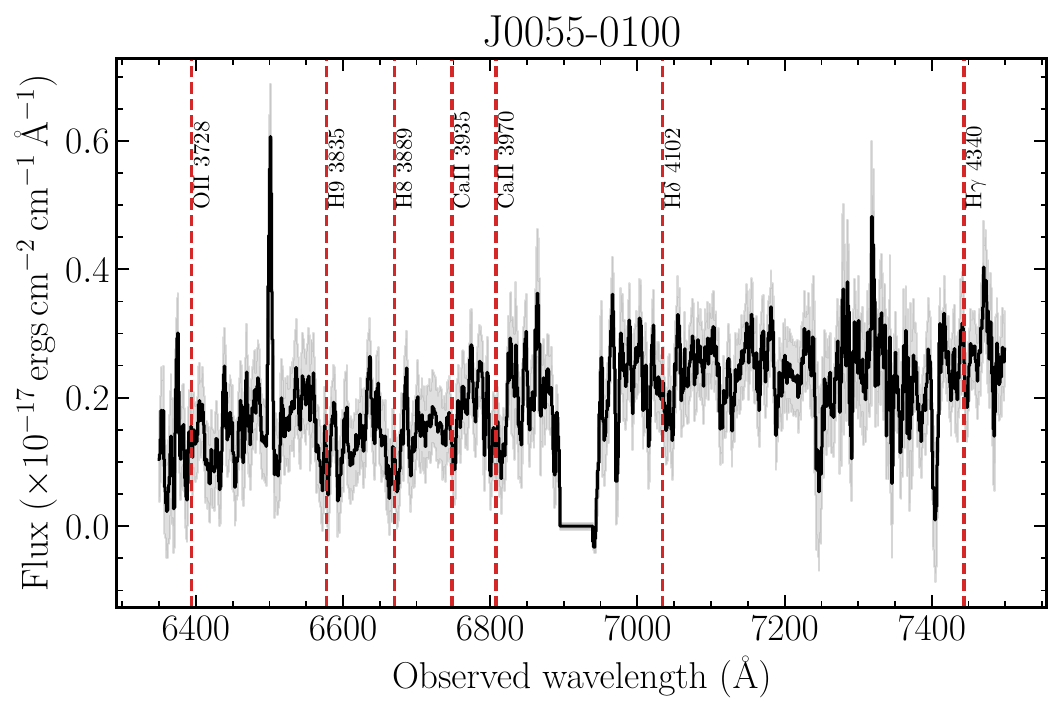}
  \end{subfigure}
  \begin{subfigure}{0.475\textwidth}
    \centering\includegraphics[width=\textwidth]{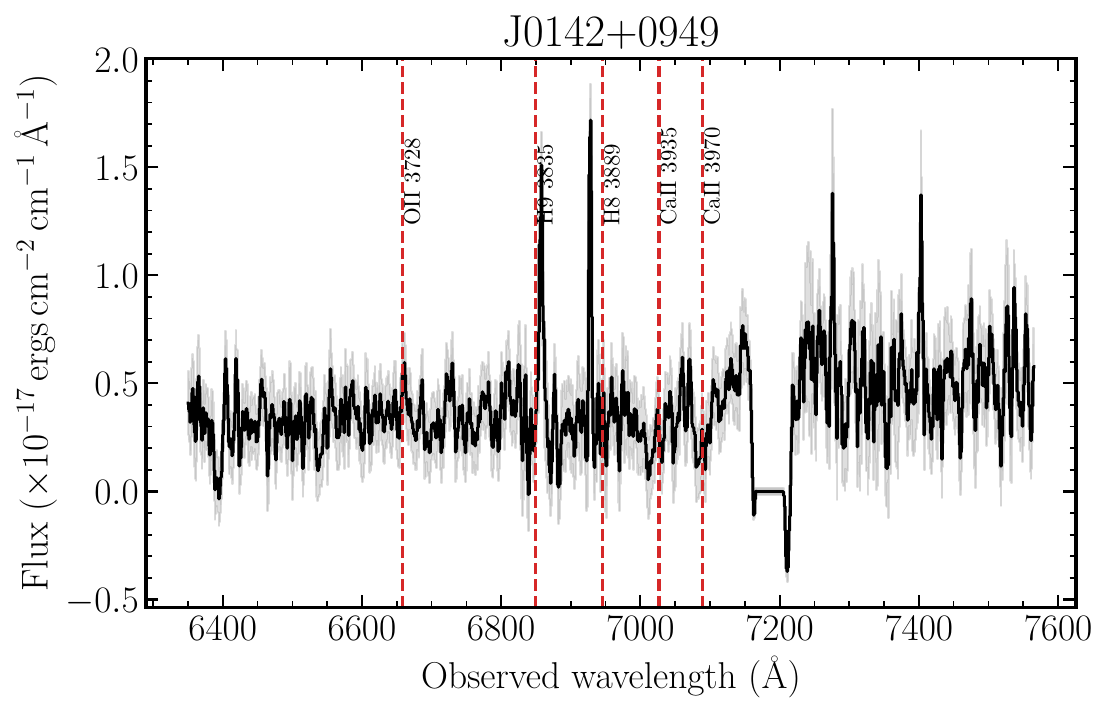}
  \end{subfigure}

  \caption{The spectra of the \usmg\ host galaxies, J0055$-$0100 and J0142$+$0949. Both of them are smoothed by 5 pixels, and both of them are identified based on the \CaII\ absorption present in their spectra. The vertical red dashed lines correspond to the Hydrogen Balmer lines and \CaII\ absorption lines corresponding to $z = 0.716$ (J0055$-$0100) and $z = 0.7858$ (J0142$+$0949).}
  \label{fig:gal_spec_caii}
\end{figure*}

As mentioned in Section 4.1, we identify the \usmg\ host galaxies based on the emission lines present in the spectra of the candidate galaxies except for two cases (J0055$-$0100 and J0142+0949). In Figure \ref{fig:gal_spec_caii}, we show the spectra of these two \usmg\ host galaxies and the \CaII\ absorption present in them.

\section{Misc}
\begin{figure*}
        \centering
        \begin{subfigure}[b]{0.475\textwidth}
            \centering
            \includegraphics[width=\textwidth]{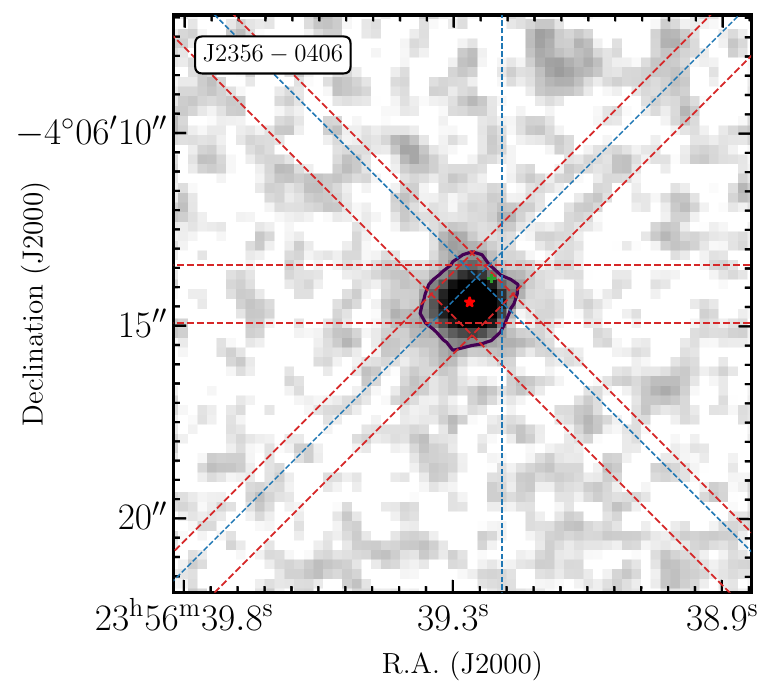}
        \end{subfigure}
        \hfill
        \begin{subfigure}[b]{0.475\textwidth}  
            \centering 
            \includegraphics[width=\textwidth]{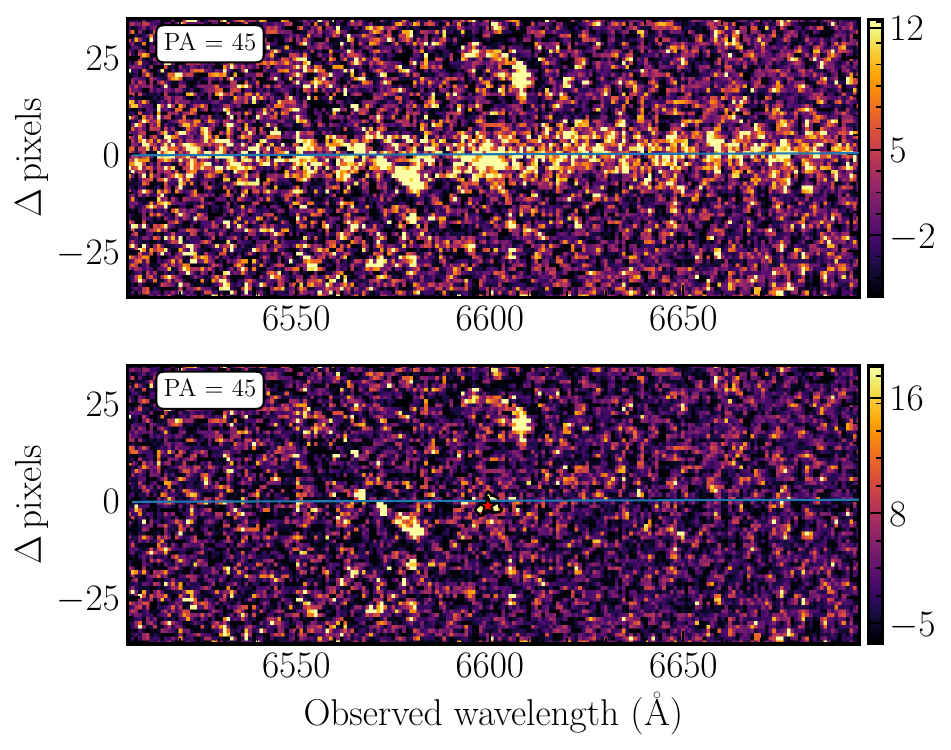}
        \end{subfigure}
        \vskip\baselineskip
        \begin{subfigure}[b]{0.475\textwidth}   
            \centering 
            \includegraphics[width=\textwidth]{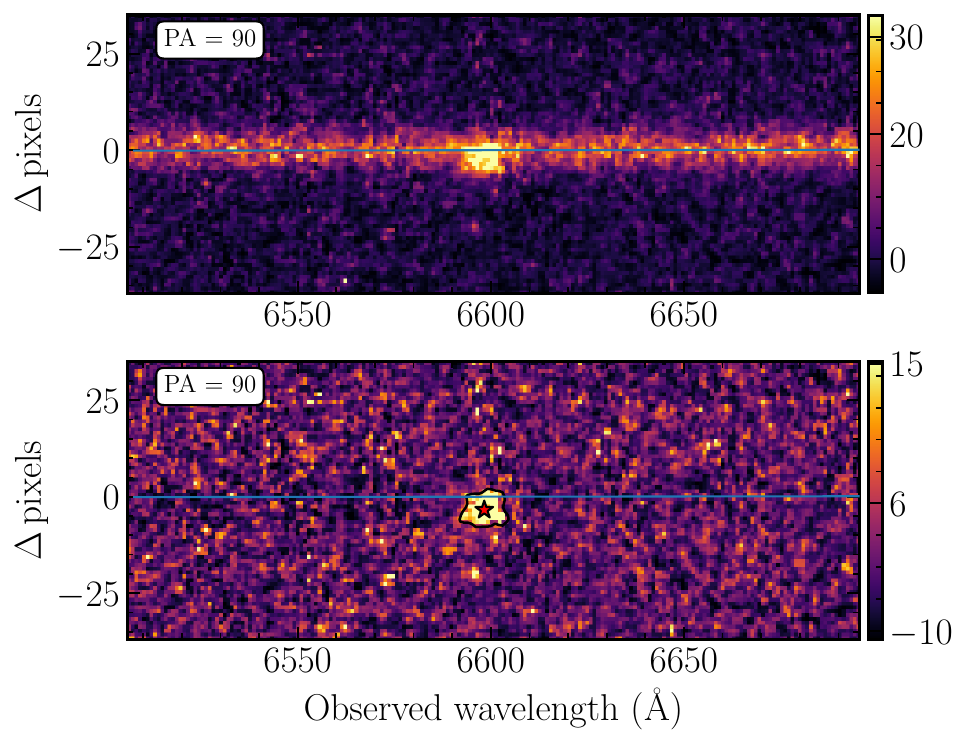}
        \end{subfigure}
        \hfill
        \begin{subfigure}[b]{0.475\textwidth}   
            \centering 
            \includegraphics[width=\textwidth]{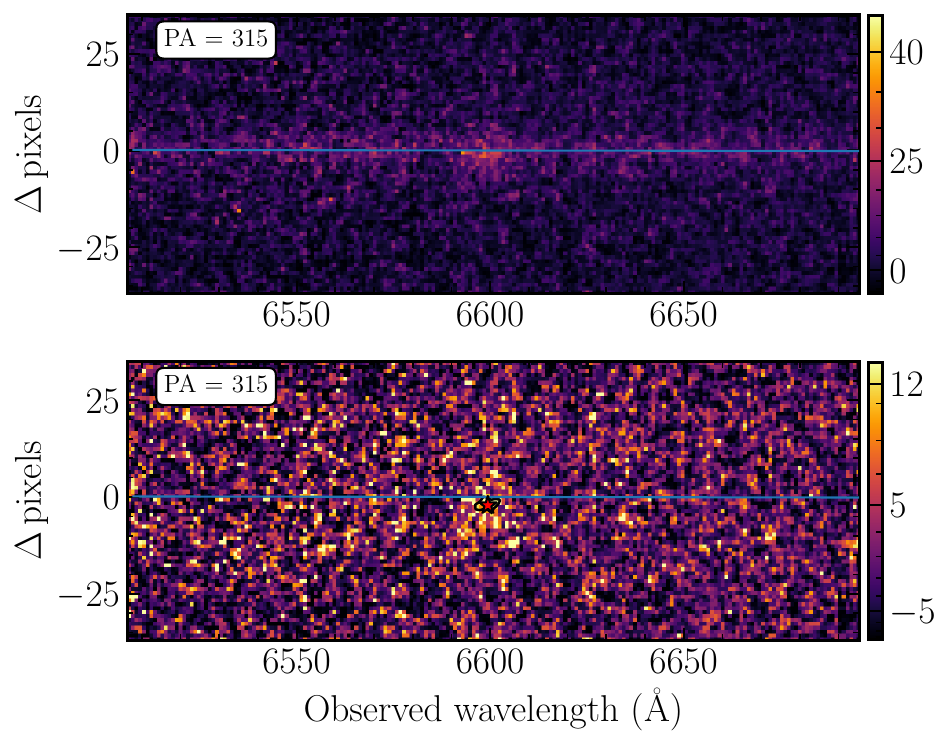}
        \end{subfigure}
        \caption{Illustration of the procedure to identify the impact parameter of the \usmg\ host galaxy towards the GOTOQ J2356$-$0406. The top left panel shows the zoomed-in observational configuration for the GOTOQ. The three pairs of red dashed lines show the three different slit configurations. The quasar centroid is marked with a red `$\star$'. The blue dashed lines correspond to the physical shift along the spatial axis of the slit of the \OII\ centroid from the central trace of the quasar. The most probable location of the \usmg\ host galaxy is marked with a green `+'. The top right, bottom left, and bottom right panels show the detection of \OII\ emission and the associated shift of the \OII\ emission from the central quasar trace. The central quasar trace is shown with a blue horizontal line, while the centroid of the \OII\ emission is shown with a red `$\star$' in each of these three panels.}
        \label{fig:J2356_rho}
\end{figure*}

\begin{figure}
    \centering
    \includegraphics[width=0.45\textwidth]{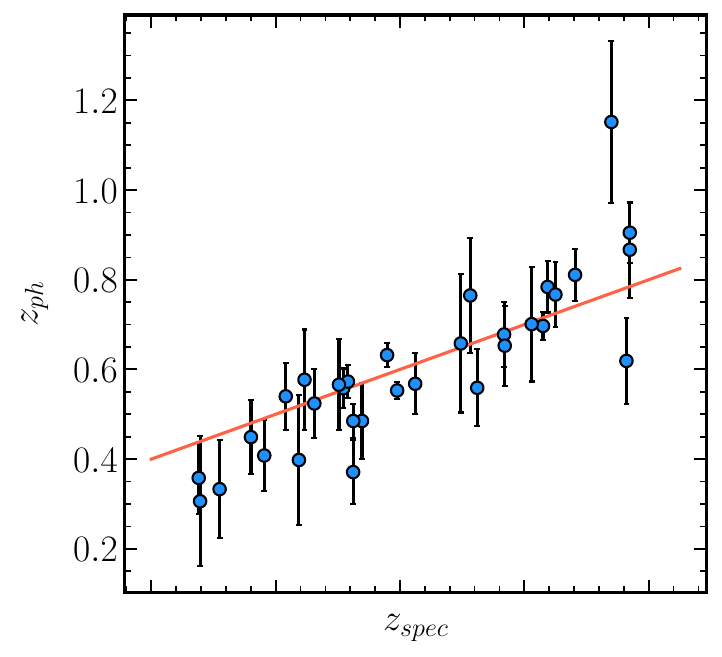}
    \caption{Comparison of the spectroscopic and photometric redshifts of the spectroscopically confirmed \usmg\ host galaxies. The orange line corresponds to the $z_{ph} = z_{sp}$}
    \label{fig:compare_photoz}
\end{figure}


\bsp	
\label{lastpage}
\end{document}